\documentclass[longauth]{aa}  

\usepackage{graphicx}
\usepackage{comment}

\usepackage{txfonts}
\usepackage{multirow}

\usepackage[colorlinks=true,linkcolor=blue]{hyperref}
 \hypersetup{
    colorlinks,
    linkcolor={blue},
    citecolor={blue},
    urlcolor={blue}
}
\newcommand{\orcid}[1]{\textsuperscript{\href{https://orcid.org/#1}{\includegraphics[width=8pt]{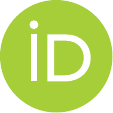}}}}
 \def\be{\begin{equation}}
\def\ee{\end{equation}}
 \def\papI{\href{https://academic.oup.com/mnras/article/487/4/5209/5498307}{Paper I}}
\def\papII{\href{https://academic.oup.com/mnras/article/487/4/5235/5513479}{Paper II}}
\def\papIII{\href{https://iopscience.iop.org/article/10.3847/1538-4357/ad9391}{Paper III}}
\def\Ims{\mathrm{Im}}
\def\Res{\mathrm{Re}}

\usepackage{amssymb}
\usepackage{mathrsfs}
\usepackage{lipsum, babel}
\usepackage{setspace}

\begin{document}

   \title{The DESI DR1 peculiar velocity survey: \\ 
   growth rate measurements from the galaxy 
power spectrum}

   \author{
   F.~Qin\inst{1}\orcid{0000-0001-7950-7864} \fnmsep\thanks{Corresponding author: \email{qin@cppm.in2p3.fr}}, 
C.~Blake	\inst{2},
C.~Howlett	\inst{3},
R. J.~Turner	\inst{2},
K.~Lodha	\inst{4,5},
J.~Bautista	\inst{1},
Y.~Lai	\inst{3,6},
A.~J.~Amsellem	\inst{7},
J.~Aguilar	\inst{8},
S.~Ahlen	\inst{9},
D.~Bianchi	\inst{10,11},
D.~Brooks	\inst{12},
S.~BenZvi	\inst{13},
A.~Carr	\inst{4},
E.~Chaussidon	\inst{8},
T.~Claybaugh	\inst{8},
A.~Cuceu	\inst{8},
A.~de la Macorra	\inst{14},
K.~Douglass	\inst{13},
P.~Doel	\inst{12},
S.~Ferraro	\inst{8,15},
A.~Font-Ribera	\inst{16},
J.~E.~Forero-Romero	\inst{17,18},
E.~Gaztañaga	\inst{19,20,21},
S.~Gontcho A Gontcho	\inst{8,22},
G.~Gutierrez	\inst{23},
J.~Guy	\inst{8},
H.~K.~Herrera-Alcantar	\inst{24,25},
K.~Honscheid	\inst{26,27,28},
D.~Huterer	\inst{29,30},
M.~Ishak	\inst{31},
R.~Joyce	\inst{32},
A.~G.~Kim	\inst{8},
D.~Kirkby	\inst{33},
T.~Kisner	\inst{8},
A.~Kremin	\inst{8},
O.~Lahav	\inst{12},
C.~Lamman	\inst{28},
M.~Landriau	\inst{8},
L.~Le~Guillou	\inst{34},
M.~E.~Levi	\inst{8},
M.~Manera	\inst{35,16},
A.~Meisner	\inst{32},
R.~Miquel	\inst{36,16},
J.~Moustakas	\inst{37},
A.~Muñoz-Gutiérrez	\inst{38},
S.~Nadathur	\inst{20},
N.~Palanque-Delabrouille	\inst{25,8},
W.~J.~Percival	\inst{39,40,41},
C.~Poppett	\inst{8,42,15},
F.~Prada	\inst{43},
I.~P\'erez-R\`afols	\inst{44},
C.~Ross	\inst{3},
G.~Rossi	\inst{45},
E.~Sanchez	\inst{46},
D.~Schlegel	\inst{8},
K.~Said	\inst{3},
M.~Schubnell	\inst{29,30},
H.~Seo	\inst{47},
J.~Silber	\inst{8},
D.~Sprayberry	\inst{32},
G.~Tarl\'{e}	\inst{30},
B.~A.~Weaver	\inst{32},
P.~Zarrouk	\inst{34},
R.~Zhou	\inst{8},
H.~Zou	\inst{48}  
}

   \institute{
   Aix-Marseille University, CNRS/IN2P3, CPPM, Marseille 13288, France. \and
Centre for Astrophysics \& Supercomputing, Swinburne University of Technology, P.O. Box 218, Hawthorn, VIC 3122, Australia	. \and
School of Mathematics and Physics, University of Queensland, Brisbane, QLD 4072, Australia	. \and
Korea Astronomy and Space Science Institute, 776, Daedeokdae-ro, Yuseong-gu, Daejeon 34055, Republic of Korea	. \and
University of Science and Technology, 217 Gajeong-ro, Yuseong-gu, Daejeon 34113, Republic of Korea	. \and
Steward Observatory, University of Arizona, 933 N. Cherry Avenue, Tucson, AZ 85721, USA	. \and
Department of Physics, Carnegie Mellon University, 5000 Forbes Avenue, Pittsburgh, PA 15213, USA\and
Lawrence Berkeley National Laboratory, 1 Cyclotron Road, Berkeley, CA 94720, USA	. \and
Department of Physics, Boston University, 590 Commonwealth Avenue, Boston, MA 02215 USA	. \and
Dipartimento di Fisica ``Aldo Pontremoli'', Universit\`a degli Studi di Milano, Via Celoria 16, I-20133 Milano, Italy	. \and
INAF-Osservatorio Astronomico di Brera, Via Brera 28, 20122 Milano, Italy	. \and
Department of Physics \& Astronomy, University College London, Gower Street, London, WC1E 6BT, UK	. \and
Department of Physics \& Astronomy, University of Rochester, 206 Bausch and Lomb Hall, P.O. Box 270171, Rochester, NY 14627-0171, USA	. \and
Instituto de F\'{\i}sica, Universidad Nacional Aut\'{o}noma de M\'{e}xico,  Circuito de la Investigaci\'{o}n Cient\'{\i}fica, Ciudad Universitaria, Cd. de M\'{e}xico  C.~P.~04510,  M\'{e}xico	. \and
University of California, Berkeley, 110 Sproul Hall \#5800 Berkeley, CA 94720, USA	. \and
Institut de F\'{i}sica d’Altes Energies (IFAE), The Barcelona Institute of Science and Technology, Edifici Cn, Campus UAB, 08193, Bellaterra (Barcelona), Spain	. \and
Departamento de F\'isica, Universidad de los Andes, Cra. 1 No. 18A-10, Edificio Ip, CP 111711, Bogot\'a, Colombia	. \and
Observatorio Astron\'omico, Universidad de los Andes, Cra. 1 No. 18A-10, Edificio H, CP 111711 Bogot\'a, Colombia	. \and
Institut d'Estudis Espacials de Catalunya (IEEC), c/ Esteve Terradas 1, Edifici RDIT, Campus PMT-UPC, 08860 Castelldefels, Spain	. \and
Institute of Cosmology and Gravitation, University of Portsmouth, Dennis Sciama Building, Portsmouth, PO1 3FX, UK	. \and
Institute of Space Sciences, ICE-CSIC, Campus UAB, Carrer de Can Magrans s/n, 08913 Bellaterra, Barcelona, Spain	. \and
University of Virginia, Department of Astronomy, Charlottesville, VA 22904, USA	. \and
Fermi National Accelerator Laboratory, PO Box 500, Batavia, IL 60510, USA	. \and
Institut d'Astrophysique de Paris. 98 bis boulevard Arago. 75014 Paris, France	. \and
IRFU, CEA, Universit\'{e} Paris-Saclay, F-91191 Gif-sur-Yvette, France	. \and
Center for Cosmology and AstroParticle Physics, The Ohio State University, 191 West Woodruff Avenue, Columbus, OH 43210, USA \and
Department of Physics, The Ohio State University, 191 West Woodruff Avenue, Columbus, OH 43210, USA	. \and
The Ohio State University, Columbus, 43210 OH, USA	. \and
Department of Physics, University of Michigan, 450 Church Street, Ann Arbor, MI 48109, USA	. \and
University of Michigan, 500 S. State Street, Ann Arbor, MI 48109, USA	. \and
Department of Physics, The University of Texas at Dallas, 800 W. Campbell Rd., Richardson, TX 75080, USA	. \and
NSF NOIRLab, 950 N. Cherry Ave., Tucson, AZ 85719, USA	. \and
Department of Physics and Astronomy, University of California, Irvine, 92697, USA	. \and
Sorbonne Universit\'{e}, CNRS/IN2P3, Laboratoire de Physique Nucl\'{e}aire et de Hautes Energies (LPNHE), FR-75005 Paris, France	. \and
Departament de F\'{i}sica, Serra H\'{u}nter, Universitat Aut\`{o}noma de Barcelona, 08193 Bellaterra (Barcelona), Spain	. \and
Instituci\'{o} Catalana de Recerca i Estudis Avan\c{c}ats, Passeig de Llu\'{\i}s Companys, 23, 08010 Barcelona, Spain	. \and
Department of Physics and Astronomy, Siena University, 515 Loudon Road, Loudonville, NY 12211, USA	. \and
Instituto de F\'{\i}sica, Universidad Nacional Aut\'{o}noma de M\'{e}xico,  Circuito de la Investigaci\'{o}n Cient\'{\i}fica, Ciudad Universitaria, Cd. de M\'{e}xico  C.~P.~04510,  M\'{e}xico	. \and
Department of Physics and Astronomy, University of Waterloo, 200 University Ave W, Waterloo, ON N2L 3G1, Canada	. \and
Perimeter Institute for Theoretical Physics, 31 Caroline St. North, Waterloo, ON N2L 2Y5, Canada	. \and
Waterloo Centre for Astrophysics, University of Waterloo, 200 University Ave W, Waterloo, ON N2L 3G1, Canada	. \and
Space Sciences Laboratory, University of California, Berkeley, 7 Gauss Way, Berkeley, CA  94720, USA	. \and
Instituto de Astrof\'{i}sica de Andaluc\'{i}a (CSIC), Glorieta de la Astronom\'{i}a, s/n, E-18008 Granada, Spain	. \and
Departament de F\'isica, EEBE, Universitat Polit\`ecnica de Catalunya, c/Eduard Maristany 10, 08930 Barcelona, Spain	. \and
Department of Physics and Astronomy, Sejong University, 209 Neungdong-ro, Gwangjin-gu, Seoul 05006, Republic of Korea	. \and
CIEMAT, Avenida Complutense 40, E-28040 Madrid, Spain	. \and
Department of Physics \& Astronomy, Ohio University, 139 University Terrace, Athens, OH 45701, USA	. \and
National Astronomical Observatories, Chinese Academy of Sciences, A20 Datun Road, Chaoyang District, Beijing, 100101, P.~R.~China	.   \\
   }

   \date{Received November 15, 2025; accepted Feburuary 16, 2026}

  \abstract
  { \normalsize  The large-scale structure of the Universe and its evolution encapsulate a wealth of cosmological information. A powerful means of unlocking this knowledge lies in measuring the auto-power spectrum and/or the cross-power spectrum of the galaxy density and momentum fields, followed by the estimation of cosmological parameters based on these spectrum measurements. In this study, we generalize the cross-power spectrum model to accommodate scenarios where the density and momentum fields are derived from distinct galaxy surveys. The growth rate of the large-scale structures of the Universe, commonly represented as $f\sigma_8$, is extracted by jointly fitting the monopole and quadrupole moments of the auto-density power spectrum, the monopole of the auto-momentum power spectrum, and the dipole of the cross-power spectrum. Our estimators, theoretical models and parameter-fitting framework have been tested using mocks, confirming their robustness and accuracy in retrieving the fiducial growth rate from simulation. These techniques are then applied to analyze the power spectrum of the DESI Bright Galaxy Survey and Peculiar Velocity Survey, and
  the fit result of the growth rate is $f\sigma_8=0.440^{+0.080}_{-0.096}$ at effective redshift $z_{\rm eff}=0.07$.  
 By synthesizing the fitting outcomes from correlation functions, maximum likelihood estimation and power spectrum,  yields a consensus value of $f\sigma_8(z_{\rm eff}=0.07) = 0.450 ^{+0.055}_{-0.055}$, and correspondingly we obtain  $\gamma=0.580^{+0.110}_{-0.110}$, $\Omega_\mathrm{m}=0.301^{+0.011}_{-0.011}$ and $\sigma_8=0.834^{+0.032}_{-0.032}$. The measured $f\sigma_8$ and $\gamma$ are consistent with the prediction of the $\Lambda$ Cold Dark Matter Model and General Relativity.   
 The public code of power spectrum measurements, power spectrum theoretical models and window function convolution are available at \url{https://github.com/FeiQin-cosmologist/Galaxy_Power_Spectrum}, and see Appendix \ref{sec:appsdf1233} for more links. The examples for the utilization of the code can be found in this Jupyter Notebook
\url{https://github.com/FeiQin-cosmologist/Galaxy_Power_Spectrum/blob/main/CosmPSPy/Code/Examp_PS.ipynb}

  }

 \titlerunning{Growth Rate Measurements from the Galaxy 
Power Spectrum}
 \authorrunning{F. Qin et al.}
   \maketitle

\section{Introduction}
 
One of the most profound scientific pursuits in cosmology is to unravel the evolutionary processes that have shaped the large-scale structures of our Universe. Galaxies exhibit peculiar motions—superimposed on their Hubble recession velocities—due to gravitational fluctuations arising from local variations in the matter density field. These peculiar velocities serve as a crucial observational probe, offering deep insights into the underlying dark matter distribution and the dynamic history of cosmic expansion and structure formation. Consequently, they have garnered growing interest as an essential diagnostic tool for assessing the accuracy and predictive power of cosmological models.

An approach to assess the cosmological models involves measuring the low-order kinematic moments of the cosmic flow field based on the radial components of galaxies’ peculiar velocities, and subsequently comparing these estimates with the predictions of the cosmological models 
\citep{Kaiser1988CosFlow,Jaffe1995,Strauss1995,Parnovsky2001,Watkins2009,Feldman2010,Hong2014,Nusser2014,Scrimgeour2016,Zhang2017,Peery2018,Qin2018,Qin2019CosFlow,Qin2021CosFlowBoxCox,Qin2021CosFlowCF4,Whitford2023,Watkins2023,Heinesen2023,Lopes2024}. However, due to the significant cosmic variance inherent in low-redshift velocity surveys, the constraints imposed on cosmological models through this method are typically not tight.

An alternative approach involves estimating cosmological parameters and subsequently comparing them with predictions of cosmological models. In the field of cosmology, researchers commonly aim to investigate the rate at which large-scale structures evolve, as well as the ratio between galaxy density and dark matter density. The former quantity is associated with the structure linear growth rate $f$, while the latter relates to the galaxy biasing parameter $b$. Furthermore, in practical applications, normalized parameters $f\sigma_8$ and $b\sigma_8$ have attracted attention as measurable quantities. $\sigma_8$ refers to the root mean square of mass density fluctuations within spheres of 8 Mpc $h^{-1}$. Various methodologies have been proposed in prior studies to extract these parameters from galaxy survey data \citep{Ishak2019,Ishak2025}.

One technique involves the reconstruction of the cosmological density and velocity fields 
\citep{Nusser1991,Zaroubi1995,Croft1997,Kudlicki2000,Branchini2002,Pike2005,Erdogdu2006, Kitaura2012,Wang2012Recons,ShiFeng2018}, or the cosmography of the local Universe \citep{Courtois2013,Dupuy2023}. The parameters $f\sigma_8$ and $b\sigma_8$ can be estimated by comparing the observed peculiar velocities of galaxies with the reconstructed velocity field 
\citep{Springob2014,Carrick2015, Ma2012,Said2020,Boruah2020,Boruah2021,Boruah2022,Hollinger2024}. However, systematic errors associated with this method are difficult to quantify, and previous studies have indicated that the errors tend to be underestimated \citep{Turner2023Recons,Blake2024recons}. Recently, artificial intelligence techniques have been increasingly applied to field reconstruction  
\citep{Sungwook2021,Mao2021,WU2021AI,Qin2023Recon,Ganeshaiah2023,Wu2023AI,Lilow2024,yuyuWang2024,Shi2025}, offering novel opportunities in this research area. Moreover, since the covariance matrix of the density and velocity fields can be formulated in terms of $f\sigma_8$ and $b\sigma_8$, these parameters can alternatively be estimated by maximizing the likelihood function under the assumption of a Gaussian distribution  
\citep{Johnson2014, Adams2017, Lai2023,Carreres2023,Rocher2023,Ravoux2025}. However, this method remains computationally demanding. Another commonly used approach is the two-point statistic method, which has played a significant role in cosmology over the past decades. Based on this method, $f\sigma_8$ and $b\sigma_8$ can be derived from the galaxy two-point correlation functions, the velocity-correlation functions and the galaxy-velocity cross-correlation functions  
\citep{Kaiser1984correlation,Gorski1989,Howlett2015correlationfunction,Dupuy2019,YuyuWang2018,YuyuWang2021,Turner2021,Turner2023correlation,Qin2022,Qin2023HaloProf,Lyall2023,Lyall2024,Blake2024correlation,DESIPV_Nguyen,DESIPV_Turner}. Nevertheless, these approaches tend to be computationally intensive too.

Instead of studying the two-point correlations in configuration space, we can also analyze them in Fourier space, i.e. analyzing the so-called power spectrum. The concept of the cosmological power spectrum was first introduced by \cite{Kaiser1987}. The first measurement of the (galaxy) density power spectrum was carried out by \cite{ Kaiser1991PS}. The methodology for estimating the density power spectrum was later refined by \cite{Feldman1994}, forming the standard approach widely used today. In addition to these foundational contributions, \cite{Yamamoto2006} and \cite{Bianchi2015} further developed estimators for the multipoles of the density power spectrum.  Other notable applications of the density power spectrum include studies by \cite{Yamamoto2003,Sanchez2008,Blake2010,Blake2011,Beutler2012,Johnson2014,Scoccimarro2015,Blake2018,Beutler2019,Wang2018PSBOXCOX,Ivanov2020,Beutler2021}.

However, the linear growth rate of the structure,  $f\sigma_8$ is only primarily sensitive to higher-order components of the density power spectrum, resulting in relatively weak constraints when inferred from this method alone. To improve the estimation of $f\sigma_8$, the momentum power spectrum—initially explored in \cite{Park2000}—has recently gained widespread use in cosmology. \cite{Howlett2019} (here after \papI) adapted the momentum power spectrum method to be implemented in a manner analogous to the conventional density power spectrum approach. \cite{Qin2019PS} (here after \papII) successfully applied the momentum power spectrum to constrain $f\sigma_8$ using data from the combined 6dFGS peculiar velocity survey (6dFGSv, \citealt{Campbell2014}) and 2MASS Tully-Fisher (2MTF, \citealt{Hong2019}) survey. Building upon their research, \cite{Qin2025} (here after \papIII) advanced the existing methodology by extending it to the cross-power spectrum of density and momentum fields—both derived from a unified survey catalog. Their work achieved a successful estimation of $f\sigma_8$ through the peculiar velocity data from the Sloan Digital Sky Survey peculiar velocity catalog (SDSSv, \citealt{Howlett2022}). This paper marks the fourth in the ongoing series. Herein, we further refine and expand our analytical framework to model and measure the cross-power spectrum in more complex cases, where the density and momentum fields are drawn from distinct galaxy survey catalogs—each exhibiting unique survey geometries, selection criteria, and sample characteristics. Additional studies employing the momentum power spectrum include those by \cite{Park2006,Appleby2023,Shi2024}.

We constrain $f\sigma_8$ using Data Release 1 (DR1) from the Bright Galaxy Survey  (BGS, \citealt{Hahn2023}) and  Peculiar Velocity Survey (DESI-PV, \citealt{DESIPV_Carr,DESIPV_Ross,DESIPV_Douglass})  of the Dark Energy Spectroscopic Instrument (DESI, \citealt{DESI2024.I.DR1,DESI.DR2.BAO.cosmo}). The DESI-PV DR1 stands as the largest peculiar velocity survey ever assembled to date. This paper forms part of a comprehensive series results from DESI-PV DR1 \citep{DESIPV_Turner,DESIPV_Lai,DESIPV_Ross,DESIPV_Douglass,DESIPV_Bautista,DESIPV_Carr,DESIPV_Nguyen}.

This paper is organized as follows. Section \ref{sec:datamock} presents the data and mocks employed for measuring the power spectrum. Section \ref{sec:psest} introduces the estimators utilized to compute the power spectrum. Section \ref{sec:psmodel} delves into the theoretical models of the power spectrum. In Section \ref{sec:testmock}, we outline the fitting methodologies and validate our estimators and models through extensive testing on mocks. Section \ref{sec:results} showcases the final parameter fitting results derived from the  survey data. In Section \ref{sec:discus}, we discuss the parameter constraints in light of predictions and fits from other data sets. Lastly, Section \ref{sec:conclustion} offers a comprehensive conclusion summarizing the key findings of the study.

In this paper, we will adopt the AbacusSummit  \citep{ABACUS2021}  cosmological  parameters\footnote{abacus\_cosm000: \url{https://abacussummit.readthedocs.io/en/latest/cosmologies.html}} $n_s = 0.96490$, $A_s=2.101198\times10^{-9} $,  $\Omega_m=0.315192$,  $\Omega_bh^2=0.02237$, $\Omega_{c}h^2=0.12000$,  $\sigma_8=0.8113545$ and   $H_0=100 h$ km s$^{-1}$ Mpc$^{-1}$, where $h=0.67360$. We adopt a flat $\Lambda$ Cold Dark Matter ($\Lambda$CDM) fiducial cosmological model.

\section{Data and mocks}\label{sec:datamock}
 
\subsection{The Dark Energy Spectroscopic Instrument}

Located at the Kitt Peak National Observatory in Arizona, USA, the Dark Energy Spectroscopic Instrument is a multi-fiber spectrograph mounted on the 4-meter Mayall Telescope,  designed to explore the large-scale
structure of the universe and the
expansion history of the universe  \citep{Snowmass2013.Levi,DESI2016a.Science,DESI2016b.Instr,DESI2022.KP1.Instr,Corrector.Miller.2023,FiberSystem.Poppett.2024}. The DESI focal plane spans 8 square degrees and is equipped with 5,000 optical fibers, each of which can be positioned onto individual targets within an observational field using robotic positioners \citep{FocalPlane.Silber.2023}. 
The spectrographs span the near-ultraviolet to near-infrared spectrum (3,600 to 9,800 \AA), delivering a spectral resolution that  ascends from 2,000 in the blue camera to 5,000 in the red camera. The observing strategy is detailed in \cite{Spectro.Pipeline.Guy.2023} and \cite{SurveyOps.Schlafly.2023}. Following its Early Data Release (EDR, \citealt{DESI2023b.KP1.EDR,DESI2023a.KP1.SV}), the survey has now progressed to DR1  \citep{DESI2024.I.DR1,DESI2024.VII.KP7B}. For this study, the galaxy sample is drawn from the DESI DR1 catalog. We use the DESI DR1 large-scale structure catalog\footnote{\url{https://data.desi.lbl.gov/doc/releases/dr1/}} \citep{DESI2024.II.KP3}.

\subsubsection{The BGS data}

\begin{figure}
 \includegraphics[width=\columnwidth]{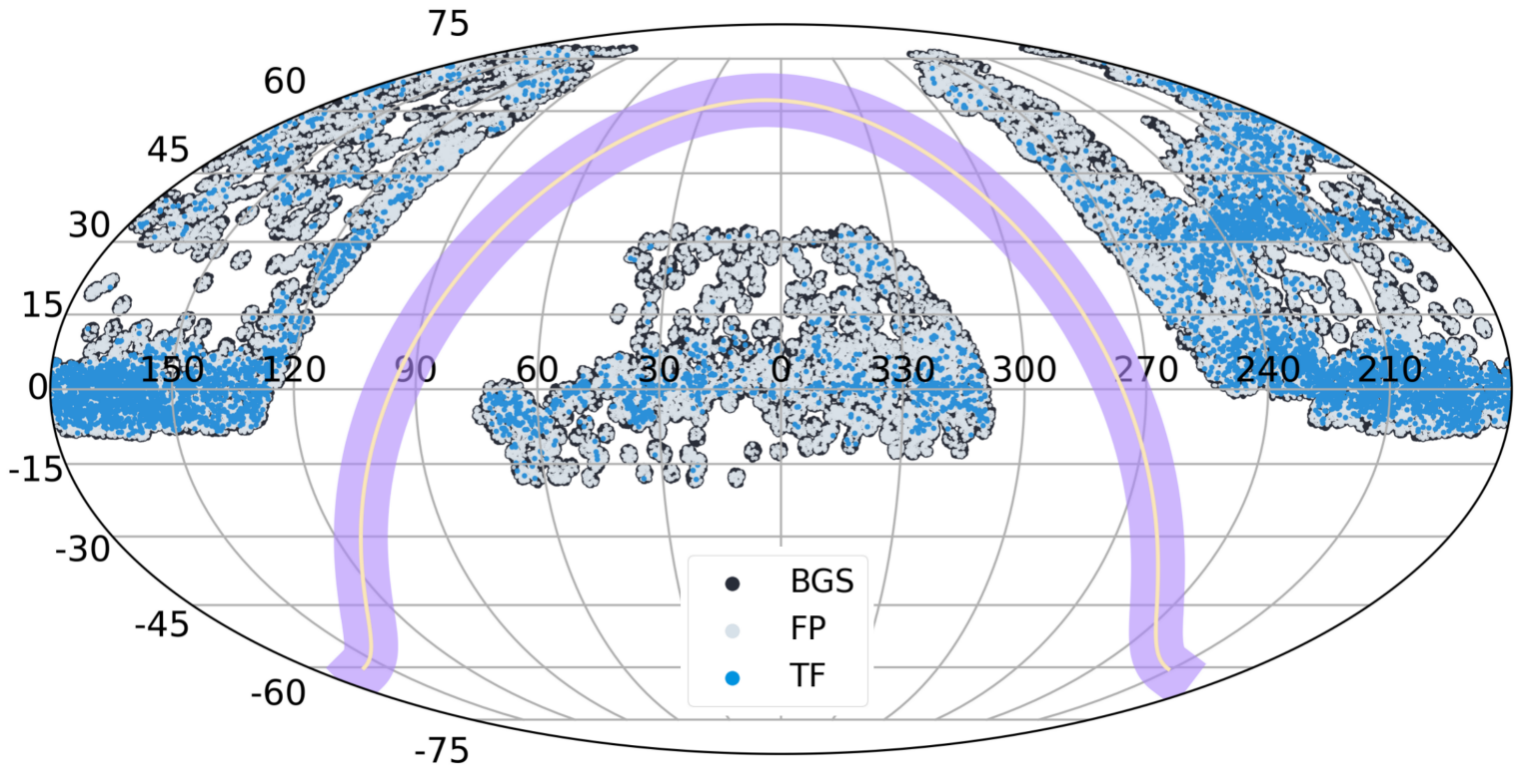}
 \caption{The sky coverage in equatorial coordinates of the galaxies analyzed within this study. The black dots delineate the sky distribution of BGS galaxies, the gray dots represent that of FP galaxies, and the blue dots illustrate the distribution of TF galaxies. The purple band signifies the galactic plane.
}
 \label{pltsky}
 \end{figure}

\begin{figure} 
\includegraphics[width=\columnwidth]{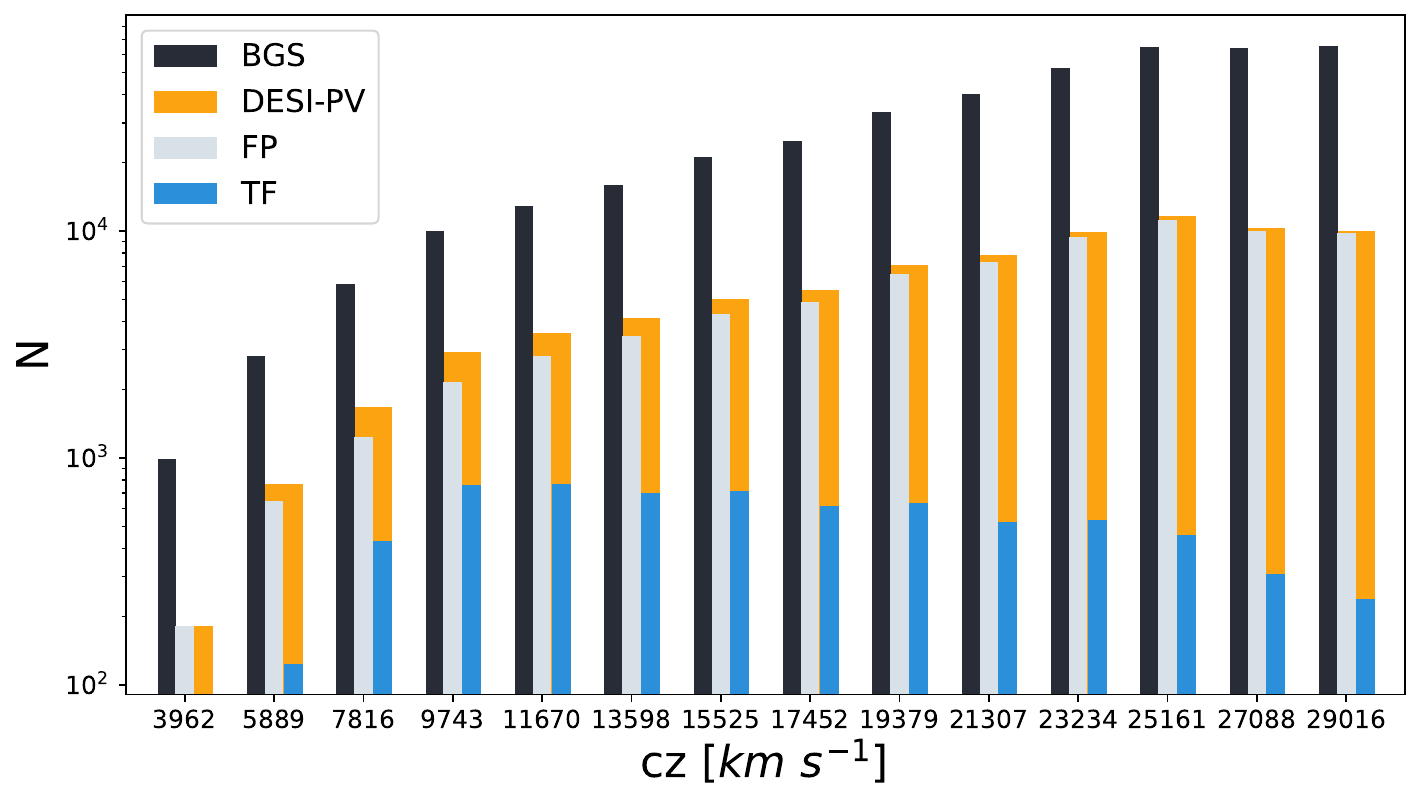}
 \caption{The redshift distribution   of the galaxies employed in this paper. The y-axis is presented on logarithmic scales to enhance the clarity of the data. The black bars depict the redshift distribution of the BGS galaxies, the gray bars delineate the redshift distribution of FP galaxies, and the blue bars illustrate the redshift distribution of TF galaxies. Additionally, the yellow bars represent the redshift distribution of the entire DESI-PV sample, which constitutes a combination of FP and TF galaxies. }
 \label{pltcz}
 \end{figure}

\begin{figure} 
\centering
 \includegraphics[width=\columnwidth]{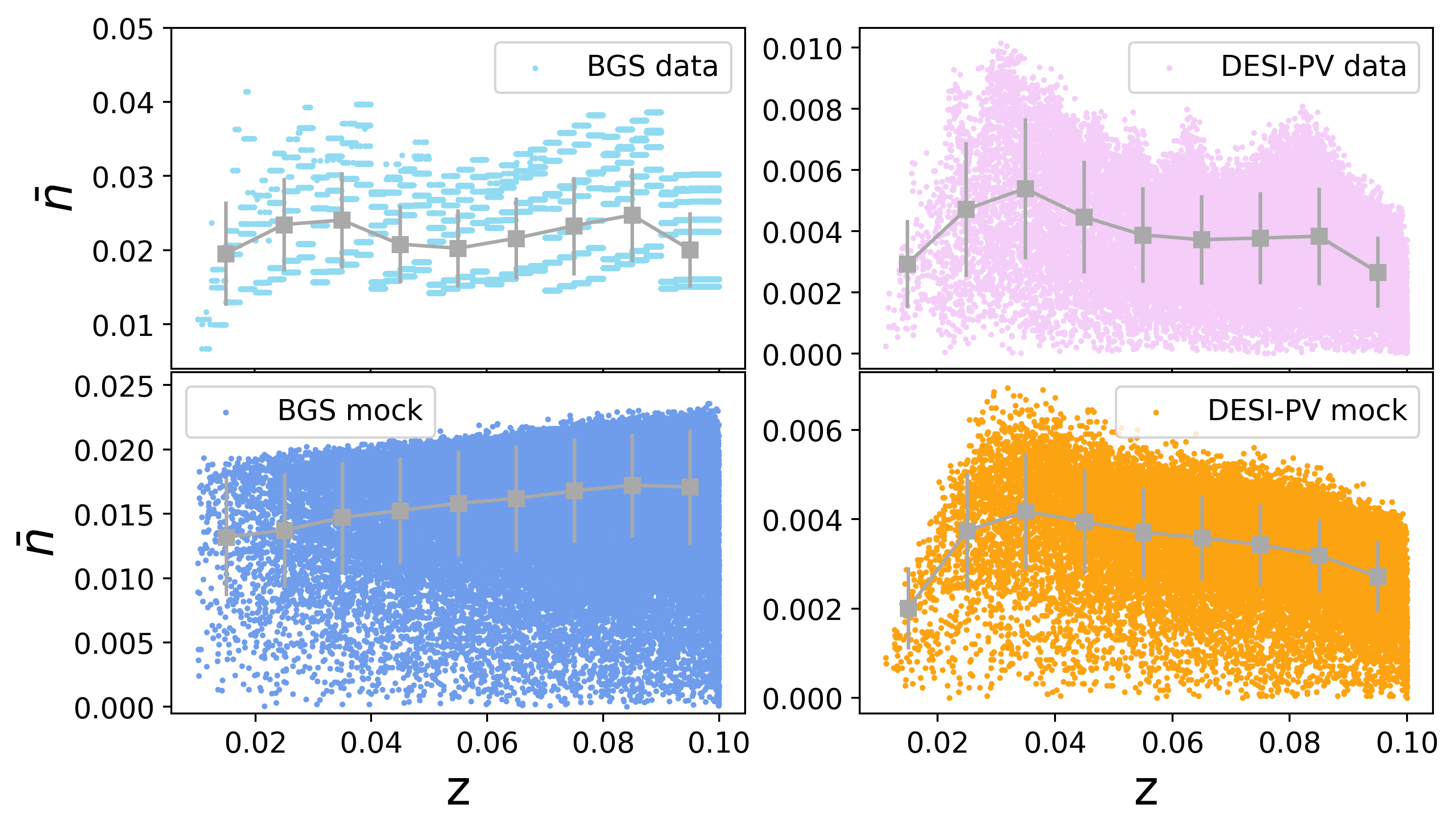}
 \caption{
 The galaxy mean number density, denoted as $\bar{n}({\bf r})$, as a function of redshift $z$, for both the BGS and DESI-PV data and their corresponding mocks. Notably, the data points are derived from their respective random catalogs to ensure a smooth representation. The gray square with error bar represents the mean value and standard deviation of the data points within each redshift bin.
 }
 \label{pltnb}
\end{figure}

The Bright Galaxy Survey (BGS, \citealt{Hahn2023}) Bright sample has undergone  refinement by \cite{DESIPV_Turner}, who introduced tailored selection criteria to enhance its suitability for analyses involving peculiar velocities. The final BGS data catalogue employed in our study encompass a total of 415,523 galaxies, with an r-band absolute magnitude threshold set at -17.7.
 The sky distribution of these galaxies is illustrated by the black dots in Fig.\ref{pltsky}. 
 Furthermore, the redshifts of these galaxies, measured in the CMB frame, span the range from $z = 0.01$ to $z = 0.1$, as depicted by the black bars in Fig.\ref{pltcz}. The galaxy mean number density $\bar{n}({\bf r})$ is shown in the top-left panel of Fig.\ref{pltnb}. The calculation of $\bar{n}({\bf r})$  is detailed in \cite{DESIPV_Bautista}. 
 In this paper, the density field is derived from the BGS catalogue.

\subsubsection{The DESI-PV data}

Due to the peculiar motions of galaxies, the apparent distance of a galaxy, denoted as $d_z$, deviates from its true comoving distance, $d_h$. This discrepancy is measureable and can be quantified through the logarithmic distance ratio, defined as
$\eta \equiv \log_{10}\frac{d_z}{d_h}$. 
It can be converted to the line-of-sight peculiar velocity using the velocity estimator Eq.\ref{watvp}.
Within the DESI-PV survey, two distinct populations of galaxies are observed:
spiral galaxies, also known as late-type galaxies, and elliptical galaxies, referred to as early-type galaxies. The $\eta$ for late-type galaxies is derived from the Tully-Fisher (TF, \citealt{Tully1977}) relation, whereas for early-type galaxies, it is inferred from the Fundamental Plane (FP, \citealt{Djorgovski1987}, \citealt{Dressler1987}) relation. In DESI-PV,
the Tully-Fisher sample \citep{DESIPV_Douglass} comprises a total of 6,806 late-type galaxies. The sky distribution of these galaxies is illustrated by the blue dots in Fig.\ref{pltsky}. Furthermore, the redshifts of these galaxies span the range from $z = 0.015$ to $z = 0.1$, as depicted by the blue bars in Fig.\ref{pltcz}. In contrast, the Fundamental Plane sample \citep{DESIPV_Ross} consists of 73,822 early-type galaxies. Their sky distribution  is marked by gray dots in Fig.\ref{pltsky}, and their redshifts span the interval from $z = 0.01$ to $z = 0.1$, as shown in the gray bars of Fig.\ref{pltcz}.

The DESI-PV catalog is a combination of the TF sample and the FP sample. The zero-point calibration bridging these two datasets has been rigorously investigated by \cite{DESIPV_Carr}, ensuring a physical alignment. The redshift distribution of this combined sample is illustrated by the yellow bars in Fig.\ref{pltcz}. Meanwhile, the galaxy mean number density $\bar{n}({\bf r})$ is  depicted in the top-right panel of Fig.\ref{pltnb}. 
In this study, the momentum field is  constructed from the DESI-PV catalog.
In accordance with the reasoning articulated in \cite{DESIPV_Turner}, we impose a redshift cut of $ z_{\text{cut}} = 0.05 $ on the TF galaxies when measuring the power spectrum—excluding those galaxies lying beyond this threshold—to mitigate systematic biases inherent in the TF velocity measurements. This identical redshift cut is consistently applied across both mocks and random catalogs, ensuring coherence and comparability throughout the analysis. Following \cite{DESIPV_Turner}, a 4$\sigma$-clipping has been applied to the measured $\eta$ values of the DESI-PV data too.

\subsection{Mocks}

A considerable number of mocks is indispensable for validating the algorithm and accurately estimating the uncertainties in the power spectrum and associated parameters. 
\cite{DESIPV_Bautista} offers a comprehensive set of 675 BGS mocks, meticulously designed to mirror the characteristics of the BGS data. In parallel, 675 FP mocks have been constructed to emulate the FP data, alongside 675 TF mocks tailored to align with the TF data. Each set of BGS, FP, and TF mocks shares the same observer, resulting in a total of 675 mock sets. These mocks are derived from the AbacusSummit simulation, and the fiducial value of $f\sigma_8$ for these mocks is 0.466 at an effective redshift of $z  = 0.2$.

The FP and TF mocks are generated with the specific aim of replicating the selection function and survey geometry, the Fundamental Plane relation, the Tully-Fisher relation, and the galaxy clustering observed in the   FP and TF data, respectively. The DESI-PV mocks are formed by combining the TF and FP mocks. The galaxy mean number density $\bar{n}({\bf r})$ of DESI-PV mocks is illustrated in the bottom-right panel of Fig.\ref{pltnb}. The galaxy mean number density $\bar{n}({\bf r})$ of BGS mocks is illustrated in the bottom-left panel of Fig.\ref{pltnb}.
 
\subsection{The random catalogs}\label{sec:rand}

The random catalogues for both the   data and mocks are also supplied by \cite{DESIPV_Bautista}. 
In order to precisely extract cosmological parameters from the power spectrum, it is imperative to appropriately account for the survey geometry of both BGS and DESI-PV. Employing fast Fourier transforms to measure the power spectrum in the context of a non-periodic and incomplete survey geometry leads to a convolution of the true power spectrum with the survey’s window function. To address this, 
the random catalogue will be employed to estimate the window function, thereby enabling the application of an equivalent convolution to our theoretical model during the data fitting process. 
The random catalogues for   FP and TF are generated independently, with the objective of more faithfully replicating the distinct observational features of the TF and FP surveys, respectively. The random catalogue for DESI-PV is constructed by combining the random catalogues of TF and FP.

\section{The estimation of the power spectrum}\label{sec:psest}

 \subsection{The field functions}\label{sec:Defin}

In this paper, we investigate the redshift-space power spectrum of galaxies, encompassing both the auto-power spectrum and cross-power spectrum of the galaxy density and momentum fields. The density field is derived from the BGS catalog, whereas the momentum field is obtained from the DESI-PV catalog. 
Within our paper series, the density (contrast) field is defined by
\be
\delta({\bf r})\equiv\frac{\rho({\bf r})-\bar{\rho}}{\bar{\rho}}
\ee
where $\rho({\bf r})$ is the mass density at position ${\bf r}=[r_x,r_y,r_z]$, $\bar{\rho}$ represents the average mass density of the Universe. The line-of-sight momentum field is defined by
\be 
p({\bf r})\equiv[1+\delta({\bf r})]v({\bf r}) 
\ee 
with $v({\bf r})$ is denoting the line-of-sight peculiar velocity at position ${\bf r}$. 
The auto-density power spectrum $P^{\delta}$, the auto-momentum power spectrum $P^{p}$ and the cross-power spectrum of the density and momentum fields $P^{\delta p}$ are originally formulated as two-point correlations of $\delta$  and/or $p$ in Fourier  space, as expressed in
{\setstretch{0.5}
\be  \label{defmomps456}
(2\pi)^3\delta^D({\bf k}-{\bf k}')P^{\delta}({\bf k})\equiv\langle  \delta({\bf k}) \delta^*({\bf k}') \rangle
\ee 
\be \label{defmomps}
(2\pi)^3\delta^D({\bf k}-{\bf k}')P^{p}({\bf k})\equiv\langle p({\bf k})p^*({\bf k}')\rangle
\ee 
\be \label{defmomps789}
(2\pi)^3\delta^D({\bf k}-{\bf k}')P^{\delta p}({\bf k})\equiv\langle \delta({\bf k})  p^*({\bf k}') \rangle
\ee }
respectively, where $\delta^D({\bf k}-{\bf k}')$ indicates the Dirac $\delta$-function, `*' denotes the complex conjugate. ${\bf k}$ is the wave vector.  

We favor the momentum power spectrum over the velocity power spectrum because, in principle, the velocity field is a continuous entity defined everywhere in space—even in regions devoid of galaxies. However, our measurements are limited to locations where galaxies (and thus mass) are present, effectively yielding a mass-weighted, or momentum-based power spectrum. In regions devoid of galaxies, we record zero velocity not because the velocity is truly zero, but due to the absence of tracers. Additionally, the momentum power spectrum retains the same fundamental information as the velocity power spectrum, and with additional non-linear contributions introduced by the $(1+\delta)$ term that be incorporated into the theoretical modeling. There exist methods to probe the underlying continuous velocity field, however, they come with notable drawbacks. For example, the reconstruction of velocity field, it  inherently dependent on cosmological assumptions. Alternatively, one may employ a Voronoi tessellation to discretize the data, thereby avoiding artificial zero-velocity measurements caused by sparse galaxy distribution. However, this method introduces complications in the Fourier space due to the irregularity of the resulting grid.

Following the methodology outlined in \cite{Feldman1994}, which constitutes the standard approach widely adopted today, the estimator for the density field is presented in 
\be\label{fieldd}
F^{\delta}({\bf r})\equiv \frac{w_{\delta}({\bf r})\left[ n_{\delta}({\bf r})-\alpha n_s({\bf r}) \right]}{A_{\delta}} ~.
\ee
In this formulation $w_{\delta}({\bf r})$, $n_{\delta}({\bf r})$ and $n_s({\bf r})$ represent the weight factor, the number density of galaxies and the number density of random points at position ${\bf r}$.  The number of random points is $\alpha$ times greater than the number of galaxies. $A_{\delta}$ serves as the normalization factor, defined by
\be \label{normPS123autodd}
A_{\delta}^2=\int w_{\delta}^2({\bf r})\bar{n}_{\delta}^2({\bf r}) d^3r~,
\ee 
which ensures that the amplitude of the measured power aligns with it in a universe unaffected by survey selection effects. The computation of the mean galaxy number density $\bar{n}_{\delta}({\bf r})$ at position ${\bf r}$ is detailed in \cite{DESIPV_Bautista} or see Fig.\ref{pltnb}. As indicated in Eq.\ref{fieldd}, to estimate the density field, it is necessary to subtract the random catalog, this is because we measure the density contrast field $\delta$, which can only be accomplished if we possess knowledge of, and are able to subtract, the mean number of galaxies within each grid cell which estimated from randoms of that cell.

Similarly, following the derivation in \papI, the estimator for the momentum field is given by
\be\label{fieldp}
F^p({\bf r})\equiv \frac{w_p({\bf r})n_p({\bf r})v({\bf r})}{A_p} ~,
\ee
where $w_{p}({\bf r})$ and $n_{p}({\bf r})$ denote the weight factor and the number density of galaxies at position ${\bf r}$, respectively.    $A_p$ functions as the normalization factor, defined by
\be \label{normPS123auto}
A_{p}^2=\int w_{p}^2({\bf r})\bar{n}_{p}^2({\bf r}) d^3r~,
\ee 
which again adjusts the amplitude of the measured power to conform with it in an idealized universe without survey selection biases. The computation of the mean galaxy number density $\bar{n}_p({\bf r})$ at position ${\bf r}$ is also detailed in \cite{DESIPV_Bautista} or see Fig.\ref{pltnb}. Notably, no random catalog is subtracted when estimating  the momentum field. This is because that the mean velocity at any spatial point is inherently zero; consequently, subtracting a random catalogue during the construction of the momentum field would be equivalent to subtracting zero.

In this paper, we define the Fourier transform of the field functions presented in Eq.\ref{fieldd} and \ref{fieldp} through the following formulation 
\be\label{A0}
F({\bf k})\equiv\frac{1}{V}\int  F({\bf r}) e^{i {\bf k} \cdot {\bf r} }    d^3r,
\ee
and inversely, 
\be\label{A0inv}
F({\bf r})\equiv\frac{1}{(2\pi)^3}\int  F({\bf k}) e^{-i {\bf k} \cdot {\bf r} }    d^3k
\ee

\subsection{The estimators of the power spectrum multipoles} \label{sec:PSest}

The observed redshift of a galaxy is comprised of two dominant contributions: the Hubble recessional redshift resulting from the cosmic expansion, and the peculiar velocity redshift arising from localized gravitational influences. Consequently, the inferred position of a galaxy based on its observed redshift does not represent its true comoving position; instead, it reveals a distorted position in what is referred to as `redshift-space position'. This phenomenon is widely recognized as redshift-space distortion (RSD).

Redshift-space distortions disrupt the spherical symmetry of the power spectrum with respect to the line of sight. To capture this anisotropic behavior, a prevalent methodology involves decomposing the redshift-space power spectrum $P({\bf k})$ into a series of Legendre polynomials $L_{\ell}(\mu)$, i.e. 
\be\label{plkest}
P({\bf k})= \sum_{\ell}P_{\ell}(k )L_{\ell}(\mu)~,~~(\ell=0,1,2,3,4,...)
\ee
In this decomposition, the angular dependence of the power spectrum $P({\bf k})$ is encapsulated within the Legendre polynomials $L_{\ell}(\mu)$, while the amplitude information of $P({\bf k})$ are encoded in the multipole moments $P_{\ell}(k )$, commonly referred to as power spectrum multipoles. Here, $\mu={\bf\hat{r}}\cdot {\bf\hat{k}}=\mathrm{cos}\,\theta$ denotes the cosine of the angle between the unit wave vector ${\bf\hat{k}}$ and the line-of-sight unit vector ${\bf\hat{r}}$.

Following the formalism established by \cite{Yamamoto2006} and \papI, the estimators for the auto-density and auto-momentum power spectrum multipoles are defined by  
\be \label{autops}
P^{\delta}_{\ell}(k)\equiv|F^{\delta} (k)F^{\delta*}_{\ell}(k)|-\mathscr{N}_{\ell}^{\delta} 
\ee 
and
\be \label{autopss}
P^{p}_{\ell}(k)\equiv|F^{p}(k)F^{p*}_{\ell}(k)|-\mathscr{N}_{\ell}^{p} 
\ee 
respectively. The index `$\delta$' and `$p$' corresponds to the auto-density and auto-momentum power spectrum, respectively. The associated shot-noise contributions are described by 
\be\label{Pnoised}
\mathscr{N}^{\delta}_{\ell}= 
 \frac{(1+\alpha)(2\ell+1)}{A^2_{\delta}V}\int w^2_{\delta}({\bf r})\bar{n}_{\delta}({\bf r})L_{\ell}(\mu) d^3r 
\ee 
\be\label{Pnoisep}
\mathscr{N}^{p}_{\ell}=\frac{2\ell
+1}{A^2_pV} \int w_{p}^2({\bf r})\bar{n}_p({\bf r})\langle v^2({\bf r}) \rangle L_{\ell}(\mu)d^3r 
\ee 
respectively.

As outlined in Appendix \ref{sec:CRSest} (or see \papIII), the estimator for the cross-power spectrum multipoles is defined  through \footnote{There is a small typos in Equation 19 of \papIII, while the code used in \papIII ~is correct.}
\be  \label{crsps}
P^{\delta p}_{\ell}(k)\equiv\frac{1}{2}\mathrm{Im}\{|F^{p}(k)F_{\ell}^{\delta *}(k)|-|F^{\delta}(k)F_{\ell}^{p *}(k)| \}-\mathscr{N}^{\delta p}_{\ell} 
\ee 
with the corresponding shot-noise term detailed in
\be\label{Pnoisedp}
\mathscr{N}^{\delta p}_{\ell}= \frac{2\ell+1}{A_{\delta}A_pV}\int w_{\delta}({\bf r})w_{p}({\bf r})\min\{\bar{n}_{\delta}({\bf r}),\bar{n}_{p}({\bf r})\}\langle v({\bf r}) \rangle L_{\ell}(\mu) d^3r 
\ee
In this context, `$\delta p$' specifically refers to the density-momentum cross-power spectrum, `$\mathrm{Im}$' denotes imaginary part of a complex number. 
Given that both the time evolution of gravitational interactions and the initial conditions of the Universe remain invariant under the transformation $\delta \rightarrow -\delta$ and $v \rightarrow -v$, the cross-power spectrum between the density and momentum fields must also exhibit invariance under such transformations. This leads directly to the identity $P^{\delta p}_{\ell}=-P^{\delta p *}_{\ell}$, which holds exclusively for purely imaginary functions. See Appendix \ref{sec:symetric} for more details. 

The shot noise observed in the auto-power spectrum originates from the discrete nature of galaxies as tracers of the underlying density and momentum fields. In the case of the cross-power spectrum, shot noise arises due to the potential overlap of galaxies used in the estimation of both the density and momentum fields. This form of noise is proportional to the sample containing fewer objects \citep{Smith2009}.

The expressions for the estimators of auto-power spectrum multipoles have been thoroughly examined by \cite{Bianchi2015}. Nevertheless, our formulations of the field function definitions and Fourier transform definition deviate slightly from that of   \cite{Bianchi2015}. Consequently, in below, we shall restate the corresponding expressions for clarity and consistency. For convenience, we adopt the methodological framework proposed by \cite{Bianchi2015} and introduce the following function
\be \label{sdgdef123}
T_{\ell}({\bf k})\equiv \int ({\bf\hat{k}}\cdot {\bf\hat{r}})^{\ell} F({\bf r}) e^{i {\bf k} \cdot {\bf r} }    d^3r 
\ee
where $\ell=0,1,2,3,4,...$. Consequently, based on Eq.\ref{autops} or Eq.\ref{autopss}, the corresponding even-multipoles of the auto-power spectrum is expressed in (see Appendix \ref{sec:CRSest3} for more discussion)
\be\label{P0k}
P_0(k)=  \frac{1}{V} \int \frac{d\Omega_k}{4\pi}[F({\bf k}) T_0^*({\bf k}) ]-\mathscr{N}_0
\ee
\be\label{P2k}
P_2(k)= \frac{5}{2V }\int \frac{d\Omega_k}{4\pi}F({\bf k})\left[ 3T^*_2({\bf k})-T^*_0({\bf k})  \right] 
\ee
\be\label{P4k}
P_4(k)= \frac{9}{8V }\int \frac{d\Omega_k}{4\pi}F({\bf k})\left[ 35T^*_4({\bf k})-30T^*_2({\bf k})+3T^*_0({\bf k})  \right] 
\ee
where $d\Omega_k$ denotes the solid angle. See Appendix \ref{sec:Tl} for the expressions of $T_0$, $T_2$ and $T_4$.  It is important to note that, unlike in \cite{Bianchi2015}, the normalization factors $A_{\delta}$ and $A_p$ are inherently incorporated within the field functions in our formulation, and our definition of Fourier transformation diverges from them too. For multipoles with $\ell>0$, the shot noise becomes negligible and can be safely ignored  
 \citep{Blake2010,Howlett2019}. If ignoring the wide-angle or general relativistic effects, only the even-multipoles of the auto-power spectrum are non-zero  
\citep{Bianchi2015,Beutler2020,Castorina2020}. 
 
 In this paper,  building upon Eq.\ref{crsps} and Eq.\ref{sdgdef123},  
we derive the associated odd-multipoles for the cross-power spectrum, as expressed in
\be\label{P1k}
P_1(k)=\frac{3}{V}\int \frac{d\Omega_k}{4\pi}\frac{\mathrm{Im}\{F^p({\bf k}) T^{\delta *}_1({\bf k})-F^{\delta}({\bf k}) T^{p*}_1({\bf k}) \}}{2}  ,
\ee
\be\label{P3k}
\begin{split}
P_3(k)=\frac{7}{2 V}\int \frac{d\Omega_k}{4\pi} \bigg[ 5\times\frac{\mathrm{Im}\{F^p({\bf k}) T^{\delta *}_3({\bf k})-F^{\delta}({\bf k}) T^{p*}_3({\bf k}) \}}{2}\\
-3\times\frac{\mathrm{Im}\{F^p({\bf k}) T^{\delta *}_1({\bf k})-F^{\delta}({\bf k}) T^{p*}_1({\bf k}) \}}{2}  \bigg]  ,
\end{split}
\ee
See Appendix \ref{sec:Tl} for the expressions of $T_1$ and $T_3$. 
Due to similar reasons, only the odd-multipoles exhibit non-vanishing contributions to the cross-power spectrum, and the influence of shot noise diminishes to a negligible extent for $\ell>0$. 

The \textsc{PYTHON} code for our power spectrum estimators can be download from the website links provided in Appendix \ref{sec:appsdf1233} or see \url{https://github.com/FeiQin-cosmologist/Galaxy_Power_Spectrum/blob/main/CosmPSPy/Code/PSestFun.py}.

\subsection{The FKP weights}\label{sec:FKP}

In \cite{Feldman1994}, the authors introduced a weighting scheme for the density power spectrum aimed at minimizing the fractional variance of its estimation. This method is widely known as the Feldman-Kaiser-Peacock (FKP) weighting scheme and is given by 
\be\label{wopt}
w_{\delta}({\bf r},k)=\frac{1}{1+\bar{n}_{\delta}({\bf r})P^{\delta}_{FKP}(k)}, 
\ee
Building upon this work, \papI ~proposed an analogous weighting strategy tailored specifically for the momentum field, given by
\be\label{woptp}
w_p({\bf r},k)= \frac{1}{\langle v^2({\bf r}) \rangle+\bar{n}_p({\bf r})P^{p}_{FKP}(k)},
\ee
where $\langle v^2({\bf r}) \rangle$ accounts for both intrinsic scatter and measurement errors associated with peculiar velocity estimates, it can be calculated using Eq.\ref{stdvs}.

The FKP weights are derived under the assumption that the power spectrum follows a Gaussian distribution. However, FKP weights may not represent the optimal choice when the power spectrum is not Gaussian. As demonstrated in Section 6 of \papI, varying the value of $P_{\mathrm{FKP}}(k)$ alters the effective survey depth of the galaxy survey. Different choices of FKP weights influence the trade-off between upweighting higher redshift data—thereby increasing cosmic volume and reducing power spectrum variance—and upweighting nearby objects where velocity measurement errors are smaller. Following \papI~ and \papII~ and \papIII, we set $P^{\delta}_{\mathrm{FKP}}=1600h^{-3}\,\mathrm{Mpc^{3}} $ and $P^{p}_{\mathrm{FKP}}=5\times 10^9h^{-3}\,\mathrm{Mpc^{3}\,km^{2}\,s^{-2}}$ for the density and momentum fields, respectively, to achieve optimal measurement accuracy.

\subsection{Grid correction and Nyquist frequency}

To compute the field functions of Eq.\ref{fieldd} and \ref{fieldp} from galaxies, it is necessary to assign the galaxies onto a three-dimensional grid. This gridding process effectively convolves the field values with a top-hat window function along each spatial direction, resulting in each Fourier mode being multiplied by a Sinc function. To mitigate this effect, following the methodology outlined in \cite{Abate2008}, \cite{Johnson2014} and \cite{Adams2017}, we apply a correction factor by dividing each voxel corresponding to a wave vector ${\bf k}=[k_x,k_y,k_z]$ by the following Sinc functions for the Fourier transformed  field functions (i.e. $F({\bf k})$ and $T_{\ell}({\bf k})$)
\be 
\Gamma({\bf k})\equiv \frac{8}{L_xL_yL_z}
\frac{\sin\bigg(\frac{k_xL_x}{2}\bigg)}{k_x}
\frac{\sin\bigg(\frac{k_yL_y}{2}\bigg)}{k_y}
\frac{\sin\bigg(\frac{k_zL_z}{2}\bigg)}{k_z}
\ee
where $L_x$, $L_y$ and $L_z$ denote the size of the voxel along $x$, $y$ and $z$ directions, respectively. In this study, we construct the galaxy grid using a voxel size of $L_x=L_y=L_z=2$ Mpc $h^{-1}$ within a cubic box of size $L=800$ Mpc $h^{-1}$ centered on the observer. This configuration provides sufficient spatial coverage to include the galaxies observed in the DESI survey.

Furthermore, we consider only those Fourier modes corresponding to wavenumbers $k$ that exceed the Nyquist frequency, defined by
\be 
K_{Ny}=\min\bigg\{ \frac{\pi}{L_x},\frac{\pi}{L_y},\frac{\pi}{L_z}\bigg\} .
\ee 
This ensures that the sampling rate meets the minimum requirement for an undistorted estimation of the power spectrum.

\section{The theoretical model of the power spectrum}\label{sec:psmodel}

\subsection{The models of power spectrum}

\cite{Kaiser1987} developed a linear model for the density power spectrum, commonly referred to as the Kaiser Formula (see Appendix \ref{sec:app1}), which has been extensively utilized over the past decades. Building upon this foundation, \cite{Koda2014} extended the analysis to the power spectrum of peculiar velocities, and incorporated empirical damping terms into the linear models in order to more accurately account for non-linear motions of galaxies. However, these linear and quasi-linear models are effective only for $k<0.15$ $h$ Mpc $^{-1}$, where measurement errors tend to be relatively larger.

To more effectively capture non-linear effects and extend the analysis into regions with smaller measurement errors, we employ one-loop perturbation theory models for the power spectrum in this paper. These redshift-space power spectrum models were initially developed by \cite{Vlah2012denPS,Vlah2013denPSHaloBias,Okumura2014} and summarized in Appendix A of \papI~ and updated in Section 4 of \papIII. Following the theoretical framework outlined in \cite{Vlah2012denPS,Vlah2013denPSHaloBias} and \papI, the density power spectrum model is described by
\be \label{psmodd}
\begin{split}
P^{\delta}(k,\mu)&=P_{00} + \mu^2(2P_{01}+P_{02}+P_{11}) \\
& +\mu^4(P_{03}+P_{04}+P_{12}+P_{13}+\frac{1}{4}P_{22}) ~.
\end{split} 
\ee 
Based on the theoretical formulation presented in \cite{Okumura2014} and \papI, the momentum power spectrum model is given by
\be \label{psmodp} 
P^{p}(k,\mu)=\frac{[a(z)H(z)]^2}{k^{2}}\big[P_{11}+\mu^2(2P_{12}+3P_{13}+P_{22})\big] ~.
\ee 
Additionally, the cross-power spectrum model is also derived in \cite{Okumura2014}. However, according to \cite{Chen2025pairwisPS} (Equation D.2 therein), there exist minor inaccuracies in the derivation of the cross-power spectrum expression in \cite{Okumura2014} (Equation 2.19 therein), and consequently in \papIII. In this paper, we derive the corrected version which is given in
\be  \label{psmoddp}  
\begin{split}
P^{\delta p}(k,\mu)&=\frac{a(z)H(z)}{k} \mu\Bigg[P_{01} + P_{02} + P_{11} \\
&+ \mu^2\bigg(\frac{3}{2}P_{03} + 2 P_{04} + \frac{3}{2} P_{12} + 2P_{13} + \frac{1}{2}P_{22}\bigg) \Bigg]~.  
\end{split}
\ee 
Notably,  this is the  imaginary part of cross-power spectrum model. The new formulations for $P_{mn}$, $(m,n=0,1,2,3,4)$ is presented in Section \ref{sec:loopterms}.  
In the aforementioned equations, the scale factor is determined from 
\be \label{ahz}
a(z)\equiv\frac{1}{1+z}. 
\ee 
Assuming $\Lambda$CDM, the Hubble parameter is given by 
\be \label{euqEZ}
\frac{H(z)}{H_0}=E(z)\equiv\sqrt{\frac{\Omega_{m}}{a^3}+\frac{1-\Omega_{m}-\Omega_{\Lambda}}{a^2}+\Omega_{\Lambda}}~,
\ee 
where $H_0$, $\Omega_{m}$ and $\Omega_{\Lambda}$ are the Hubble constant,  matter density parameter and dark energy density parameter in the present-day Universe at effective redshift $z_{\rm eff}=0$.

The models described in Eq.\ref{psmodd}, \ref{psmodp} and \ref{psmoddp} represent highly general frameworks, with the sole underlying assumption being the local plane-parallel approximation. These models do not presuppose the $\Lambda$CDM paradigm. To derive the $\Lambda$CDM-specific models, one simply needs to employ $\Lambda$CDM theory in the computation of $E(z)$ (in Eq.\ref{euqEZ}), and the growth factor $D(a)$ (in Eq.\ref{dafun}) as well as the linear matter power spectrum $P_L$ (as mentioned in Section \ref{sec:loopterms}). However, you are entirely at liberty to utilize alternative theoretical approaches to calculate $P_L$, $D(a)$ and $E(z)$, thereby generating power spectrum models corresponding to other cosmological models.

To obtain the theoretical models for the power spectrum multipoles, one can substitute Eq.\ref{psmodd}, \ref{psmodp} and \ref{psmoddp} into the following
\be\label{psmod}
P_l(k)=\frac{(2\ell+1)}{2} \int^1_{-1}P(k,\mu)L_l(\mu)d\mu~,
\ee 
which essentially performs a  Legendre transformation corresponding to Eq.\ref{plkest}. The aforementioned theoretical models for the power spectrum can be readily converted into models for the galaxy two-point correlation function, the velocity correlation function, and the galaxy-velocity cross-correlation function using the \textsc{hankl} package\footnote{The \textsc{hankl} PYTHON package: \url{https://hankl.readthedocs.io/en/latest/install.html}}  \citep{Karamanis2021}, see \cite{DESIPV_Turner}
and Appendix \ref{sec:app1}  for more discussion. The momentum correlation is equivalent to the velocity correlation in linear scales, as demonstrated by Equation 2 of \papI, therefore, the model of velocity correlation function can be converted from the model of momentum power spectrum approximately.  

 The \textsc{PYTHON} code for our power spectrum models can  be download from the website links provided in Appendix \ref{sec:appsdf1233} or see \url{https://github.com/FeiQin-cosmologist/Galaxy_Power_Spectrum/blob/main/CosmPSPy/Code/PSmodFun.py}. 

\subsection{The new loop terms for the cross power spectrum}\label{sec:loopterms}

The loop terms $P_{mn}$, $(m,n=0,1,2,3,4)$ in Eq.\ref{psmodd}, \ref{psmodp} and \ref{psmoddp} were initially formulated by \cite{Vlah2012denPS,Vlah2013denPSHaloBias,Okumura2014}, \papI~ and \papIII. Their models, however, were specifically tailored for situations in which both density and momentum fields originate from the same galaxy survey dataset. 

In this paper, we generalize $P_{mn}$ of cross-power spectrum Eq.\ref{psmoddp} to accommodate cases where the density and momentum fields stem from different surveys with differing observational properties. This generalization implies that the cross-power spectrum Eq.\ref{psmoddp}  will incorporate multiple biasing parameters. The corresponding new formulations for $P_{mn}$  (of Eq.\ref{psmoddp}) are detailed in the subsequent equations
{\setstretch{0.3}
\be  \label{loop00}
\begin{split}
P_{00} & =  b_1^{\delta}b_1^{p} D^2 P_L + b_1^{\delta}D^4 \Big(b_2^{p}K_{00}+b_s^{p}K^s_{00} +2b_1^p(I_{00}+3J_{00}k^2P_L)\Big)\\
&+\frac{1}{2}D^4\Big(b_2^{\delta}b_2^pK_{01}+2b_1^{p}(b_2^{\delta}K_{00}+b_s^{\delta}K^s_{00})+b_s^{\delta}b_s^pK^s_{01}\\
&+b_2^pb_s^{\delta}K^s_{02} +b_2^{\delta}b_s^pK^s_{02}+4b^p_{3nl}P_L \sigma_3^2\Big)
\end{split}
\ee
\be  \label{loop01}
\begin{split}
P_{01}&=fD^2\Big(b_1^{\delta}(P_L+2D^2(I_{01}+b_1^pI_{10}+3(J_{01}+b_1^pJ_{10})k^2P_L))\\
&-D^2(b_2^{\delta}(K_{10}+b_1^pK_{11})+b_s^{\delta}K^s_{10}+b_1^pb_s^{\delta}K^s_{11}+b^p_{3nl}P_L\sigma_3^2)\Big)
\end{split}
\ee
\be  \label{loop11}
\begin{split}
P_{11} & =  f^2D^2\bigg(\mu^2P_L+D^2\Big(b_1^{\delta}b_1^pI_{31}+(2I_{11}+b_1^{\delta}b_1^pI_{13}+2b_1^{\delta}I_{22}\\
&+2b_1^pI_{22}+6((b_1^{\delta}+b_1^p)J_{10}+J_{11})k^2P_L)\mu^2\Big)\bigg)
\end{split}
\ee
\be  
\begin{split}
P_{13}&=-f^4k^2D^2\Big(\sigma_{\delta,vS}^2\mu^2\big(P_L+D^2(2I_{11}+2(b_1^{\delta}+b^p_1)I_{22} \\
&+6k^2P_L(J_{11}+(b_1^{\delta}+b_1^p)J_{10}))\big) +\sigma^2_{\delta,vT}b_1^{\delta}b_1^{p} D^2(\mu^2I_{13}+I_{31})\Big)
\end{split}
\ee
\be  \label{loop04}
\begin{split}
P_{04}&=-\frac{1}{2}f^4b_1^{\delta}k^2\sigma_{\delta,vT}^2D^4(I_{02}+\mu^2I_{20}+2k^2P_L(J_{02}+\mu^2J_{20}))\\
&+\frac{1}{4}f^4k^4b_1^{\delta}b_1^pP_{00} ( \sigma_{\delta,vT}^4+\sigma_4^2)
\end{split}
\ee}
where the index `$\delta$' and `$p$' denotes the parameters for density and momentum fields, respectively. The terms $P_{00}$ in the expression of $P_{04}$ should be calculated from Eq.\ref{loop00} of this paper. Following the derivation outlined in \papIII, we arrive at
{\setstretch{0.3}
\be\label{loop02}
\begin{split}
P_{02} & = f^{2}b_{1}^{\delta}D^{4}(I_{02} + \mu^{2}I_{20} + 2k^{2}P_{L}(J_{02} + \mu^{2}J_{20}))- f^2k^2\sigma^2_{\delta,vT}P_{00}\\
& + f^{2}D^{4}\Big(b_{2}^{\delta}(K_{20}+\mu^{2}K_{30}) + b_{s}^{\delta}(K^{s}_{20}+\mu^{2}K^{s}_{30})\Big)
\end{split}
\ee
\be  
\begin{split}
P_{12} & = f^{3}D^{4}\Big(I_{12} + \mu^{2}I_{21} - b_{1}^{\delta}(I_{03} + \mu^{2}I_{30})  + 2k^{2}P_{L}(J_{02} + \mu^{2}J_{20})\Big) \\
&- f^2k^2\sigma^2_{\delta,vT}P_{01}+2f^3 D^4k^2\sigma^2_{\delta,vT}\Big(I_{01}+I_{10} +3k^2P_{L}(J_{01}+J_{10})\Big)  \\
\end{split}
\ee }
where $P_{00}$ and $P_{01}$ in the above equations should be obtained from Eq.\ref{loop00} and \ref{loop01} of this paper, separately. By following the derivation presented in \papI, we also obtain 
{\setstretch{0.5}
\be  
P_{03}=-f^2k^2\sigma^2_{\delta,vS}P_{01} 
\ee
\be  \label{loop22}
\begin{split}
P_{22}&=\frac{1}{4}f^4D^4(I_{23}+2\mu^2 I_{32}+\mu^4I_{33})+f^4k^4\sigma_{\delta,vT}^4P_{00}-f^2k^2\sigma_{\delta,vT}^2\\
&\times\Big(2P_{02}-f^2D^4\big(b_2^{\delta}(K_{20}+\mu^2K_{30})+b_s^{\delta}(K^s_{20}+\mu^2K^s_{30})\big)\Big)
\end{split}
\ee }
in the above, the $P_{00}$, $P_{01}$ and $P_{02}$ should be derived from Eq.\ref{loop00}, \ref{loop01} and \ref{loop02} of this paper, respectively. The term $D$ represents the linear growth factor. Under the assumption of General Relativity (GR) and $\Lambda$CDM, its analytical expression is given by \citep{Heath1977,Howlett2015correlationfunction}
\be \label{dafun}
D(a)=\frac{D_{gr}(a)}{D_{gr,0}}
\ee 
where
\be\label{theDgr}
D_{gr}(a)=E(a)\int^a_0\frac{da'}{a'^3H(a')^3}, 
\ee 
and where $D_{gr,0}$ symbolizes the $D_{gr}(a)$ of the present-day Universe at redshift $z_{\rm eff}=0$, $E(a)$ is given in Eq.\ref{euqEZ}. 
The terms $I_{mn}$, $K_{mn}$, $K^s_{mn}$, $\sigma_3$ and $\sigma_4$ are derived from integrations of the linear matter power spectrum $P_L$, following the approach detailed in \cite{Vlah2012denPS} and \papI . In this study, we utilize the \textsc{camb}\footnote{CAMB:~ \url{https://camb.info/}~.} package \citep{Lewis:1999bs} to calculate $P_L(k)$ assuming $\Lambda$CDM, with the resulting $P_L(k)$ illustrated in Fig.\ref{pltPL}.

The parameters of the power spectrum models are summarized in Table  \ref{bkflb}. 
Within the context of the cross-power spectrum formalism Eq.\ref{psmoddp}, 
the fully parameterized model comprises 10 parameters.  
On the other hand, 
by equating the biasing parameters and velocity dispersion parameters  of the density field with those of the momentum field, the equations Eq.\ref{loop00} to \ref{loop22} simplify to the expressions detailed in Appendix A of \papI. As a result, we obtain the $P_{mn}$ for the auto-density and auto-momentum power spectrum models (i.e. Eq.\ref{psmodd} and Eq.\ref{psmodp}) conform to the formulations proposed by \papI, \papII ~and \papIII. As argued in \cite{McDonald2009,Saito2014} and \papI, there are two higher-order galaxy biasing parameters can be determined from the linear bias using local Lagrangian relations:
{\setstretch{0.3}
\be  
b_s=-\frac{4}{7}(b_1-1)
~,~b_{3nl}=\frac{32}{315}(b_1-1) ~.
\ee }
They can also be set as free parameters when fitting the models to the measurements to gain more flexibility of the models. 

Furthermore, setting all loop terms $I_{mn}$, $K_{mn}$, $K^s_{mn}$, $\sigma_3$ and $\sigma_4$ to zeros, along with setting the higher-order biasing and velocity dispersion parameters to zeros, leads to a reduction of the power spectrum models in Eq.\ref{psmodd}, \ref{psmodp} and \ref{psmoddp} to the Kaiser formulas, as demonstrated in Fig.\ref{pltKaiser} of Appendix \ref{sec:app1}.

\begin{table*}   \centering
\caption{ The parameters of the power spectrum models.  }
\begin{tabular}{|c|c|c|}
\hline
\hline
 Parameter    & Explanation    \\
\hline
  $f\in[0,+\infty)$  & The linear growth rate of the large-
scale structures, $f\equiv  \frac{d\ln\,D(a)}{d\ln\,a}$  .  \\
\hline
  $b_{1}\in[0,+\infty)$ & The linear biasing parameter.    \\
\hline
  $b_{2}\in(-\infty,+\infty)$ & The second-order local biasing parameter.   \\
\hline
  $b_{s}\in(-\infty,+\infty)$ & The second-order non-local biasing parameter.   \\
\hline
  $b_{3nl}\in(-\infty,+\infty)$ & The third-order biasing parameter.   \\
\hline
  $\sigma_{vT}\in[0,+\infty) ~[Mpc~h^{-1}]$ & The non-linear velocity dispersion for $P_{02}$, $P_{04}$, $P_{12}$, $P_{22}$ and the vector part of $P_{13}$.    \\
\hline
  $\sigma_{vS}\in[0,+\infty) ~[Mpc~h^{-1}]$ & The non-linear velocity dispersion for $P_{03}$ and the scaler part of $P_{13}$.    \\ 
      \hline
\end{tabular}
\tablefoot{For the density field and the momentum field, each of them have the following set of parameters.}
 \label{bkflb}
\end{table*}

\subsection{The window function convolution}\label{sec:windoconv}

In actual astronomical observations, our view is confined to galaxies residing within a limited spatial volume and with a particular   sky completeness function. Consequently, the power spectrum measured from galaxy surveys reflects only the characteristics of this restricted volume, rather than representing the entire universe. However, theoretical formulations of the power spectrum are typically based on the infinitely extended Universe. Therefore, in order to reliably infer cosmological parameters from galaxy surveys, it is crucial to incorporate the impacts of survey geometry and observational completeness when comparing theoretical models with measured power spectrum. This necessitates the application of a convolution operation involving the window function, as formally described by
\be 
{\bf P}^{c}(k)=  {\bf W} \cdot  {\bf P}(k')
\ee 
where the model power spectrum multipoles $ P_{\ell'}(k')$ $(\ell'=0,1,2,3,4,...)$ are concatenated into a single vector ${\bf P}(k')=[ P_{0}(k'),P_{1}(k'),P_{2}(k'),P_{3}(k'),P_{4}(k'),...]$. 
The term $P^{c}(k)$ denotes the convolved model  power spectrum multipoles, which serves as the counterpart for direct comparison with the measured power spectrum multipole. The window function convolution matrix ${\bf W}$, characterized by dimensions $(N_{\ell}\times N_{k})\times(N_{\ell'}\times N_{k'})$. Specifically, $N_k$ denotes the number of $k$-bins associated with each measured power spectrum multipole, while $N_{\ell}$ represents the total count of measured power spectrum multipoles. Furthermore, $N_{k'}$ indicates the number of $k'$ values employed in computing each model power spectrum multipole, and $N_{\ell'}$ refers to the overall number of theoretical power spectrum multipoles considered.

The window function convolution matrix ${\bf W}$ is constructed using random catalogs, following the approach detailed in Appendix \ref{sec:CRSest2}. 
Each entry of the matrix ${\bf W}$ is defined according to 
{\setstretch{0.5}
\be  \label{winmat}
{\bf W}_{\ell}(k, k')  =\frac{4\pi}{A^2}\int d\Omega_k\sum^{\ell}_{m=-\ell}Y^{\ell}_{m}(\hat{{\bf k}})\sum_{\ell'}\sum^{\ell'}_{m'=-\ell'}  \frac{I^{\ell\ell'}_{mm'}(\hat{{\bf k}},\hat{{\bf k}}')}{(2\ell'+1)V}~, 
\ee }
where 
{\setstretch{0.5}
\be \label{Allmm}
I^{\ell\ell'}_{mm'}(\hat{{\bf k}},\hat{{\bf k}}')=\frac{1}{(2\pi)^3} \int G( {{\bf k}}- {{\bf k}}') \tilde{S}^{\ell\ell' \ast}_{mm'}( {\bf k}-{\bf k}')  Y^{\ell'\ast}_{m'}(\hat{{\bf k}}') d^3k'~,
\ee 
\be \label{Smmll}
\tilde{S}^{\ell\ell'  }_{mm'}( {{\bf k}}- {{\bf k}}' )=\frac{1}{V}\int Y^{\ell}_{m}(\hat{{\bf r}})Y^{\ell' \ast}_{m'}(\hat{{\bf r}}) \bar{n}({\bf r}) w({\bf r})  e^{i{\bf r}\cdot ({{\bf k}}- {{\bf k}}')}d^3r~,
\ee 
\be \label{nbfft}
G( {{\bf k}}- {{\bf k}}')=\int w({\bf r}) \bar{n}({\bf r})  e^{i {\bf r} \cdot ({{\bf k}}- {{\bf k}}')  }  d^3r~,
\ee }
where $Y^{\ell}_{m}(\hat{{\bf r}})$ corresponds to the spherical harmonics, $w({\bf r})$ symbolizes the FKP weight introduced in Section \ref{sec:FKP}, and $\bar{n}$ stands for the mean galaxy number density outlined in Section \ref{sec:datamock} and \ref{sec:Defin}.  

For the auto-power spectrum, $w({\bf r})$ and $\bar{n}({\bf r})$ should be computed individually for both the density and momentum fields, and $A^2$ is the computed from Eq.\ref{normPS123autodd} or \ref{normPS123auto} respectively. However, when considering the cross-power spectrum, and in accordance with the arguments articulated in Appendix \ref{sec:CRSest2}, the following substitutions must be made to the normalization factor in Eq.~\ref{winmat}
{\setstretch{0.5}
\be 
A^2 \rightarrow   A_pA_{\delta}
\ee }
along with the following corresponding modifications to Eq.~\ref{Allmm} 
{\setstretch{0.5}
\be \label{symictrinS}
\begin{split}
&G(\hat{{\bf k}}-\hat{{\bf k}}')\tilde{S}^{\ell\ell' \ast}_{mm'}(\hat{{\bf k}}-\hat{{\bf k}}')
\rightarrow \\
 &\frac{1}{2}\left[G^{p}(\hat{{\bf k}}-\hat{{\bf k}}')\tilde{S}^{\delta, \ell\ell' \ast}_{mm'}(\hat{{\bf k}}-\hat{{\bf k}}')+G^{\delta}(\hat{{\bf k}}-\hat{{\bf k}}')\tilde{S}^{p,\ell\ell' \ast}_{mm'}(\hat{{\bf k}}-\hat{{\bf k}}') \right] 
 \end{split}
\ee }
where $G^{\delta} $ and $G^{p} $ are derived from Eq.~\ref{nbfft}, applied separately to the density and momentum fields, whereas the terms $\tilde{S}^{\delta, \ell\ell' \ast}_{mm'}$ and $\tilde{S}^{p,\ell\ell'}_{mm'}$ are obtained through Eq.~\ref{Smmll}, also evaluated independently for the density and momentum fields.

The window function convolution matrix is computed by assign the random points onto a three-dimensional grid.  In this paper, we set  $n_x=n_y=n_z=140$ for a cubic box size $L=700$ Mpc $h^{-1}$, in order to get a grid without empty voxels in the region where random points are full-filled.

\begin{figure*} 
 \includegraphics[width=62mm]{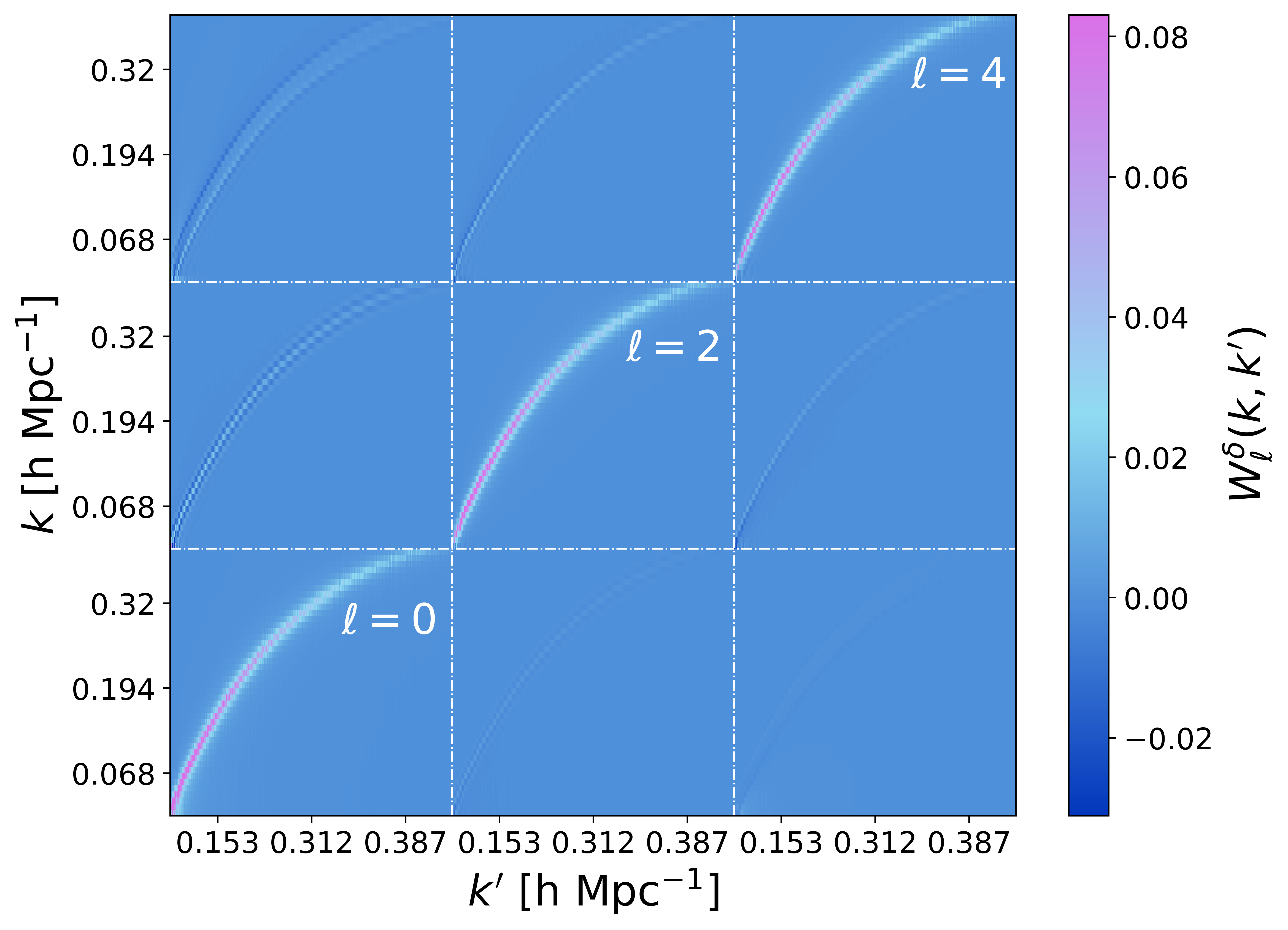}
 \includegraphics[width=62mm]{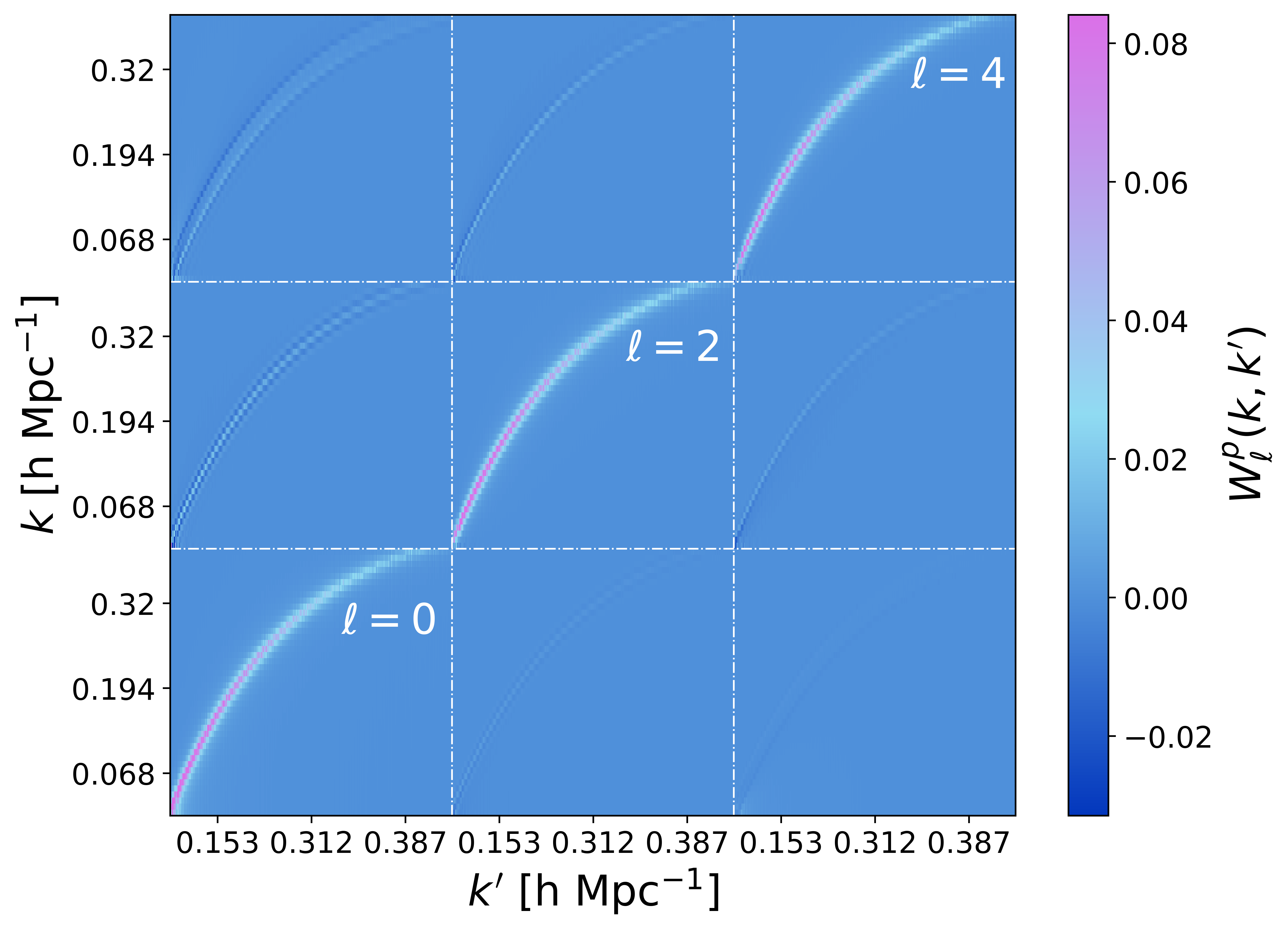}
 \includegraphics[width=61mm]{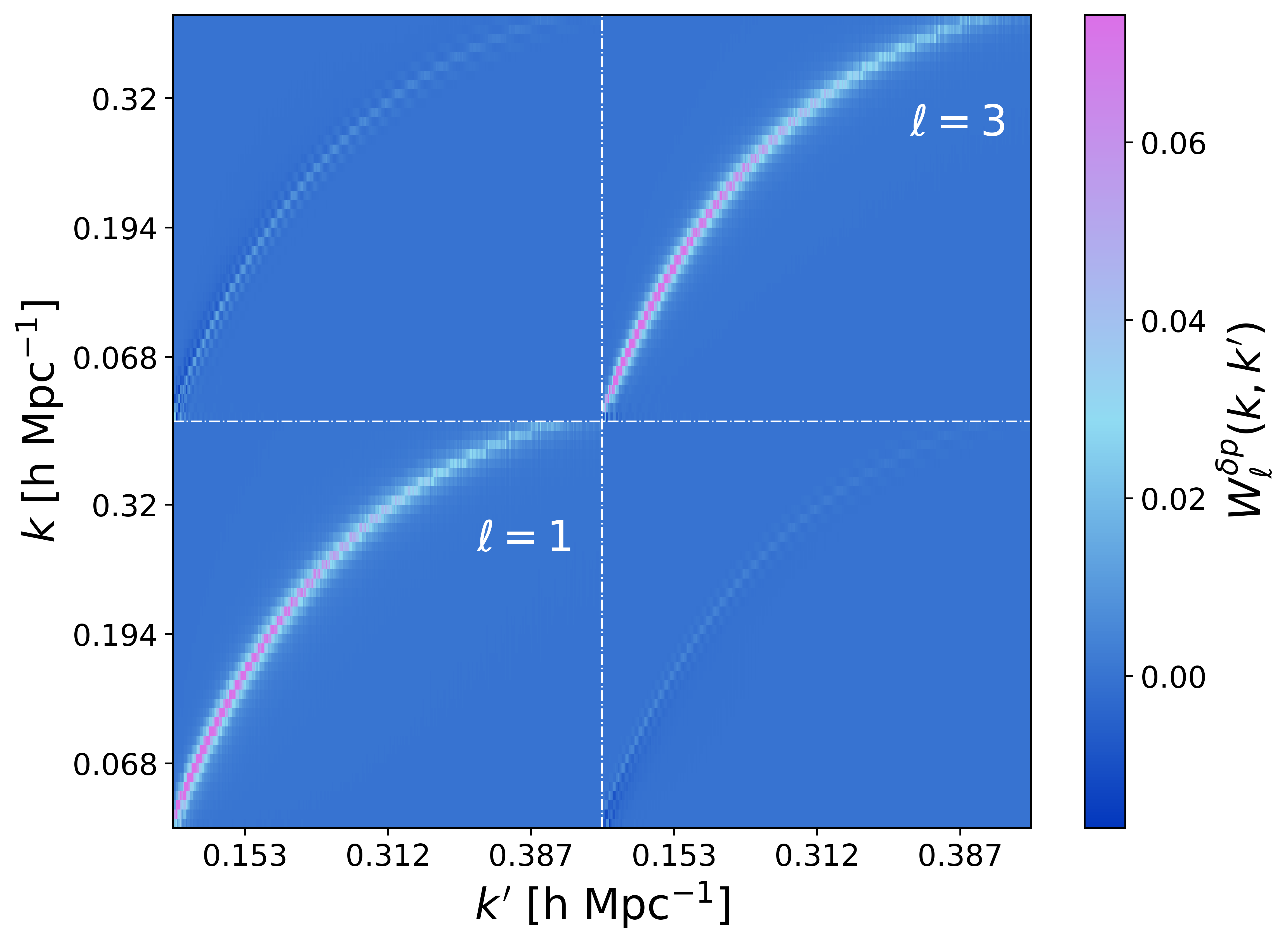}
 \caption{ The window function convolution matrix. The horizontal-axis is presented on logarithmic scales to enhance the clarity of the data. The left panel displays the convolution matrix for the density power spectrum of the BGS data. The middle panel illustrates the convolution matrix for the momentum power spectrum of the DESI-PV data. The right panel exhibits the convolution matrix for the cross power spectrum between the BGS and DESI-PV data.}
 \label{pltconv}
\end{figure*}

\begin{figure} 
\centering
\includegraphics[width=\columnwidth]{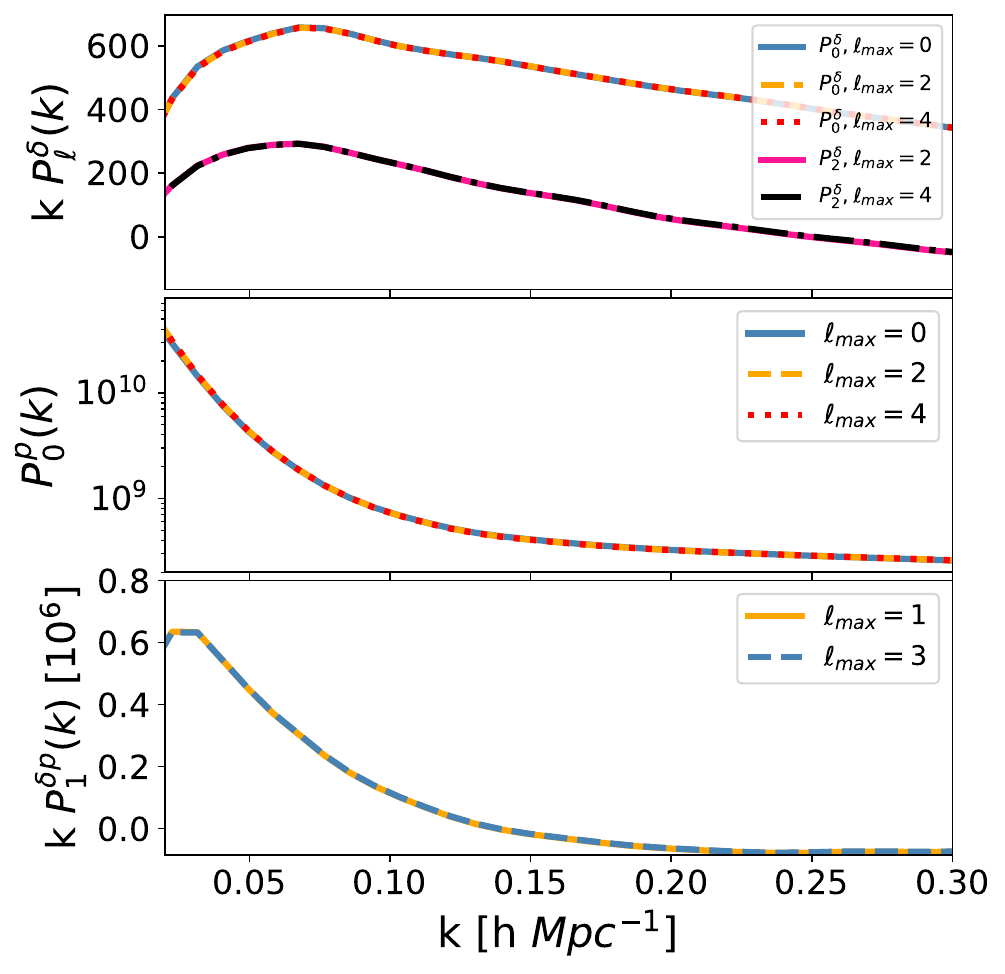}
 \caption{ {  
 The window function-convolved model power spectrum computed with different $\ell_{max}$ values. The parameters of the power spectrum models are taken from Table.\ref{tabs2last2}. }
  } 
 \label{pltdgfz33}
 \end{figure}

As depicted in Fig.\ref{pltconv}, the images illustrates the window function convolution matrix across different types of power spectrum, including the density power spectrum of the BGS data (
left panel), the momentum power spectrum of the DESI-PV data (middle panel), and their cross-power spectrum (
right panel). 
Notably, as revealed in Fig.\ref{pltconv}, the off-diagonal blocks of the ${\bf W}$ exhibit non-zero values, signifying that the convolution of any given multipole receives contributions from other multipoles. This interdependence implies that higher-order multipoles must be incorporated into the modeling framework, even if they are not directly involved in the final fitting procedure. In this work, we include all multipoles up to $\ell_{max}=4$ to ensure comprehensive and accurate representation. {  
As shown in Fig.\ref{pltdgfz33}, the differences arising from varying $\ell_{max}$ values are negligible, indicating that $\ell=4$ is sufficiently large to ensure convergence.} We  employ $k'\in[0,0.4075]~h$ Mpc$^{-1}$ (with 300 bins) in computing
each model power spectrum multipole, which should be slightly larger than the $k\in[0,0.3]~h$ Mpc$^{-1}$ which associated with each measured power
spectrum multipole, in order to fully cover the diagonal information of each block of the convolution matrix.

The \textsc{PYTHON} code for window function convolution can  be download from the website links provided in Appendix \ref{sec:appsdf1233} or see \url{https://github.com/FeiQin-cosmologist/Galaxy_Power_Spectrum/blob/main/CosmPSPy/Code/PSmodFun.py}.

\section{Tests on mocks}\label{sec:testmock}
\subsection{Fitting method}

In this study, we concatenate the density, momentum and cross power spectrum to effectively constrain the growth rate $f\sigma_8$. By concatenating the models of these three spectrum, i.e. Eq.\ref{psmodd}, Eq.\ref{psmodp} and Eq.\ref{psmoddp}, we derive a comprehensive framework comprising 13 parameters in total. These encompass the growth rate $f\sigma_8$, the biasing parameters $[b_1\sigma_8, b_2\sigma_8, b_s\sigma_8, b_{3nl}\sigma_8]$ for both the density and momentum fields, as well as the velocity dispersion parameters $[\sigma_{vS}, \sigma_{vT}]$ for the density and momentum fields, respectively. However, in practical applications, it is not necessary to treat all 13 parameters as independent variables during model fitting with measurements, because these models don't sensitive equally to every parameter. Through the following testing using mocks, we discover that a streamlined set of only five key parameters $[~f\sigma_8,~b_{1}\sigma_8,~b_{2}\sigma_8,~\sigma_{vT},~\sigma_{vS}~]$ suffices to maintain robustness and precision in the fitting process (see Appendix \ref{sec:moretest} for more discussion). This leads us to adopt a unified linear biasing parameter $b_1\sigma_8$ for both the density and momentum fields, a shared second-order biasing parameter $b_{2}\sigma_8$ for both fields. Furthermore, we employ shared velocity dispersion parameters $[\sigma_{vT},~\sigma_{vS}]$ across both the density and momentum fields.

In our study, the covariance matrix of the power spectrum is derived not from the actual Universe, but from a finite number of mocks. This limitation introduces an uncertainty into the inverse of the covariance matrix. This issue is referred to as the Hartlap effect \citep{Hartlap2007}. To address this challenge, we adopt the methodology proposed by \cite{Sellentin2016}, which different from the conventional approach of directly minimizing the $\chi^2$. Instead, we employ a likelihood function based on the following modified $t$-distribution to derive more robust estimates of the parameters 
$\theta=[f\sigma_{8},b_{1}\sigma_{8},b_{2}\sigma_{8},   \sigma^2_{vT},\sigma^2_{vS}]$: 
{\setstretch{0.5}
\be\label{tdis}
\mathcal{L}(P|\theta) \propto |\boldsymbol{\mathsf{C}}|^{-\frac{1}{2}} \left[  1+\frac{ \chi^2(P|\theta)}{N-1}  \right]^{-\frac{N}{2}},
\ee }
where $N=675$ is the number of mocks. Within this framework, the $\chi^2$ is formulated according to
$ 
 \chi^2(\boldsymbol{P}_{o}|\theta)=
\left[ {\bf P}_{o}-{\bf P}^{c}_{m}(\theta)\right]\boldsymbol{\mathsf{C}}^{-1}\left[{\bf P}_{o}-{\bf P}^{c}_{m}(\theta)\right]^T 
$,  
where ${\bf P}_{o}$ signifies the measured power spectrum, ${\bf P}^{c}_{m}$ denotes the  model power spectrum convolved with the survey window function, and $\boldsymbol{\mathsf{C}}$ represents the covariance matrix.

In this research, cosmological parameters are extracted through a  combination of the auto-density power spectrum monopole $P^{\delta}_0$ and quadrupole $P^{\delta}_2$, the auto-momentum power spectrum monopole $P^{p}_0$, as well as the cross-power spectrum dipole $P^{\delta p}_1$. Specifically, both ${\bf P}_{o}=[P^{\delta}_0,P^{\delta}_2, P^{p}_0,P^{\delta p}_1]$ and ${\bf P}^c_{m}=[P^{\delta~c}_0,P^{\delta~c}_2,P^{p~c}_0,P^{\delta p~c}_1]$ are represented as $4N_k$ vectors, while the covariance matrix $\boldsymbol{\mathsf{C}}$ spans a dimensionality of $4N_k\times 4N_k$. Additional measured multipole components of the power spectrum of the  survey data are found to be close to zero and exhibit high noise levels; therefore, they are excluded from the parameter estimation process in this work.

Parameter inference is conducted using the Metropolis-Hastings Markov Chain Monte Carlo (MCMC) methodology, under uniformly flat priors defined over the interval  $f\sigma_8\in(0,1.5]$, $b_1\sigma_8\in[0 ,3]$, $b_{2}\sigma_{8}\in[-5,5]$, $\sigma^2_{vT}\in(0,350^2]$  and $\sigma^2_{vS}\in(0,350^2]$.  
Our growth rate constraints are obtained   through the measurement of power spectrum in bins of $k\in[0.025,~0.3]$ with bin widths of $0.009 h\,\mathrm{Mpc^{-1}}$.

\subsection{Fitting $f\sigma_8$ of the mocks}\label{sec:5.2}

\begin{figure*}  
\includegraphics[width=80mm]{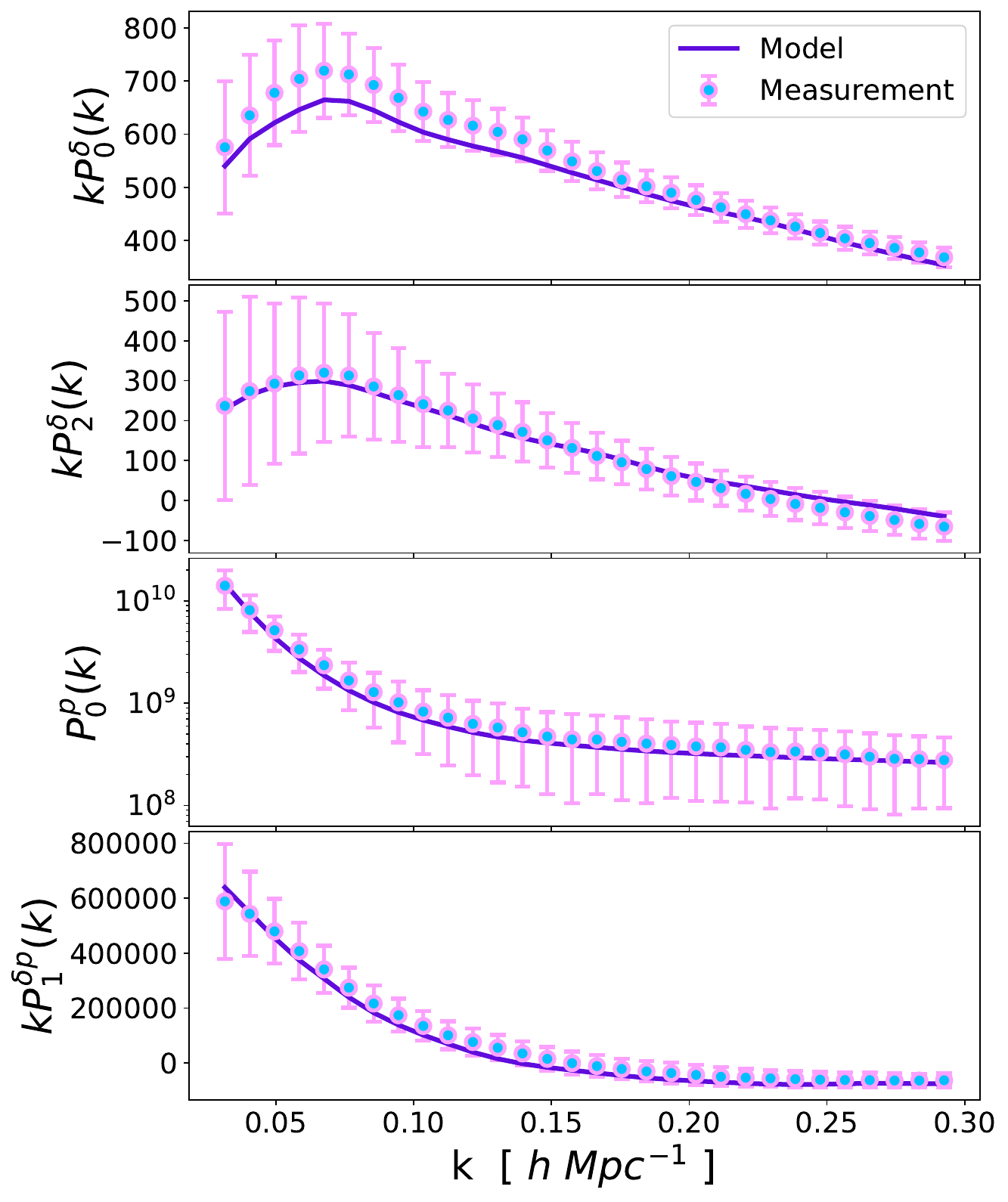}
\includegraphics[width=100mm]{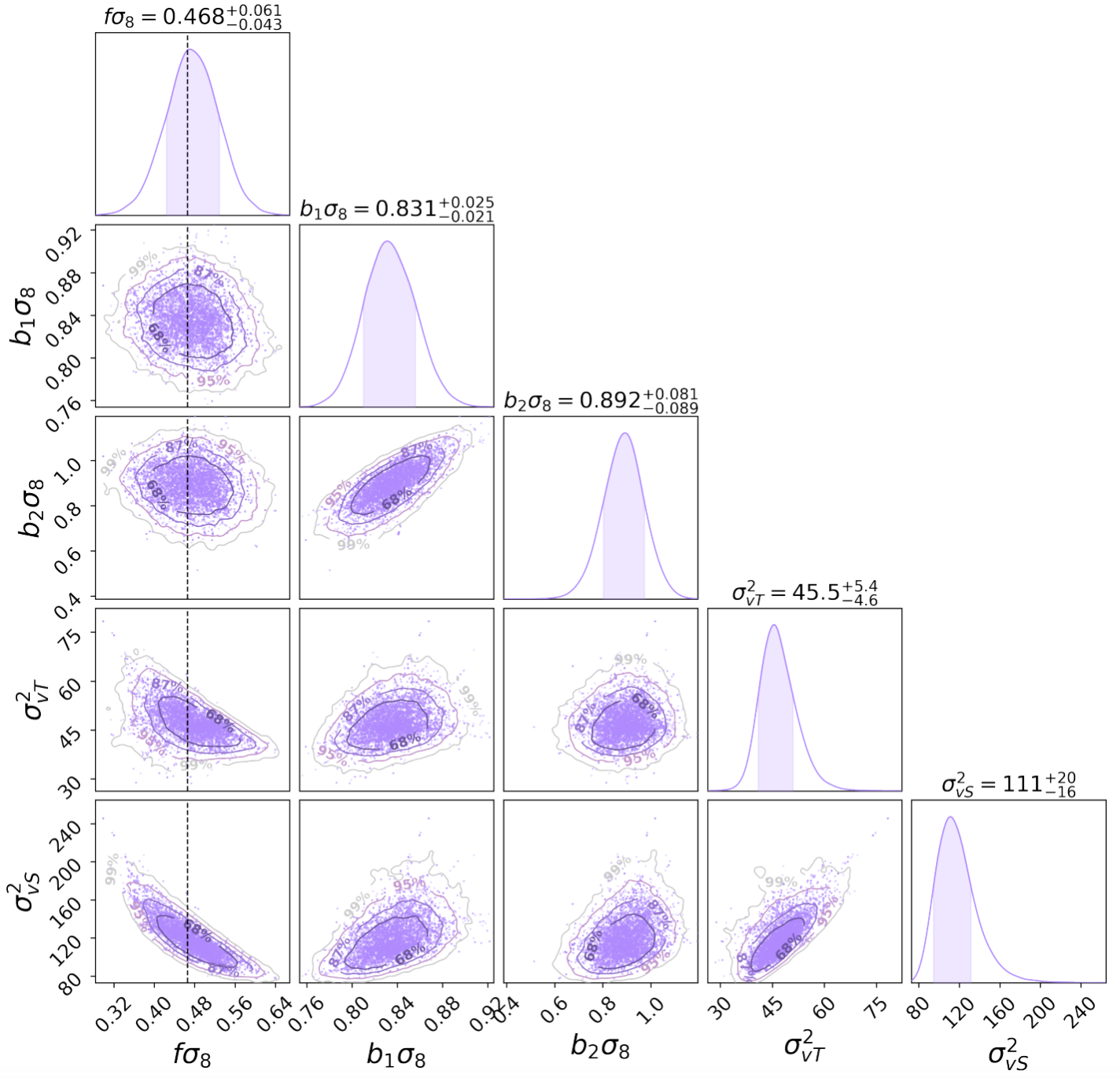}
 \caption{ 
 The power spectrum and parameter fitting outcomes from the BGS and DESI-PV mocks are displayed. In the left panels, the filled circles represent the averaged measurements of the density monopole, density quadrupole, momentum monopole, and cross dipole power spectrum, respectively, across 675 mocks. The error bars reflect the uncertainty associated with a single realization. The   
 fitted model power spectrum are overlaid as curves. On the right, the marginalized histograms and two-dimensional contours of the MCMC samples for the cosmological parameters are shown, with the MCMC fit results annotated at the top of each histogram (or see Table \ref{tabs2last2}). The 2D contours delineate the 1, 1.5, 2, and 2.5$\sigma$ confidence levels, while the shaded regions in the histograms indicate the $1\sigma$ confidence interval. The vertical dashed line marks the fiducial value $f\sigma_8 = 0.466$.  }
 \label{pltfsig8mock}
\end{figure*}

We begin by fitting the power spectrum of the mock average. As illustrated in Fig. \ref{pltfsig8mock}, the filled circles represent the average of the power spectrum measured from 675 mocks, while the solid curves depict the  
best-fit theoretical models. The density power spectrum is measured from the BGS mocks, whereas the momentum power spectrum is obtained from the DESI-PV mocks. Each cross-power spectrum is computed using a BGS mock and its corresponding DESI-PV mock (which has the same observer as that BGS mock). 

The MCMC resulting parameter estimates are summarized in Table \ref{tabs2last2} and illustrated in the right panel of Fig.\ref{pltfsig8mock}. Notably, the estimated  value of $f\sigma_{8}=0.468^{+0.061}_{-0.043}$, which closely aligns with the mock fiducial value of $f\sigma_{8,\rm fid}=0.466$ at effective redshift $z_{\rm eff}=0.2$.   This agreement demonstrates that our fitting methodology effectively recovers the true growth rate embedded in the mocks, and that the power spectrum models perform robustly up to the non-linear scale of $k_{\text{max}}$=0.3 $h$ Mpc$^{-1}$ under current conditions. In this analysis, we exclude the momentum power spectrum quadrupole, $P^p_{2}$, as it is heavily dominated by noise in  survey data. To further assess the adaptability of our model, we also conducted fits incorporating $P^p_{2}$ using the mock average, where $P^p_{2}$ exhibits sufficient smoothness to yield a reliable fit. For a more detailed exploration, please refer to Appendix \ref{gtrnew}. 

   The $\chi^2$ values reported in Table \ref{tabs2last2} are computed using the mean of the 675 catalogs, with uncertainties scaled according to a single realization. The reduced $\chi^2$ is 0.08 which is much sammler than one since we are fitting the mock mean with a single-mock covariance. However, when multiplying the $\chi^2$ by a factor of 675, the resulting value becomes excessively large. This discrepancy arises because, although the models demonstrate sufficient accuracy for our current cosmological volume, its precision is inadequate at a scale 675 times larger than our survey volume. Attempting to fit real observational data in such a vast cosmological volume would therefore yield an unsatisfactory fit. Fortunately, we are not yet at a stage where such discrepancies constitute a critical systematic concern.

\begin{table}
\centering
\caption{The MCMC-fit cosmological parameter estimates  derived from mock average.}
\resizebox{\columnwidth}{!}{
\begin{tabular}{cccccc}
\hline\noalign{\vskip 5pt}
$f\sigma_8$ & 
$b_1\sigma_8$& 
$b_2\sigma_8$ & 
$\sigma^2_{vT}$ &
$\sigma^2_{vS}$ &
$\chi^2/$d.o.f 
\\
\noalign{\vskip 5pt}
\hline
\noalign{\vskip 5pt}
$0.468^{+0.061}_{-0.043}$  & 
$0.831^{+0.025}_{-0.021}$ & 
$0.892^{+0.081}_{-0.089}$ & 
$45.5^{+5.4}_{-4.6}$ & 
$111^{+20}_{-16}$ & 
$9.625/(120-5)$ \\
\noalign{\vskip 5pt}
\hline
\end{tabular}
}
\tablefoot{These parameter estimates are derived from the BGS and DESI-PV mocks are presented here. These estimates are obtained by fitting the average of 675 BGS mocks and 675 DESI-PV mocks. The degrees of freedom (d.o.f.) are calculated based on 120 data points and 5 free parameters.}
\label{tabs2last2}
\end{table}

\begin{figure} 
\includegraphics[width=\columnwidth]{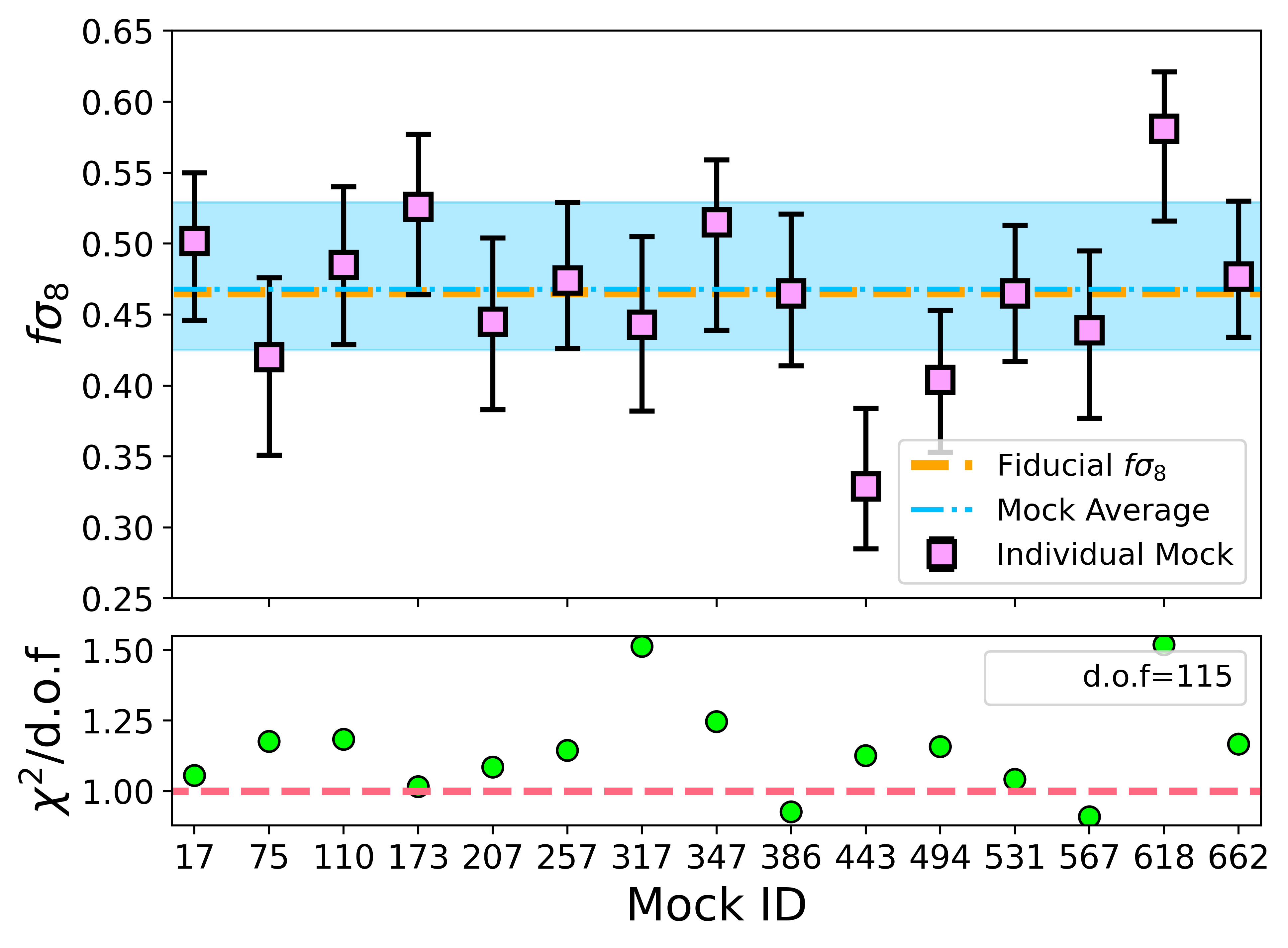}
 \caption{The $f\sigma_8$ fit results (represented by filled squares in the top panel) from 15 randomly selected BGS mocks and their corresponding DESI-PV mock counterparts. In the top panel, the yellow dashed line indicates the fiducial value $f\sigma_8 = 0.466$ of mocks. The blue band represents the $f\sigma_8$ value from Table \ref{tabs2last2}. The bottom panel illustrates the reduced $\chi^2$ values for each measurement, depicted by green filled circles. }
 \label{pltfsig7ck}
 \end{figure}

We proceed to fit $f\sigma_8$ of 15 randomly selected DESI-PV mocks and their corresponding BGS mock counterparts. As illustrated in the top panel of Fig.\ref{pltfsig7ck}, the filled squares represent the MCMC estimates of $f\sigma_8$ derived from each of the 15 mock sets, all of which align closely with the fiducial $f\sigma_8$ value (indicated by the yellow dashed line). The blue shaded band displays the $f\sigma_8$ constraint extracted from Table \ref{tabs2last2} (i.e. mock average), providing an independent reference for comparison. The reduced $\chi^2$ values for all 15 measurements are found to be near unity, underscoring the robustness of the model and confirming that it provides an excellent statistical fit to the mock data.

\begin{figure} 
\includegraphics[width=\columnwidth]{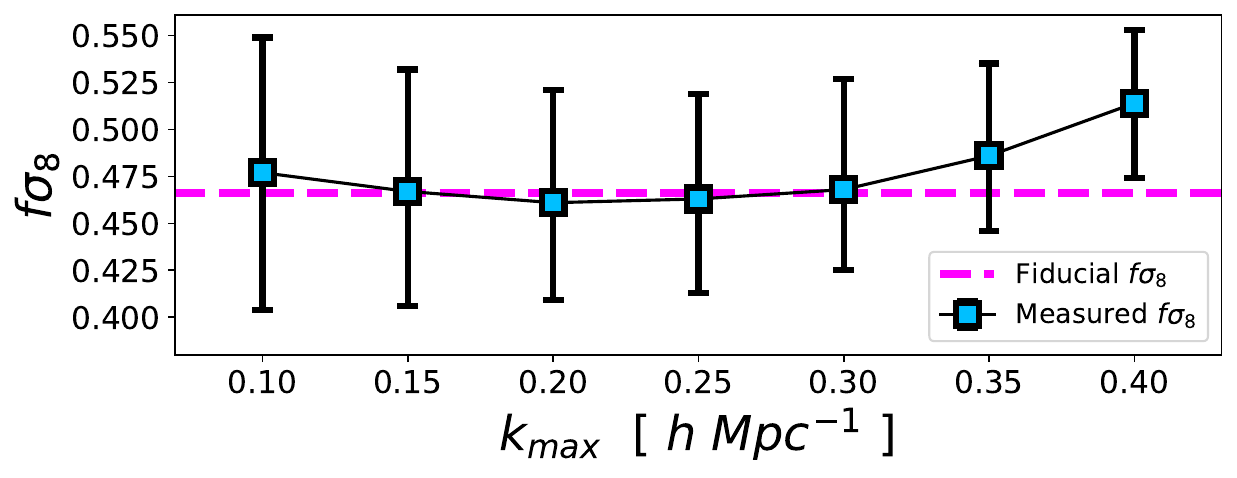}
 \caption{  { 
 The blue filled-squares show the estimated $f\sigma_8$ as a function of cut-off wave number $k_{max}$. The pink dashed-
line is the fiducial value 0.466. 
  } }
 \label{plt09ftt33}
 \end{figure}

{  
Fig.\ref{plt09ftt33} presents the growth rate constraints derived from fitting the power spectrum using different wave number cut-offs $k_{max}$, demonstrating that our models yield stable fits and reliable for $k_{max} < 0.35~h$ Mpc$^{-1}$.
}

\section{Results}\label{sec:results}

 \subsection{The systematic error of the FP data}

\begin{figure} 
\centering
 \includegraphics[width=\columnwidth]{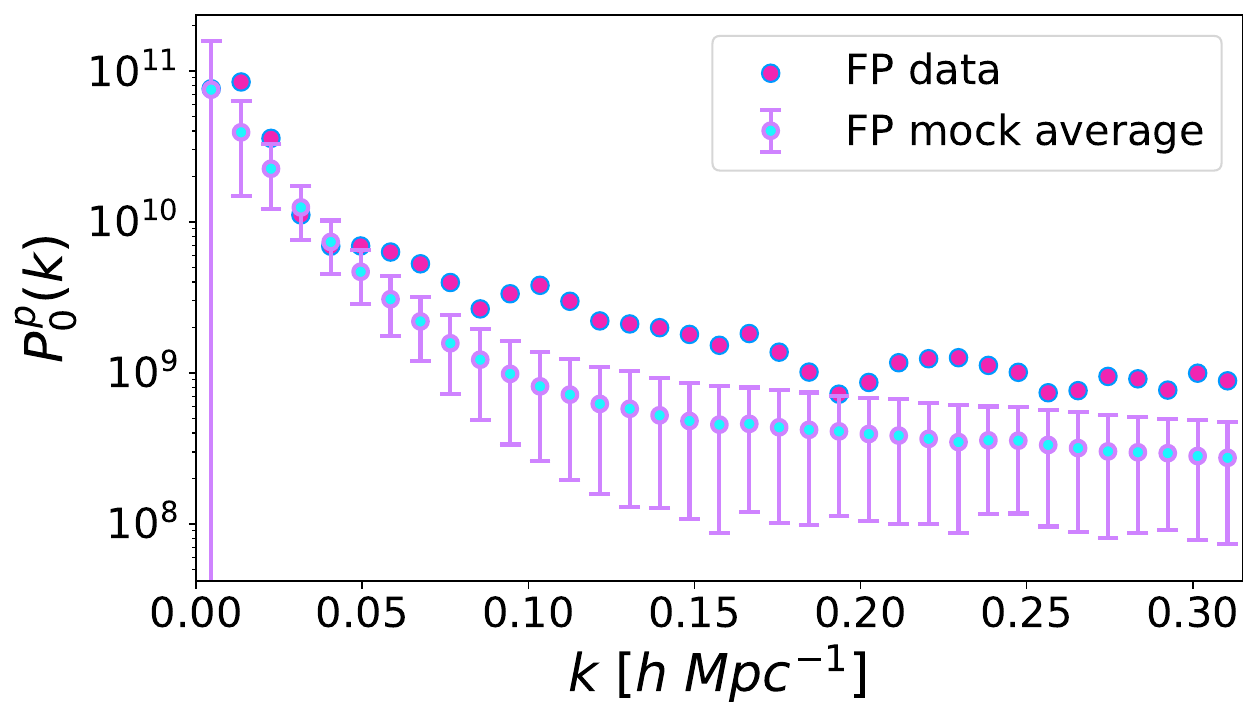}
 \caption{ We compare the momentum power spectrum derived from the FP data (represented by magenta filled circles) with the averaged spectrum obtained from FP mocks (illustrated by blue filled circles accompanied by error bars).   } 
 \label{pltcompFP}
\end{figure}

In Fig.\ref{pltcompFP}, we compare the measured momentum power spectrum of the FP survey data with the average momentum power spectrum derived from 675 FP mocks. As the plot illustrates, the momentum power spectrum from the FP data begins to deviate - specifically, it becomes elevated relative to the mocks starting at $k = 0.05\ h\ \text{Mpc}^{-1}$, and this discrepancy becomes even more pronounced at $k = 0.1\ h\ \text{Mpc}^{-1}$, where the data exhibits significantly higher values than the mock predictions. This bias originates from a systematic bias in the measured $\eta$ values of the FP data. In particular, within larger galaxy groups, the measured $\eta$ tends to be more negative, leading to this anomalous behavior in the momentum power spectrum.
This bias was initially identified by \cite{Howlett2022} through an analysis of the SDSSv data and has since been confirmed in DESI FP data, as reported by \cite{DESIPV_Ross}. However, the underlying physical cause of this bias remains elusive and not yet fully understood. In future  endeavors, a comprehensive galaxy group catalog of DESI will be meticulously constructed, serving as a pivotal foundation for the calibration of this bias. 

\begin{figure} 
 \includegraphics[width=\columnwidth]{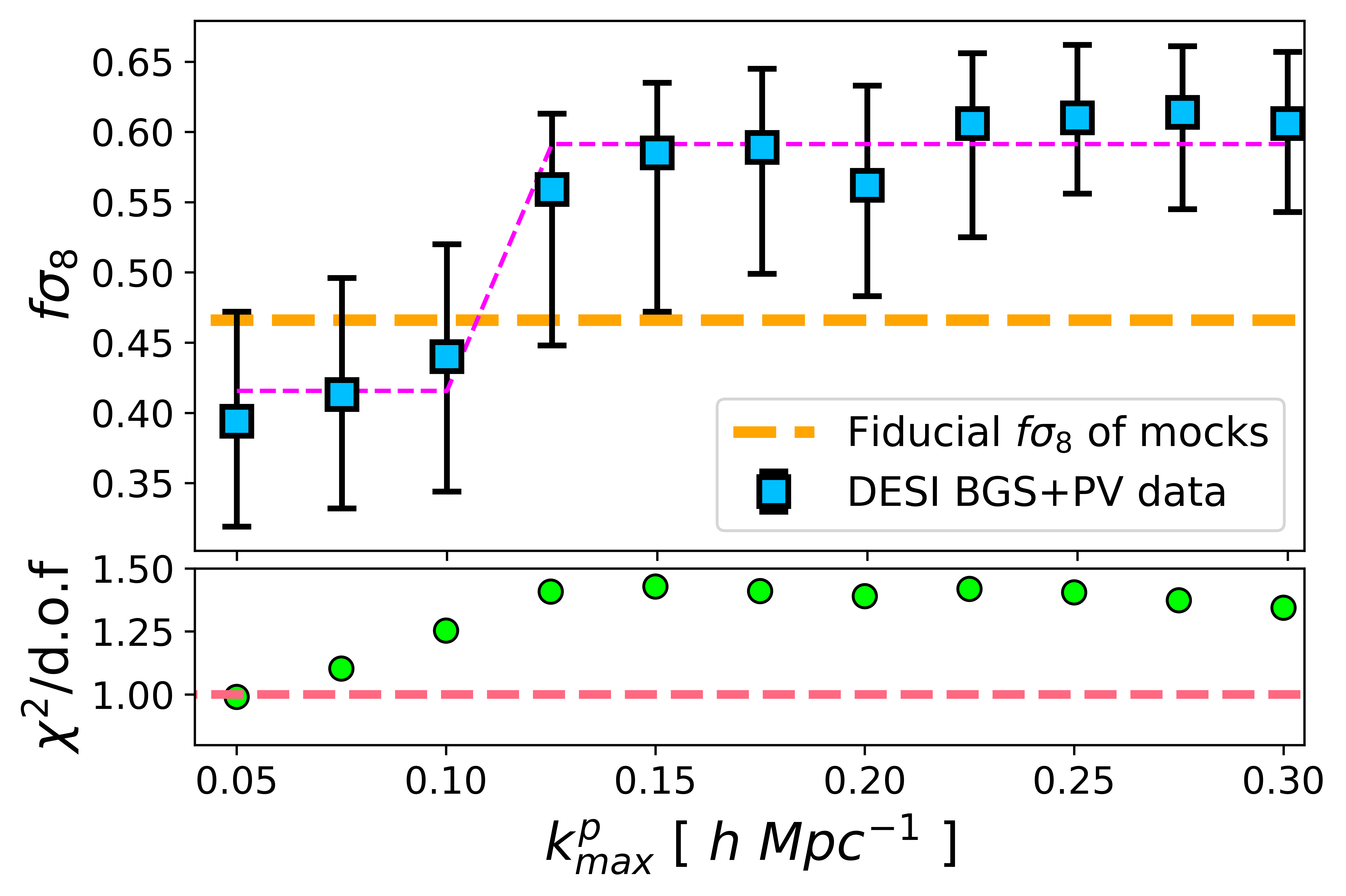}
 \caption{  In the top panel, the blue-filled squares show the estimated growth rate $f\sigma_8$ as a
function of the cutoff wavenumber $k^p_{max}$ of momentum power spectrum. We fix $k_{\text{max}} = 0.3\ h\ \text{Mpc}^{-1}$ for the density and cross power spectrum.  They are  derived from   the 
BGS and DESI-PV  survey data.  The magenta-colored dashed line indicates the average of the blue-filled squares. The yellow dashed line indicates the fiducial value $f\sigma_8 = 0.466$ of mocks.   The bottom panel illustrates the reduced $\chi^2$ values for each measurement, depicted by green filled circles. } 
 \label{pltcompFPfsig}
\end{figure}

This systematic deviation also affects the estimation of $f\sigma_8$, as the DESI-PV catalog is predominantly composed of FP galaxies rather than TF galaxies. As shown in the top panel of Fig.\ref{pltcompFPfsig}, the blue-filled squares represent the estimated values of the growth rate $f\sigma_8$ (derived from the power spectrum of the BGS and DESI-PV data) as a function of the cutoff wavenumber $k^p_{\text{max}}$ of momentum power spectrum. We fix $k_{\text{max}} = 0.3\ h\ \text{Mpc}^{-1}$ for the density and cross power spectrum. Notably, at $k^p_{\text{max}} = 0.125\ h\ \text{Mpc}^{-1}$, the measured $f\sigma_8$ abruptly rises from approximately 0.4 to around 0.6.  

To mitigate this bias, we have to adopt a conservative approach by restricting the momentum power spectrum analysis to $k^p_{\text{max}} = 0.1\ h\ \text{Mpc}^{-1}$ when fitting $f\sigma_8$ from  survey data, while retaining $k_{\text{max}} = 0.3\ h\ \text{Mpc}^{-1}$ for the density and cross power spectrum. The information encoded in the momentum power spectrum is predominantly concentrated on the larger cosmological scales, specifically at wavenumbers below $k \le 0.1\ h\ \text{Mpc}^{-1}$ \citep{Howlett2019,DESIPV_Bautista}. Consequently, adopting a conservative cutoff at $k^p_{\text{max}} = 0.1\ h\ \text{Mpc}^{-1}$ enables a robust analysis of the data without sacrificing significant signal content and does not impact our main conclusion.

It is important to emphasize that $k^p_{\text{max}} = 0.1\ h\ \text{Mpc}^{-1}$ reflects the limit of reliability for the measured momentum power spectrum, rather than a breakdown of the theoretical model. Indeed, the model of momentum power spectrum remains robust and accurate up to $k_{\text{max}} = 0.3\ h\ \text{Mpc}^{-1}$ as demonstrated in Section \ref{sec:5.2}.
In future studies, with access to an unbiased velocity survey  
, we anticipate being able to extend the momentum power spectrum analysis to $k_{\text{max}} = 0.3\ h\ \text{Mpc}^{-1}$, thereby enhancing the precision and constraining power of our cosmological inferences.

\subsection{Result of $f\sigma_8$}

\begin{figure*}  
\centering
\includegraphics[width=80mm]{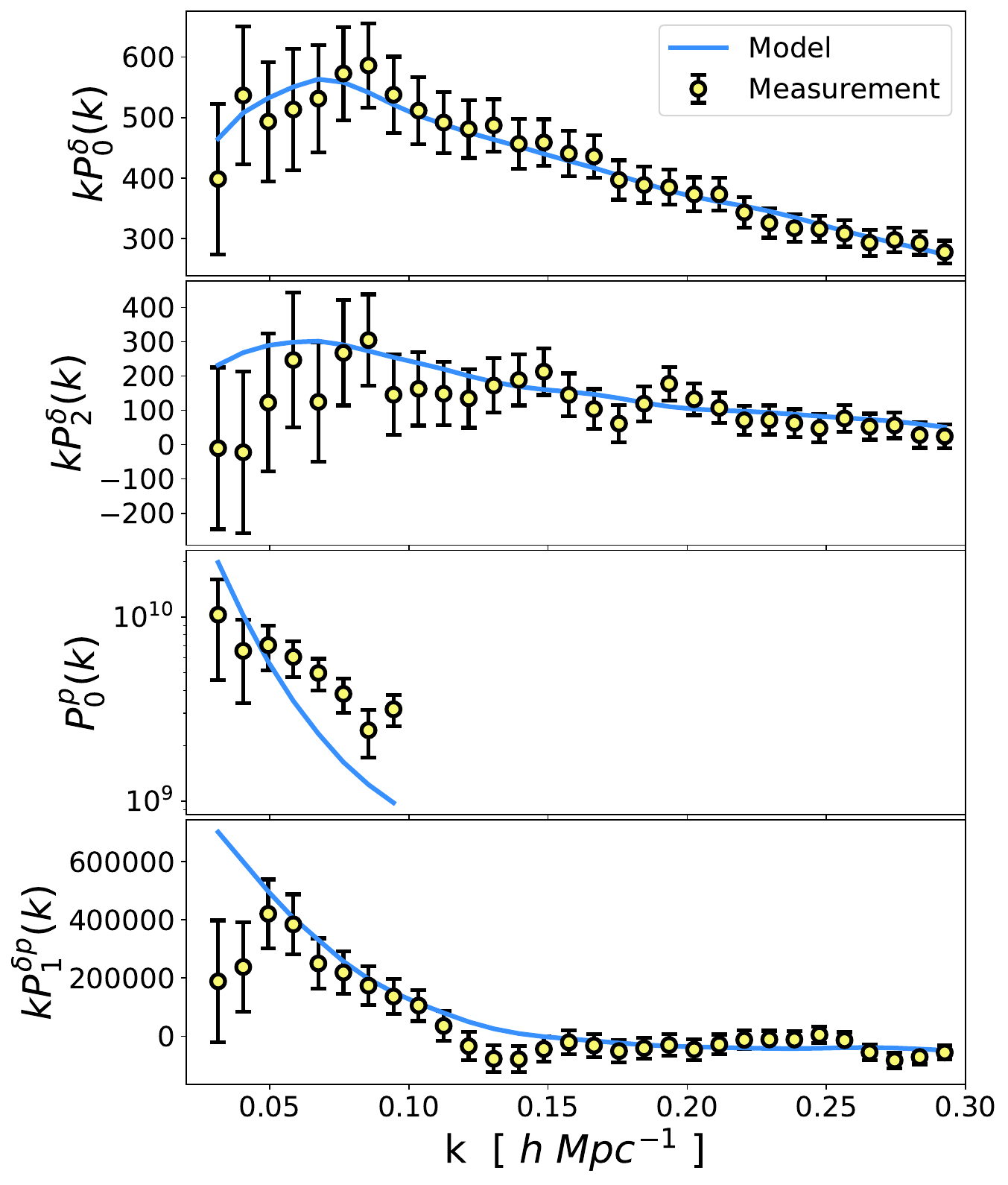}
\includegraphics[width=100mm]{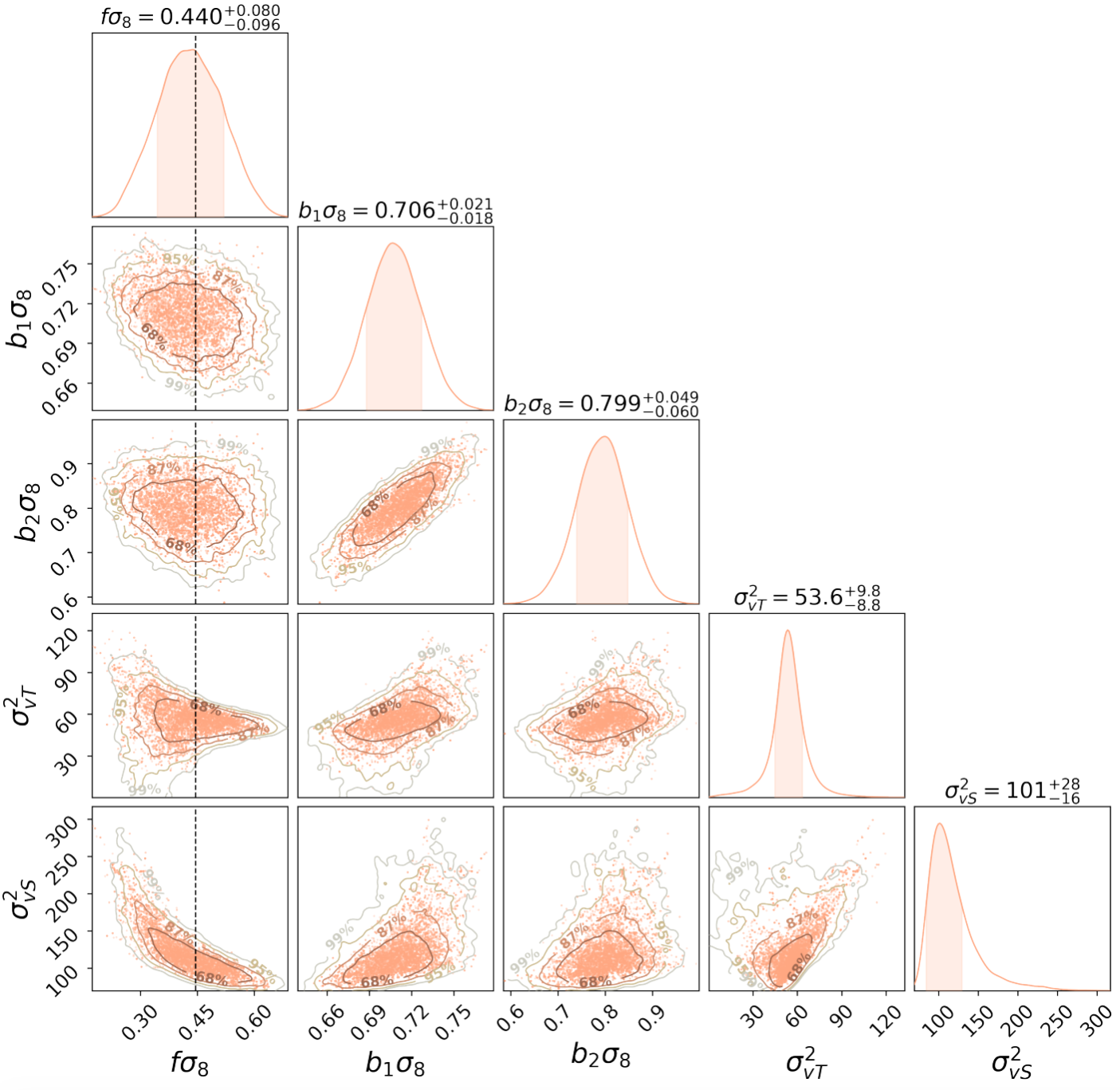}
 \caption{ Same as Fig.\ref{pltfsig8mock}, but for the BGS data and DESI-PV data. In the right panel,  the vertical dashed line marks the GR+\cite{Planck2020} prediction $f\sigma_8=0.446$. 
 The corresponding marginalized parameter values are summarized in Table \ref{tabs2lassst2}. }
 \label{pltfsig8survey}
\end{figure*}

As depicted in Fig.\ref{pltfsig8survey}, the power spectrum fitting results for the BGS and DESI-PV data are presented. In the left-hand-side panels, the top, middle, and bottom subplots display the measured density monopole and quadrupole, momentum monopole, and cross dipole power spectrum, respectively, with   filled circles representing the measurements of the data. The curves illustrate the theoretical model spectrum that have been fitted to the measured data points. The goodness-of-fit is quantified by a reduced chi-squared value of $\chi^2/\mathrm{d.o.f}=116.563/(98-5)=1.2534$, indicating a satisfactory agreement between model and data. The derived cosmological parameter constraints are summarized in Table~\ref{tabs2lassst2} and  illustrated in the right panel of Fig.\ref{pltfsig8survey}. Notably, the   estimate for the growth rate is 
$   
f\sigma_8(z_{\rm eff} = 0.07)=0.440^{+0.080}_{-0.096} 
$
at the effective redshift $z_{\rm eff} = 0.07$.

To compute the theoretically predicted value of $f\sigma_8$, we follow the formalism outlined in \cite{Howlett2015correlationfunction}, wherein the growth rate as a function of redshift is expressed as
\be  \label{fsig8th}
f(a)\sigma_8(a)=\Omega_{m}(a)^{\gamma}\sigma_{8}\frac{D_{gr}(a*)}{D_{gr,0}}\frac{D_{\gamma}(a)}{D_{\gamma}(a*)},
\ee 
with
\be  
\frac{D_{\gamma}(a)}{D_{\gamma}(a*)}=\mathrm{exp}\left( \int^a_{a*}\Omega_{\rm m}(a')^{\gamma}d{\rm ln}a' \right)~,~~\Omega_{\rm m}(a)=\frac{\Omega_{\rm m}}{a^3E(a)^2}.
\ee 
and $D_{gr}(a)$ obtained from Eq.\ref{theDgr}. Under the framework of General Relativity (GR), the growth index $\gamma$ is fixed at 0.55 \citep{Linder:2005in}. Consequently, with the \cite{Planck2020} cosmology, the GR predicts a growth rate of $f\sigma_8 = 0.446$ at $z_{\rm eff} = 0.07$, which aligns remarkably well with our measured value of $f\sigma_8 = 0.440^{+0.080}_{-0.096}$ at the 68\% confidence level, underscoring the consistency between theory and observation.

 The   refined estimate of the growth rate, derived by synthesizing the fitting outcomes from correlation function \citep{DESIPV_Turner}, maximum likelihood estimation (MLE,  \citealt{DESIPV_Lai}) and power spectrum of this paper, yields a consensus value of $f\sigma_8(z_{\text{eff}} = 0.07) = 0.450 ^{+0.055}_{-0.055}$, in   concordance with theoretical predictions too. For a deeper exploration of data synthesis, see \cite{DESIPV_Bautista}.

\subsection{Result of $\gamma$  }\label{sec:gamma}

We present constraints on Linder’s growth index $\gamma$ (see Eq.\ref{fsig8th}) assuming a $\Lambda$CDM background, obtained with \texttt{cobaya} \citep{Torrado:2020dgo} interfaced to \texttt{MGCAMB} \citep{Wang:2023tjj,Hojjati:2011ix} using \texttt{MG\_flag}=2. In this setting, $\gamma$ is mapped into the modified scale and redshift-dependent Poisson factor $\mu(a,k)$. In our analysis, we use DESI DR1 ShapeFit measurements in six redshift bins, distance-scale information from the post-reconstruction correlation function, and the Baryon Acoustic Oscillations (BAO) Ly$\alpha$ likelihood used in DR1 \citep{DESI2024.IV.KP6}\footnote{For details regarding the ShapeFit compression, we direct readers to section 4 of \cite{DESI2024.V.KP5}.}. We sample the background parameters in addition to $\gamma$ and report $\sigma_8$ as a derived quantity.  
To mitigate projection effects, we apply a Big-Bang Nucleosynthesi (BBN) Gaussian prior $\omega_b\sim\mathcal{N}(0.02218, 0.00055)$ and a broad Gaussian prior on $n_s$ centered on the Planck fit value with standard deviation inflated by a factor of ten, $\pi(n_s)\sim\mathcal{N}(0.9649, 0.042)$ \citep{Planck2020}. The addition of PV $f\sigma_8$ measurements with DESI DR1 ShapeFit breaks the $\sigma_8$–$\gamma$ degeneracy and sharpens the constraints on all parameters.

\begin{figure}
\centering
\includegraphics[width= \columnwidth]{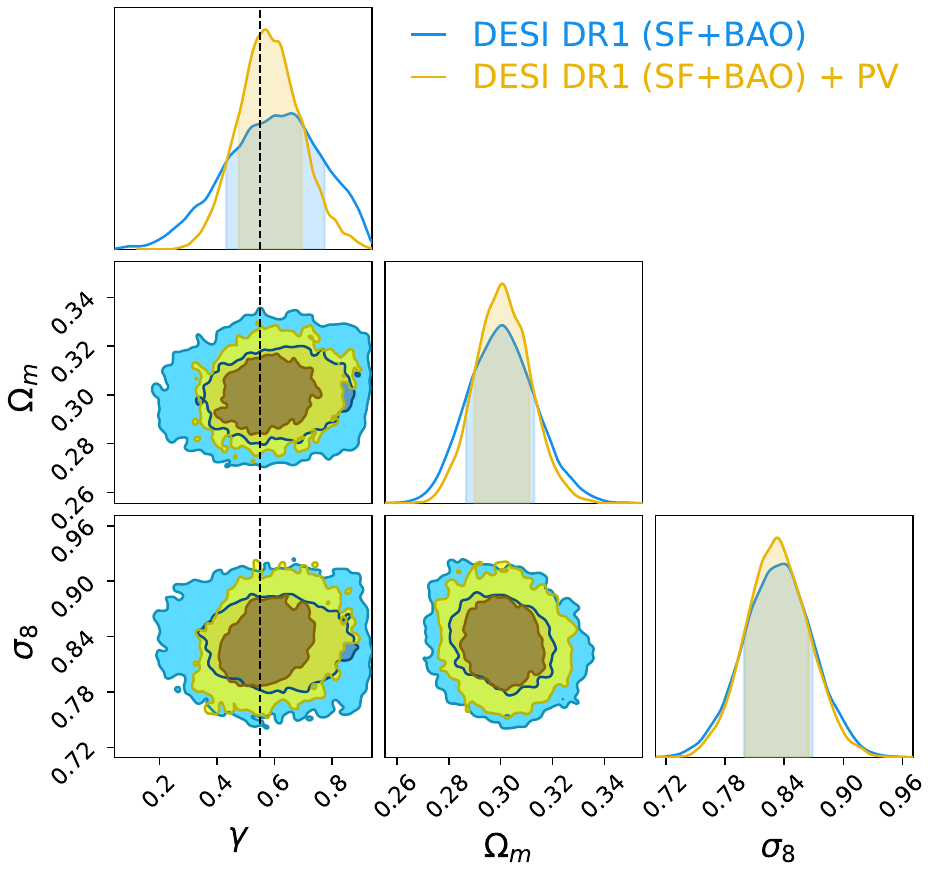}
\caption{Constraints on $(\gamma,\Omega_{\rm m}, \sigma_8)$.  DESI DR1 (SF+BAO) (blue) and its combination with PV consensus $f\sigma_8$ (orange). Shaded regions show 68\% and 95\% credible contours. The vertical dashed line marks the GR prediction $\gamma=0.55$.  
}
\label{fig:gamma_sigma8_desi}
\end{figure}

 Fig.\ref{fig:gamma_sigma8_desi} shows the  fitting results for  $(\gamma,\Omega_{\rm m}, \sigma_8)$ obtained from DESI DR1 ShapeFit (SF) + BAO, combined with  our PV consensus $f\sigma_8 = 0.450 ^{+0.055}_{-0.055}$   (orange contours). We find $\gamma=0.580^{+0.110}_{-0.110}$, 
$\Omega_\mathrm{m}=0.301^{+0.011}_{-0.011} $ and $\sigma_8=0.834^{+0.032}_{-0.032}$. Our result is consistent with GR. The PV $f \sigma_8$ information reduced the errors of $\gamma$, $\Omega_{\rm m}$ and $\sigma_8$ by $\approx 32\%$, $\approx 33\%$ and $\approx 20\%$ respectively, relative to SF+BAO alone (see blue contours or Table \ref{tab:gamma_table}).  
Our measurement of $\gamma$ falls notably below the values reported by \cite{Nguyen2023} and \cite{Calabrese2025}, who obtain higher estimate values of $\gamma \sim 0.65$.

\begin{table}
\centering
\caption{The MCMC-fit cosmological parameter estimates derived from data.}
\resizebox{\columnwidth}{!}{
\begin{tabular}{cccccc}
\hline\noalign{\vskip 5pt}
$f\sigma_8$ & 
$b_1\sigma_8$& 
$b_2\sigma_8$ & 
$\sigma^2_{vT}$ &
$\sigma^2_{vS}$ &
$\chi^2/$d.o.f 
\\
\noalign{\vskip 5pt}
\hline
\noalign{\vskip 5pt}
$0.440^{+0.080}_{-0.096}$  & 
$0.706^{+0.021}_{-0.018}$ & 
$0.799^{+0.049}_{-0.060}$ & 
$53.6^{+9.8}_{-8.8}$ & 
$101.0^{+28}_{-16}$ & 
$116.563/(98-5)$ \\
\noalign{\vskip 5pt}
\hline
\end{tabular}
}
\tablefoot{These parameter estimates are derived from
the BGS and DESI-PV    data are presented here. The degrees of freedom (d.o.f.) are calculated based on 98 data points and 5 free parameters.}
\label{tabs2lassst2}
\end{table}

\begin{table}
\centering
\caption{The estimated values of $\gamma$, $\Omega_{\rm m}$ and $\sigma_8$.}
\resizebox{\columnwidth}{!}{
\begin{tabular}{lccc}
\noalign{\vskip 3pt}
\hline\noalign{\vskip 4pt}
\multicolumn{1}{c}{\bf Dataset} & {\boldmath$\gamma$} & $\Omega_\mathrm{m}$ & $\sigma_8$ \\ \noalign{\vskip 3pt}
\hline\noalign{\vskip 3pt}
DESI DR1 SF+BAO+PV      & $0.580^{+0.110}_{-0.110} $    & $0.301^{+0.011}_{-0.011} $          & $0.834^{+0.032}_{-0.032}$\\\noalign{\vskip 4pt}
DESI DR1 SF+BAO           & $0.610^{+0.160}_{-0.160}$              & $0.301^{+0.013}_{-0.017}$   & $0.832^{+0.039}_{-0.039}$ \\\noalign{\vskip 3pt}
\hline
\end{tabular}
}
\label{tab:gamma_table}
\end{table}

\section{Discussion}\label{sec:discus}

\begin{figure*}  
 \centering
\includegraphics[width=185mm]{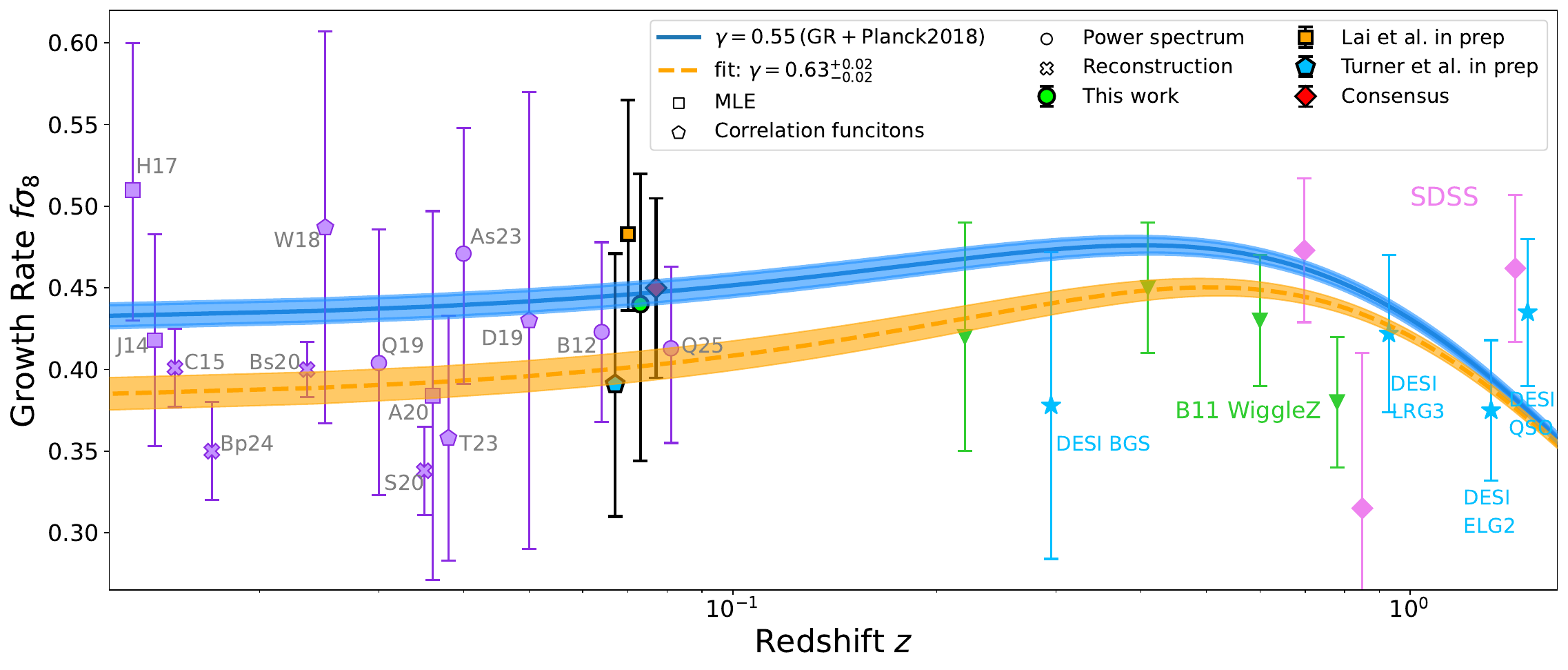}
 \caption{The growth rate $f\sigma_8$ as a function of redshift $z$ is presented. The blue curve represents the theoretical prediction  derived under the assumption of GR ($\gamma=0.55$) and a $\Lambda$CDM cosmology calibrated using \cite{Planck2020}. The green-filled circle denotes the measurement obtained in this study. The orange-filled square corresponds to the result from \cite{DESIPV_Lai} (using MLE), based on the same dataset, while the blue-filled pentagon reflects the measurement from \cite{DESIPV_Turner} (using correlation functions), also utilizing identical data. The red-filled diamond corresponds to the consensus result of these  three measurements. Additional observational constraints are illustrated by purple circles (using power spectrum), purple squares (using MLE), purple cross (using reconstruction) , and purple pentagons (using correlation functions), representing measurements from other surveys. The light-blue pentagrams  corresponds to the results from DESI DR1. The orange-dashed curve indicates the model fits to all the data points, with the shaded orange region depicting the associated uncertainties.    }
 \label{pltfsig11fz}
\end{figure*}

Fig.\ref{pltfsig11fz} displays the evolution of the growth rate $f\sigma_8$ as a function of redshift $z$. The blue curve is derived from Eq.\ref{fsig8th}, under the assumption of General Relativity and a $\Lambda$CDM cosmological model based on \cite{Planck2020}.
The green-filled circle represents the measurement derived from this paper. In comparison with the results presented in \cite{DESIPV_Turner} (blue-filled pentagon) and \cite{DESIPV_Lai} (orange-filled square), our measurement exhibits slightly broader error margins, a consequence of adopting a lower wave number cutoff $k^p_{\text{max}} = 0.1\ h\ \text{Mpc}^{-1}$ in the analysis of the momentum power spectrum, relative to the higher thresholds employed in the other two studies. The red-filled diamond denote the consensus outcome of these three measurements.  
The other measurements are:
H17: \citet{Howlett2017velocitypower};
J14: \citet{Johnson2014};
A20: \citet{Adams2020};
Q19: \citet{Qin2019PS};
Q25: \citet{Qin2025};
B12: \citet{Beutler2012};
As23: \citet{Appleby2023};
T23: \citet{Turner2023correlation};
D19: \citet{Dupuy2019};
W18: \citet{YuyuWang2018};
C15: \citet{Carrick2015};
S20: \citet{Said2020};
Bp24: \citet{Boubel2024};
Bs20: \citet{Boruah2020};
DESI: \citet{DESI2024.VII.KP7B};
SDSS: \citet{SDSS2021};
B11 WiggleZ: \citet{Blake2011}.

Although these measurements largely align with the predictions of General Relativity, a subtle discrepancy becomes apparent when they are analyzed in combination. To further investigate this, we integrate the aforementioned measured $f\sigma_{8}$ values (i.e. all the data points in Fig.\ref{pltfsig11fz}, we only use the red-filled diamond for DESI-PV) with the Planck chain\footnote{We use the chain   \url{base_plikHM_TTTEEE_lowl_lowE_lensing_1.txt}  in  \url{CosmoParams_base-plikHM_R3.01} on \url{https://irsa.ipac.caltech.edu/data/Planck/release_3/ancillary-data/} as a prior on $\Omega_{m}$.} 
and perform a fitting analysis for $\gamma$. The resulting fit value of $\gamma=0.63\pm0.02$ is then transformed into a corresponding range of $f\sigma_{8}$ values across different redshifts and is presented as the orange-colored curve in Fig.\ref{pltfsig11fz}. This result indicates a moderate tendency in the data toward a higher $\gamma$ value, implying a potentially weaker gravitational model compared to the predictions of General Relativity (blue curve in Fig.\ref{pltfsig11fz}). However, it is important to note that the  fit value and associated uncertainty for this $\gamma$ should be interpreted with caution, as there exists considerable overlap—and thus covariance—among many of the measurements in Fig.\ref{pltfsig11fz}, which has not been fully incorporated into the analysis. Nonetheless, as shown in Fig.\ref{pltfsig11fz}, a minor tension remains between current observational data and the theoretical predictions of General Relativity, a discrepancy that may be further clarified or resolved through future research and surveys.

\section{Conclusions}\label{sec:conclustion}

In this paper, we build upon the foundational research presented in \papI, \papII ~and \papIII~ by conducting a comprehensive joint analysis of the monopole and quadrupole moments of the auto-density power spectrum, the monopole of the auto-momentum power spectrum, and the dipole component of the cross-power spectrum to measure the growth rate $f\sigma_8$. The density field is derived from the BGS, whereas the momentum field is extracted from the DESI-PV. Our investigation yields the following key findings:
\begin{enumerate}
\item{We have systematically presented the power spectrum estimators and the window function convolution matrix. }
\item{We have refined the theoretical model of the cross-power spectrum and extended its applicability to scenarios in which the density and momentum fields originate from different survey catalogues. }
\item{Utilizing mocks, we demonstrate that the power spectrum models exhibit robust performance up to the non-linear scale of $k_{\rm max}$=0.3 $h$ Mpc$^{-1}$ when fitting $f\sigma_8$ using the set  $[P^{\delta}_0,P^{\delta}_2, P^{p}_0,P^{\delta p}_1]$ and treating  $[f\sigma_{8},b_{1}\sigma_{8},b_{2}\sigma_{8},   \sigma^2_{vT},\sigma^2_{vS}]$ as free parameters.}
 \item{Furthermore, our analysis of mocks reveals that the models remain reliable up to the non-linear scale of  $k_{max}$=0.25 $h$ Mpc$^{-1}$ when fitting  $f\sigma_8$ using $[P^{\delta}_0,P^{\delta}_2, P^{p}_0, P^{p}_2, P^{\delta p}_1]$ and allowing   $[f\sigma_{8}, b^{\delta}_{1}\sigma_{8}, b_{2}\sigma_{8}, b^p_{1}\sigma_{8}, \sigma^2_{\delta,vT}, \sigma^2_{vS}, \sigma^2_{p,vT}]$ to vary freely. }
\item{Our results indicate that the systematic bias of FP data enhances the measured momentum power spectrum   at $k>0.1$ $h$ Mpc$^{-1}$.  Consequently, we restrict our fitting range of momentum power spectrum to $k\le0.1$ $h$ Mpc$^{-1}$ to mitigate this effect.}
\item{We obtain a estimate of the growth rate as  $f\sigma_8=0.440^{+0.080}_{-0.096}$ at $z_{\rm eff}=0.07$. The   refined estimate of the growth rate, derived by synthesizing the fitting outcomes from \cite{DESIPV_Turner}, \cite{DESIPV_Lai} and this paper, yields a consensus value of $f\sigma_8 = 0.450 \pm 0.055$. They are in agreement with the prediction of the $\Lambda$CDM and GR models.}
\item{We obtain a estimate of $\gamma$ as  $\gamma=0.580^{+0.110}_{-0.110}$  (corresponding to the consensus $f\sigma_8$), which is in agreement with the prediction of GR. Correspondingly, we find $\Omega_\mathrm{m}=0.301^{+0.011}_{-0.011}$ and $\sigma_8=0.834^{+0.032}_{-0.032}$.}
\item{We have explored the Galilean transformation of the power spectrum and conclude that its impact is negligible within the context of our analysis.}
\item{Finally, while most individual measurements  align well with the GR prediction, a joint analysis reveals a higher $\gamma=0.63\pm0.02$, suggesting weaker gravitational effects.}
 \end{enumerate}
 
There are several promising avenues for future research building upon this work: (1) Applying our methodology to the Data Release 2 of DESI peculiar velocity survey, which is significantly more homogeneous and extensive than the Data Release 1, thereby enabling a more precise constraint on $f\sigma_8$; (2) While the current analysis assumes the local plane-parallel approximation, future efforts will focus on incorporating wide-angle effects into both our models and estimators; (3) As highlighted in Appendix \ref{sec:gali} of this paper, the momentum power spectrum and cross power spectrum lack Galilean invariance, making them susceptible to potential systematic biases in the data. To address this limitation, we aim to transition toward measuring and modeling the pair-wise momentum power spectrum, as proposed by \cite{Chen2025pairwisPS}; (4) We have not yet accounted for the gravitational redshift of galaxies. In forthcoming studies, we intend to extend our framework to measure and model this relativistic effect, following the approach outlined by \cite{Beutler2020}.

\section*{Data availability}

The data used to generate the figures in this 
paper are available at \url{https://zenodo.org/uploads/17672675}. The public code of power spectrum measurements, power spectrum theoretical models and window function convolution are available at \url{https://github.com/FeiQin-cosmologist/Galaxy_Power_Spectrum}, and see Appendix \ref{sec:appsdf1233} for more links. The examples for the utilization of the code can be found in this Jupyter Notebook
\url{https://github.com/FeiQin-cosmologist/Galaxy_Power_Spectrum/blob/main/CosmPSPy/Code/Examp_PS.ipynb}

\begin{acknowledgements}
FQ and JB is supported by the  funding from Excellence Initiative of Aix-Marseille University - A*MIDEX, a French ``Investissements d'Avenir'' program (AMX-20-CE-02 - DARKUNI).
This material is based upon work supported by the U.S. Department of Energy (DOE), Office of Science, Office of High-Energy Physics, under Contract No. DE–AC02–05CH11231, and by the National Energy Research Scientific Computing Center, a DOE Office of Science User Facility under the same contract. Additional support for DESI was provided by the U.S. National Science Foundation (NSF), Division of Astronomical Sciences under Contract No. AST-0950945 to the NSF’s National Optical-Infrared Astronomy Research Laboratory; the Science and Technology Facilities Council of the United Kingdom; the Gordon and Betty Moore Foundation; the Heising-Simons Foundation; the French Alternative Energies and Atomic Energy Commission (CEA); the National Council of Humanities, Science and Technology of Mexico (CONAHCYT); the Ministry of Science, Innovation and Universities of Spain (MICIU/AEI/10.13039/501100011033), and by the DESI Member Institutions: \url{https://www.desi.lbl.gov/collaborating-institutions}. Any opinions, findings, and conclusions or recommendations expressed in this material are those of the author(s) and do not necessarily reflect the views of the U. S. National Science Foundation, the U. S. Department of Energy, or any of the listed funding agencies. The authors are honored to be permitted to conduct scientific research on I'oligam Du'ag (Kitt Peak), a mountain with particular significance to the Tohono O’odham Nation.
\end{acknowledgements}

\bibliographystyle{aasjournal}
\bibliography{ FQinRef.bib   }

\begin{thebibliography}{}
\expandafter\ifx\csname natexlab\endcsname\relax\def\natexlab#1{#1}\fi
\providecommand{\url}[1]{\href{#1}{#1}}
\providecommand{\dodoi}[1]{doi:~\href{http://doi.org/#1}{\nolinkurl{#1}}}
\providecommand{\doeprint}[1]{\href{http://ascl.net/#1}{\nolinkurl{http://ascl.net/#1}}}
\providecommand{\doarXiv}[1]{\href{https://arxiv.org/abs/#1}{\nolinkurl{https://arxiv.org/abs/#1}}}

\bibitem[{{Abate} {et~al.}(2008){Abate}, {Bridle}, {Teodoro}, {Warren}, \&
  {Hendry}}]{Abate2008}
{Abate}, A., {Bridle}, S., {Teodoro}, L. F.~A., {Warren}, M.~S., \& {Hendry},
  M. 2008, \mnras, 389, 1739, \dodoi{10.1111/j.1365-2966.2008.13637.x}

\bibitem[{{Adams} \& {Blake}(2017)}]{Adams2017}
{Adams}, C., \& {Blake}, C. 2017, \mnras, 471, 839,
  \dodoi{10.1093/mnras/stx1529}

\bibitem[{{Adams} \& {Blake}(2020)}]{Adams2020}
---. 2020, \mnras, 494, 3275, \dodoi{10.1093/mnras/staa845}

\bibitem[{{Alam} {et~al.}(2021){Alam}, {Aubert}, {Avila}, {Balland},
  {Bautista}, {Bershady}, {Bizyaev}, {Blanton}, {Bolton}, {Bovy}, {Brinkmann},
  {Brownstein}, {Burtin}, {Chabanier}, {Chapman}, {Choi}, {Chuang}, {Comparat},
  {Cousinou}, {Cuceu}, {Dawson}, {de la Torre}, {de Mattia}, {Agathe}, {des
  Bourboux}, {Escoffier}, {Etourneau}, {Farr}, {Font-Ribera}, {Frinchaboy},
  {Fromenteau}, {Gil-Mar{\'\i}n}, {Le Goff}, {Gonzalez-Morales},
  {Gonzalez-Perez}, {Grabowski}, {Guy}, {Hawken}, {Hou}, {Kong}, {Parker},
  {Klaene}, {Kneib}, {Lin}, {Long}, {Lyke}, {de la Macorra}, {Martini},
  {Masters}, {Mohammad}, {Moon}, {Mueller}, {Mu{\~n}oz-Guti{\'e}rrez}, {Myers},
  {Nadathur}, {Neveux}, {Newman}, {Noterdaeme}, {Oravetz}, {Oravetz},
  {Palanque-Delabrouille}, {Pan}, {Paviot}, {Percival}, {P{\'e}rez-R{\`a}fols},
  {Petitjean}, {Pieri}, {Prakash}, {Raichoor}, {Ravoux}, {Rezaie}, {Rich},
  {Ross}, {Rossi}, {Ruggeri}, {Ruhlmann-Kleider}, {S{\'a}nchez}, {S{\'a}nchez},
  {S{\'a}nchez-Gallego}, {Sayres}, {Schneider}, {Seo}, {Shafieloo}, {Slosar},
  {Smith}, {Stermer}, {Tamone}, {Tinker}, {Tojeiro}, {Vargas-Maga{\~n}a},
  {Variu}, {Wang}, {Weaver}, {Weijmans}, {Y{\`e}che}, {Zarrouk}, {Zhao},
  {Zhao}, \& {Zheng}}]{SDSS2021}
{Alam}, S., {Aubert}, M., {Avila}, S., {et~al.} 2021, \prd, 103, 083533,
  \dodoi{10.1103/PhysRevD.103.083533}

\bibitem[{{Appleby} {et~al.}(2023){Appleby}, {Tonegawa}, {Park}, {Hong}, {Kim},
  \& {Yoon}}]{Appleby2023}
{Appleby}, S., {Tonegawa}, M., {Park}, C., {et~al.} 2023, \apj, 958, 180,
  \dodoi{10.3847/1538-4357/acff68}

\bibitem[{{Bautista} {et~al.}(2025){Bautista}, {Amsellem}, {Aronica}, {BenZvi},
  {Blake}, {Carr}, {Davis}, {Douglass}, {Dumerchat}, {Howlett}, {Lai},
  {Nguyen}, {Palmese}, {Qin}, {Ravoux}, {Ross}, {Said}, {Turner}, {Aguilar},
  {Ahlen}, {Bianchi}, {Brooks}, {Claybaugh}, {Cuceu}, {de la Macorra}, {Doel},
  {Font-Ribera}, {Forero-Romero}, {Gazta{\~n}aga}, {Gontcho}, {Gutierrez},
  {Herrera-Alcantar}, {Honscheid}, {Huterer}, {Ishak}, {Joyce}, {Kremin},
  {Lamman}, {Landriau}, {Le Guillou}, {Leauthaud}, {Manera}, {Meisner},
  {Miquel}, {Moustakas}, {Mu{\~n}oz-Guti{\'e}rrez}, {Nadathur}, {Percival},
  {Prada}, {P{\'e}rez-R{\`a}fols}, {Rossi}, {Sanchez}, {Schlegel}, {Schubnell},
  {Seo}, {Silber}, {Sprayberry}, {Tarl{\'e}}, {Weaver}, {Zarrouk}, {Zhou}, \&
  {Zou}}]{DESIPV_Bautista}
{Bautista}, J., {Amsellem}, A.~J., {Aronica}, V., {et~al.} 2025, arXiv
  e-prints, arXiv:2512.03228, \dodoi{10.48550/arXiv.2512.03228}

\bibitem[{{Beutler} {et~al.}(2019){Beutler}, {Castorina}, \&
  {Zhang}}]{Beutler2019}
{Beutler}, F., {Castorina}, E., \& {Zhang}, P. 2019, \jcap, 2019, 040,
  \dodoi{10.1088/1475-7516/2019/03/040}

\bibitem[{{Beutler} \& {Di Dio}(2020)}]{Beutler2020}
{Beutler}, F., \& {Di Dio}, E. 2020, \jcap, 2020, 048,
  \dodoi{10.1088/1475-7516/2020/07/048}

\bibitem[{{Beutler} \& {McDonald}(2021)}]{Beutler2021}
{Beutler}, F., \& {McDonald}, P. 2021, \jcap, 2021, 031,
  \dodoi{10.1088/1475-7516/2021/11/031}

\bibitem[{{Beutler} {et~al.}(2012){Beutler}, {Blake}, {Colless}, {Jones},
  {Staveley-Smith}, {Poole}, {Campbell}, {Parker}, {Saunders}, \&
  {Watson}}]{Beutler2012}
{Beutler}, F., {Blake}, C., {Colless}, M., {et~al.} 2012, \mnras, 423, 3430,
  \dodoi{10.1111/j.1365-2966.2012.21136.x}

\bibitem[{{Bianchi} {et~al.}(2015){Bianchi}, {Gil-Mar{\'\i}n}, {Ruggeri}, \&
  {Percival}}]{Bianchi2015}
{Bianchi}, D., {Gil-Mar{\'\i}n}, H., {Ruggeri}, R., \& {Percival}, W.~J. 2015,
  \mnras, 453, L11, \dodoi{10.1093/mnrasl/slv090}

\bibitem[{{Blake} {et~al.}(2018){Blake}, {Carter}, \& {Koda}}]{Blake2018}
{Blake}, C., {Carter}, P., \& {Koda}, J. 2018, \mnras, 479, 5168,
  \dodoi{10.1093/mnras/sty1814}

\bibitem[{{Blake} \& {Turner}(2024{\natexlab{a}})}]{Blake2024recons}
{Blake}, C., \& {Turner}, R.~J. 2024{\natexlab{a}}, The Open Journal of
  Astrophysics, 7, 87, \dodoi{10.33232/001c.124509}

\bibitem[{{Blake} \& {Turner}(2024{\natexlab{b}})}]{Blake2024correlation}
---. 2024{\natexlab{b}}, \mnras, 527, 501, \dodoi{10.1093/mnras/stad3217}

\bibitem[{{Blake} {et~al.}(2010){Blake}, {Brough}, {Colless}, {Couch}, {Croom},
  {Davis}, {Drinkwater}, {Forster}, {Glazebrook}, {Jelliffe}, {Jurek}, {Li},
  {Madore}, {Martin}, {Pimbblet}, {Poole}, {Pracy}, {Sharp}, {Wisnioski},
  {Woods}, \& {Wyder}}]{Blake2010}
{Blake}, C., {Brough}, S., {Colless}, M., {et~al.} 2010, \mnras, 406, 803,
  \dodoi{10.1111/j.1365-2966.2010.16747.x}

\bibitem[{{Blake} {et~al.}(2011){Blake}, {Brough}, {Colless}, {Contreras},
  {Couch}, {Croom}, {Davis}, {Drinkwater}, {Forster}, {Gilbank}, {Gladders},
  {Glazebrook}, {Jelliffe}, {Jurek}, {Li}, {Madore}, {Martin}, {Pimbblet},
  {Poole}, {Pracy}, {Sharp}, {Wisnioski}, {Woods}, {Wyder}, \&
  {Yee}}]{Blake2011}
---. 2011, \mnras, 415, 2876, \dodoi{10.1111/j.1365-2966.2011.18903.x}

\bibitem[{{Boruah} {et~al.}(2020){Boruah}, {Hudson}, \& {Lavaux}}]{Boruah2020}
{Boruah}, S.~S., {Hudson}, M.~J., \& {Lavaux}, G. 2020, \mnras, 498, 2703,
  \dodoi{10.1093/mnras/staa2485}

\bibitem[{{Boruah} {et~al.}(2021){Boruah}, {Hudson}, \& {Lavaux}}]{Boruah2021}
---. 2021, \mnras, 507, 2697, \dodoi{10.1093/mnras/stab2320}

\bibitem[{{Boruah} {et~al.}(2022){Boruah}, {Lavaux}, \& {Hudson}}]{Boruah2022}
{Boruah}, S.~S., {Lavaux}, G., \& {Hudson}, M.~J. 2022, \mnras, 517, 4529,
  \dodoi{10.1093/mnras/stac2985}

\bibitem[{{Boubel} {et~al.}(2024){Boubel}, {Colless}, {Said}, \&
  {Staveley-Smith}}]{Boubel2024}
{Boubel}, P., {Colless}, M., {Said}, K., \& {Staveley-Smith}, L. 2024, \mnras,
  531, 84, \dodoi{10.1093/mnras/stae1122}

\bibitem[{{Branchini} {et~al.}(2002){Branchini}, {Eldar}, \&
  {Nusser}}]{Branchini2002}
{Branchini}, E., {Eldar}, A., \& {Nusser}, A. 2002, \mnras, 335, 53,
  \dodoi{10.1046/j.1365-8711.2002.05611.x}

\bibitem[{{Calabrese} {et~al.}(2025){Calabrese}, {Hill}, {Jense}, {La Posta},
  {Abril-Cabezas}, {Addison}, {Ade}, {Aiola}, {Alford}, {Alonso}, {Amiri},
  {An}, {Atkins}, {Austermann}, {Barbavara}, {Barbieri}, {Battaglia},
  {Battistelli}, {Beall}, {Bean}, {Beheshti}, {Beringue}, {Bhandarkar},
  {Biermann}, {Bolliet}, {Bond}, {Capalbo}, {Carrero}, {Chen}, {Chesmore},
  {Cho}, {Choi}, {Clark}, {Cothard}, {Coughlin}, {Coulton}, {Crichton},
  {Crowley}, {Darwish}, {Devlin}, {Dicker}, {Duell}, {Duff}, {Duivenvoorden},
  {Dunkley}, {Dunner}, {Embil Villagra}, {Fankhanel}, {Farren}, {Ferraro},
  {Foster}, {Freundt}, {Fuzia}, {Gallardo}, {Garrido}, {Gerbino}, {Giardiello},
  {Gill}, {Givans}, {Gluscevic}, {Goldstein}, {Golec}, {Gong}, {Guan},
  {Halpern}, {Harrison}, {Hasselfield}, {He}, {Healy}, {Henderson}, {Hensley},
  {Herv{\'\i}as-Caimapo}, {Hilton}, {Hilton}, {Hincks}, {Hlo{\v{z}}ek}, {Ho},
  {Hood}, {Hornecker}, {Huber}, {Hubmayr}, {Huffenberger}, {Hughes}, {Ikape},
  {Irwin}, {Isopi}, {Joshi}, {Keller}, {Kim}, {Knowles}, {Koopman}, {Kosowsky},
  {Kramer}, {Kusiak}, {Lagu{\"e}}, {Lakey}, {Lattanzi}, {Lee}, {Li}, {Li},
  {Limon}, {Lokken}, {Louis}, {Lungu}, {MacCrann}, {MacInnis}, {Madhavacheril},
  {Maldonado}, {Maldonado}, {Mallaby-Kay}, {Marques}, {van Marrewijk},
  {McCarthy}, {McMahon}, {Mehta}, {Menanteau}, {Moodley}, {Morris},
  {Mroczkowski}, {Naess}, {Namikawa}, {Nati}, {Nerval}, {Newburgh}, {Nicola},
  {Niemack}, {Nolta}, {Orlowski-Scherer}, {Pagano}, {Page}, {Pandey},
  {Partridge}, {Perez Sarmiento}, {Prince}, {Puddu}, {Qu}, {Ragavan}, {Ried
  Guachalla}, {Rogers}, {Rojas}, {Sakuma}, {Schaan}, {Schmitt}, {Sehgal},
  {Shaikh}, {Sherwin}, {Sierra}, {Sievers}, {Sif{\'o}n}, {Simon}, {Sonka},
  {Spergel}, {Staggs}, {Storer}, {Surrao}, {Switzer}, {Tampier}, {Thiele},
  {Thornton}, {Trac}, {Tucker}, {Ullom}, {Vale}, {Van Engelen}, {Van Lanen},
  {Vargas}, {Vavagiakis}, {Wagoner}, {Wang}, {Wenzl}, {Wollack}, {Zheng}, \&
  {The Atacama Cosmology Telescope collaboration}}]{Calabrese2025}
{Calabrese}, E., {Hill}, J.~C., {Jense}, H.~T., {et~al.} 2025, \jcap, 2025,
  063, \dodoi{10.1088/1475-7516/2025/11/063}

\bibitem[{{Campbell} {et~al.}(2014){Campbell}, {Lucey}, {Colless}, {Jones},
  {Springob}, {Magoulas}, {Proctor}, {Mould}, {Read}, {Brough}, {Jarrett},
  {Merson}, {Lah}, {Beutler}, {Cluver}, \& {Parker}}]{Campbell2014}
{Campbell}, L.~A., {Lucey}, J.~R., {Colless}, M., {et~al.} 2014, \mnras, 443,
  1231, \dodoi{10.1093/mnras/stu1198}

\bibitem[{{Carr} {et~al.}(2025){Carr}, {Howlett}, {Amsellem}, {Davis}, {Said},
  {Parkinson}, {Palmese}, {Aguilar}, {Ahlen}, {Bautista}, {BenZvi}, {Bianchi},
  {Blake}, {Brooks}, {Claybaugh}, {Cuceu}, {de la Macorra}, {Doel}, {Douglass},
  {Ferraro}, {Forero-Romero}, {Gazta{\~n}aga}, {Gontcho}, {Gutierrez},
  {Herrera-Alcantar}, {Honscheid}, {Huterer}, {Ishak}, {Joyce}, {Kim},
  {Kirkby}, {Kremin}, {Lahav}, {Lamman}, {Landriau}, {Le Guillou}, {Levi},
  {Manera}, {Meisner}, {Miquel}, {Moustakas}, {Nadathur}, {Percival}, {Prada},
  {P{\'e}rez-R{\`a}fols}, {Qin}, {Ross}, {Rossi}, {Sanchez}, {Schlegel}, {Seo},
  {Sprayberry}, {Tarl{\'e}}, {Turner}, {Weaver}, {Zarrouk}, {Zhou}, \&
  {Zou}}]{DESIPV_Carr}
{Carr}, A., {Howlett}, C., {Amsellem}, A.~J., {et~al.} 2025, arXiv e-prints,
  arXiv:2512.03232, \dodoi{10.48550/arXiv.2512.03232}

\bibitem[{{Carreres} {et~al.}(2023){Carreres}, {Bautista}, {Feinstein},
  {Fouchez}, {Racine}, {Smith}, {Amenouche}, {Aubert}, {Dhawan}, {Ginolin},
  {Goobar}, {Gris}, {Lacroix}, {Nuss}, {Regnault}, {Rigault}, {Robert},
  {Rosnet}, {Sommer}, {Dekany}, {Groom}, {Sravan}, {Masci}, \&
  {Purdum}}]{Carreres2023}
{Carreres}, B., {Bautista}, J.~E., {Feinstein}, F., {et~al.} 2023, \aap, 674,
  A197, \dodoi{10.1051/0004-6361/202346173}

\bibitem[{{Carrick} {et~al.}(2015){Carrick}, {Turnbull}, {Lavaux}, \&
  {Hudson}}]{Carrick2015}
{Carrick}, J., {Turnbull}, S.~J., {Lavaux}, G., \& {Hudson}, M.~J. 2015,
  \mnras, 450, 317, \dodoi{10.1093/mnras/stv547}

\bibitem[{{Castorina} \& {White}(2020)}]{Castorina2020}
{Castorina}, E., \& {White}, M. 2020, \mnras, 499, 893,
  \dodoi{10.1093/mnras/staa2129}

\bibitem[{{Chen} {et~al.}(2025){Chen}, {Howlett}, {Lai}, \&
  {Qin}}]{Chen2025pairwisPS}
{Chen}, S.-F., {Howlett}, C., {Lai}, Y., \& {Qin}, F. 2025, arXiv e-prints,
  arXiv:2508.00066.
\newblock \doarXiv{2508.00066}

\bibitem[{{Courtois} {et~al.}(2013){Courtois}, {Pomar{\`e}de}, {Tully},
  {Hoffman}, \& {Courtois}}]{Courtois2013}
{Courtois}, H.~M., {Pomar{\`e}de}, D., {Tully}, R.~B., {Hoffman}, Y., \&
  {Courtois}, D. 2013, \aj, 146, 69, \dodoi{10.1088/0004-6256/146/3/69}

\bibitem[{{Croft} \& {Gaztanaga}(1997)}]{Croft1997}
{Croft}, R. A.~C., \& {Gaztanaga}, E. 1997, \mnras, 285, 793,
  \dodoi{10.1093/mnras/285.4.793}

\bibitem[{{Davis} \& {Scrimgeour}(2014)}]{Davis2014}
{Davis}, T.~M., \& {Scrimgeour}, M.~I. 2014, \mnras, 442, 1117,
  \dodoi{10.1093/mnras/stu920}

\bibitem[{{DESI Collaboration} {et~al.}(2016{\natexlab{a}}){DESI
  Collaboration}, {Aghamousa}, {Aguilar}, {Ahlen}, {Alam}, {Allen}, {Allende
  Prieto}, {Annis}, {Bailey}, {Balland}, {Ballester}, {Baltay}, {Beaufore},
  {Bebek}, {Beers}, {Bell}, {Bernal}, {Besuner}, {Beutler}, {Blake}, {Bleuler},
  {Blomqvist}, {Blum}, {Bolton}, {Briceno}, {Brooks}, {Brownstein},
  {Buckley-Geer}, {Burden}, {Burtin}, {Busca}, {Cahn}, {Cai}, {Cardiel-Sas},
  {Carlberg}, {Carton}, {Casas}, {Castander}, {Cervantes-Cota}, {Claybaugh},
  {Close}, {Coker}, {Cole}, {Comparat}, {Cooper}, {Cousinou}, {Crocce}, {Cuby},
  {Cunningham}, {Davis}, {Dawson}, {de la Macorra}, {De Vicente}, {Delubac},
  {Derwent}, {Dey}, {Dhungana}, {Ding}, {Doel}, {Duan}, {Ealet}, {Edelstein},
  {Eftekharzadeh}, {Eisenstein}, {Elliott}, {Escoffier}, {Evatt}, {Fagrelius},
  {Fan}, {Fanning}, {Farahi}, {Farihi}, {Favole}, {Feng}, {Fernandez},
  {Findlay}, {Finkbeiner}, {Fitzpatrick}, {Flaugher}, {Flender}, {Font-Ribera},
  {Forero-Romero}, {Fosalba}, {Frenk}, {Fumagalli}, {Gaensicke}, {Gallo},
  {Garcia-Bellido}, {Gaztanaga}, {Pietro Gentile Fusillo}, {Gerard},
  {Gershkovich}, {Giannantonio}, {Gillet}, {Gonzalez-de-Rivera},
  {Gonzalez-Perez}, {Gott}, {Graur}, {Gutierrez}, {Guy}, {Habib}, {Heetderks},
  {Heetderks}, {Heitmann}, {Hellwing}, {Herrera}, {Ho}, {Holland}, {Honscheid},
  {Huff}, {Hutchinson}, {Huterer}, {Hwang}, {Illa Laguna}, {Ishikawa},
  {Jacobs}, {Jeffrey}, {Jelinsky}, {Jennings}, {Jiang}, {Jimenez}, {Johnson},
  {Joyce}, {Jullo}, {Juneau}, {Kama}, {Karcher}, {Karkar}, {Kehoe}, {Kennamer},
  {Kent}, {Kilbinger}, {Kim}, {Kirkby}, {Kisner}, {Kitanidis}, {Kneib},
  {Koposov}, {Kovacs}, {Koyama}, {Kremin}, {Kron}, {Kronig}, {Kueter-Young},
  {Lacey}, {Lafever}, {Lahav}, {Lambert}, {Lampton}, {Landriau}, {Lang},
  {Lauer}, {Le Goff}, {Le Guillou}, {Le Van Suu}, {Lee}, {Lee}, {Leitner},
  {Lesser}, {Levi}, {L'Huillier}, {Li}, {Liang}, {Lin}, {Linder}, {Loebman},
  {Luki{\'c}}, {Ma}, {MacCrann}, {Magneville}, {Makarem}, {Manera}, {Manser},
  {Marshall}, {Martini}, {Massey}, {Matheson}, {McCauley}, {McDonald},
  {McGreer}, {Meisner}, {Metcalfe}, {Miller}, {Miquel}, {Moustakas}, {Myers},
  {Naik}, {Newman}, {Nichol}, {Nicola}, {Nicolati da Costa}, {Nie}, {Niz},
  {Norberg}, {Nord}, {Norman}, {Nugent}, {O'Brien}, {Oh}, \&
  {Olsen}}]{DESI2016a.Science}
{DESI Collaboration}, {Aghamousa}, A., {Aguilar}, J., {et~al.}
  2016{\natexlab{a}}, arXiv e-prints, arXiv:1611.00036,
  \dodoi{10.48550/arXiv.1611.00036}

\bibitem[{{DESI Collaboration} {et~al.}(2016{\natexlab{b}}){DESI
  Collaboration}, {Aghamousa}, {Aguilar}, {Ahlen}, {Alam}, {Allen}, {Allende
  Prieto}, {Annis}, {Bailey}, {Balland}, {Ballester}, {Baltay}, {Beaufore},
  {Bebek}, {Beers}, {Bell}, {Bernal}, {Besuner}, {Beutler}, {Blake}, {Bleuler},
  {Blomqvist}, {Blum}, {Bolton}, {Briceno}, {Brooks}, {Brownstein},
  {Buckley-Geer}, {Burden}, {Burtin}, {Busca}, {Cahn}, {Cai}, {Cardiel-Sas},
  {Carlberg}, {Carton}, {Casas}, {Castander}, {Cervantes-Cota}, {Claybaugh},
  {Close}, {Coker}, {Cole}, {Comparat}, {Cooper}, {Cousinou}, {Crocce}, {Cuby},
  {Cunningham}, {Davis}, {Dawson}, {de la Macorra}, {De Vicente}, {Delubac},
  {Derwent}, {Dey}, {Dhungana}, {Ding}, {Doel}, {Duan}, {Ealet}, {Edelstein},
  {Eftekharzadeh}, {Eisenstein}, {Elliott}, {Escoffier}, {Evatt}, {Fagrelius},
  {Fan}, {Fanning}, {Farahi}, {Farihi}, {Favole}, {Feng}, {Fernandez},
  {Findlay}, {Finkbeiner}, {Fitzpatrick}, {Flaugher}, {Flender}, {Font-Ribera},
  {Forero-Romero}, {Fosalba}, {Frenk}, {Fumagalli}, {Gaensicke}, {Gallo},
  {Garcia-Bellido}, {Gaztanaga}, {Pietro Gentile Fusillo}, {Gerard},
  {Gershkovich}, {Giannantonio}, {Gillet}, {Gonzalez-de-Rivera},
  {Gonzalez-Perez}, {Gott}, {Graur}, {Gutierrez}, {Guy}, {Habib}, {Heetderks},
  {Heetderks}, {Heitmann}, {Hellwing}, {Herrera}, {Ho}, {Holland}, {Honscheid},
  {Huff}, {Hutchinson}, {Huterer}, {Hwang}, {Illa Laguna}, {Ishikawa},
  {Jacobs}, {Jeffrey}, {Jelinsky}, {Jennings}, {Jiang}, {Jimenez}, {Johnson},
  {Joyce}, {Jullo}, {Juneau}, {Kama}, {Karcher}, {Karkar}, {Kehoe}, {Kennamer},
  {Kent}, {Kilbinger}, {Kim}, {Kirkby}, {Kisner}, {Kitanidis}, {Kneib},
  {Koposov}, {Kovacs}, {Koyama}, {Kremin}, {Kron}, {Kronig}, {Kueter-Young},
  {Lacey}, {Lafever}, {Lahav}, {Lambert}, {Lampton}, {Landriau}, {Lang},
  {Lauer}, {Le Goff}, {Le Guillou}, {Le Van Suu}, {Lee}, {Lee}, {Leitner},
  {Lesser}, {Levi}, {L'Huillier}, {Li}, {Liang}, {Lin}, {Linder}, {Loebman},
  {Luki{\'c}}, {Ma}, {MacCrann}, {Magneville}, {Makarem}, {Manera}, {Manser},
  {Marshall}, {Martini}, {Massey}, {Matheson}, {McCauley}, {McDonald},
  {McGreer}, {Meisner}, {Metcalfe}, {Miller}, {Miquel}, {Moustakas}, {Myers},
  {Naik}, {Newman}, {Nichol}, {Nicola}, {Nicolati da Costa}, {Nie}, {Niz},
  {Norberg}, {Nord}, {Norman}, {Nugent}, {O'Brien}, {Oh}, \&
  {Olsen}}]{DESI2016b.Instr}
---. 2016{\natexlab{b}}, arXiv e-prints, arXiv:1611.00037,
  \dodoi{10.48550/arXiv.1611.00037}

\bibitem[{{DESI Collaboration} {et~al.}(2022){DESI Collaboration}, {Abareshi},
  {Aguilar}, {Ahlen}, {Alam}, {Alexander}, {Alfarsy}, {Allen}, {Allende
  Prieto}, {Alves}, {Ameel}, {Armengaud}, {Asorey}, {Aviles}, {Bailey},
  {Balaguera-Antol{\'\i}nez}, {Ballester}, {Baltay}, {Bault}, {Beltran},
  {Benavides}, {BenZvi}, {Berti}, {Besuner}, {Beutler}, {Bianchi}, {Blake},
  {Blanc}, {Blum}, {Bolton}, {Bose}, {Bramall}, {Brieden}, {Brodzeller},
  {Brooks}, {Brownewell}, {Buckley-Geer}, {Cahn}, {Cai}, {Canning}, {Capasso},
  {Carnero Rosell}, {Carton}, {Casas}, {Castander}, {Cervantes-Cota},
  {Chabanier}, {Chaussidon}, {Chuang}, {Circosta}, {Cole}, {Cooper}, {da
  Costa}, {Cousinou}, {Cuceu}, {Davis}, {Dawson}, {de la Cruz-Noriega}, {de la
  Macorra}, {de Mattia}, {Della Costa}, {Demmer}, {Derwent}, {Dey}, {Dey},
  {Dhungana}, {Ding}, {Dobson}, {Doel}, {Donald-McCann}, {Donaldson},
  {Douglass}, {Duan}, {Dunlop}, {Edelstein}, {Eftekharzadeh}, {Eisenstein},
  {Enriquez-Vargas}, {Escoffier}, {Evatt}, {Fagrelius}, {Fan}, {Fanning},
  {Fawcett}, {Ferraro}, {Ereza}, {Flaugher}, {Font-Ribera}, {Forero-Romero},
  {Frenk}, {Fromenteau}, {G{\"a}nsicke}, {Garcia-Quintero}, {Garrison},
  {Gazta{\~n}aga}, {Gerardi}, {Gil-Mar{\'\i}n}, {Gontcho a Gontcho},
  {Gonzalez-Morales}, {Gonzalez-de-Rivera}, {Gonzalez-Perez}, {Gordon},
  {Graur}, {Green}, {Grove}, {Gruen}, {Gutierrez}, {Guy}, {Hahn}, {Harris},
  {Herrera}, {Herrera-Alcantar}, {Honscheid}, {Howlett}, {Huterer},
  {Ir{\v{s}}i{\v{c}}}, {Ishak}, {Jelinsky}, {Jiang}, {Jimenez}, {Jing},
  {Joyce}, {Jullo}, {Juneau}, {Kara{\c{c}}ayl{\i}}, {Karamanis}, {Karcher},
  {Karim}, {Kehoe}, {Kent}, {Kirkby}, {Kisner}, {Kitaura}, {Koposov},
  {Kov{\'a}cs}, {Kremin}, {Krolewski}, {L'Huillier}, {Lahav}, {Lambert},
  {Lamman}, {Lan}, {Landriau}, {Lane}, {Lang}, {Lange}, {Lasker}, {Le Guillou},
  {Leauthaud}, {Le Van Suu}, {Levi}, {Li}, {Magneville}, {Manera}, {Manser},
  {Marshall}, {Martini}, {McCollam}, {McDonald}, {Meisner},
  {Mena-Fern{\'a}ndez}, {Meneses-Rizo}, {Mezcua}, {Miller}, {Miquel},
  {Montero-Camacho}, {Moon}, {Moustakas}, {Mueller}, {Mu{\~n}oz-Guti{\'e}rrez},
  {Myers}, {Nadathur}, {Najita}, {Napolitano}, {Neilsen}, {Newman}, {Nie},
  {Ning}, {Niz}, {Norberg}, {Noriega}, {O'Brien}, {Obuljen},
  {Palanque-Delabrouille}, {Palmese}, {Zhiwei}, {Pappalardo}, {PENG},
  {Percival}, {Perruchot}, {Pogge}, {Poppett}, {Porredon}, {Prada},
  {Prochaska}, {Pucha}, {P{\'e}rez-Fern{\'a}ndez}, {P{\'e}rez-R{\`a}fols},
  {Rabinowitz}, {Raichoor}, {Ramirez-Solano}, {Ram{\'\i}rez-P{\'e}rez},
  {Ravoux}, {Reil}, {Rezaie}, {Rocher}, {Rockosi}, {Roe}, {Roodman}, {Ross},
  {Rossi}, {Ruggeri}, {Ruhlmann-Kleider}, {Sabiu}, {Safonova}, {Said},
  {Saintonge}, {Salas Catonga}, {Samushia}, {Sanchez}, {Saulder}, {Schaan},
  {Schlafly}, {Schlegel}, {Schmoll}, {Scholte}, {Schubnell}, {Secroun}, {Seo},
  {Serrano}, {Sharples}, {Sholl}, {Silber}, {Silva}, {Sirk}, {Siudek}, {Smith},
  {Sprayberry}, {Staten}, {Stupak}, {Tan}, {Tarl{\'e}}, {Tie}, {Tojeiro},
  {Ure{\~n}a-L{\'o}pez}, {Valdes}, {Valenzuela}, {Valluri},
  {Vargas-Maga{\~n}a}, {Verde}, {Walther}, {Wang}, {Wang}, {Weaver},
  {Weaverdyck}, {Wechsler}, {Wilson}, {Yang}, {Yu}, {Yuan}, {Y{\`e}che},
  {Zhang}, {Zhang}, {Zhao}, {Zhou}, {Zhou}, {Zou}, {Zou}, {Zou}, {Zu}, \& {DESI
  Collaboration}}]{DESI2022.KP1.Instr}
{DESI Collaboration}, {Abareshi}, B., {Aguilar}, J., {et~al.} 2022, \aj, 164,
  207, \dodoi{10.3847/1538-3881/ac882b}

\bibitem[{{DESI Collaboration} {et~al.}(2024{\natexlab{a}}){DESI
  Collaboration}, {Adame}, {Aguilar}, {Ahlen}, {Alam}, {Aldering}, {Alexander},
  {Alfarsy}, {Allende Prieto}, {Alvarez}, {Alves}, {Anand}, {Andrade-Oliveira},
  {Armengaud}, {Asorey}, {Avila}, {Aviles}, {Bailey},
  {Balaguera-Antol{\'\i}nez}, {Ballester}, {Baltay}, {Bault}, {Bautista},
  {Behera}, {Beltran}, {BenZvi}, {Beraldo e Silva}, {Bermejo-Climent}, {Berti},
  {Besuner}, {Beutler}, {Bianchi}, {Blake}, {Blum}, {Bolton}, {Brieden},
  {Brodzeller}, {Brooks}, {Brown}, {Buckley-Geer}, {Burtin}, {Cabayol-Garcia},
  {Cai}, {Canning}, {Cardiel-Sas}, {Carnero Rosell}, {Castander},
  {Cervantes-Cota}, {Chabanier}, {Chaussidon}, {Chaves-Montero}, {Chen},
  {Chen}, {Chuang}, {Claybaugh}, {Cole}, {Cooper}, {Cuceu}, {Davis}, {Dawson},
  {de Belsunce}, {de la Cruz}, {de la Macorra}, {Della Costa}, {de Mattia},
  {Demina}, {Demirbozan}, {DeRose}, {Dey}, {Dey}, {Dhungana}, {Ding}, {Ding},
  {Doel}, {Doshi}, {Douglass}, {Edge}, {Eftekharzadeh}, {Eisenstein},
  {Elliott}, {Ereza}, {Escoffier}, {Fagrelius}, {Fan}, {Fanning}, {Fawcett},
  {Ferraro}, {Flaugher}, {Font-Ribera}, {Forero-Romero}, {Forero-S{\'a}nchez},
  {Frenk}, {G{\"a}nsicke}, {Garc{\'\i}a}, {Garc{\'\i}a-Bellido},
  {Garcia-Quintero}, {Garrison}, {Gil-Mar{\'\i}n}, {Golden-Marx}, {Gontcho A
  Gontcho}, {Gonzalez-Morales}, {Gonzalez-Perez}, {Gordon}, {Graur}, {Green},
  {Gruen}, {Guy}, {Hadzhiyska}, {Hahn}, {Han}, {Hanif}, {Herrera-Alcantar},
  {Honscheid}, {Hou}, {Howlett}, {Huterer}, {Ir{\v{s}}i{\v{c}}}, {Ishak},
  {Jacques}, {Jana}, {Jiang}, {Jimenez}, {Jing}, {Joudaki}, {Joyce}, {Jullo},
  {Juneau}, {Kara{\c{c}}ayl{\i}}, {Karim}, {Kehoe}, {Kent}, {Khederlarian},
  {Kim}, {Kirkby}, {Kisner}, {Kitaura}, {Kizhuprakkat}, {Kneib}, {Koposov},
  {Kov{\'a}cs}, {Kremin}, {Krolewski}, {L'Huillier}, {Lahav}, {Lambert},
  {Lamman}, {Lan}, {Landriau}, {Lang}, {Lange}, {Lasker}, {Leauthaud}, {Le
  Guillou}, {Levi}, {Li}, {Linder}, {Lyons}, {Magneville}, {Manera}, {Manser},
  {Margala}, {Martini}, {McDonald}, {Medina}, {Medina-Varela}, {Meisner},
  {Mena-Fern{\'a}ndez}, {Meneses-Rizo}, {Mezcua}, {Miquel}, {Montero-Camacho},
  {Moon}, {Moore}, {Moustakas}, {Mueller}, {Mundet}, {Mu{\~n}oz-Guti{\'e}rrez},
  {Myers}, {Nadathur}, {Napolitano}, {Neveux}, {Newman}, {Nie}, {Nikutta},
  {Niz}, {Norberg}, {Noriega}, {Paillas}, {Palanque-Delabrouille}, {Palmese},
  {Pan}, {Parkinson}, {Penmetsa}, {Percival}, {P{\'e}rez-Fern{\'a}ndez},
  {P{\'e}rez-R{\`a}fols}, {Pieri}, {Poppett}, {Porredon}, {Pothier}, {Prada},
  {Pucha}, {Raichoor}, {Ram{\'\i}rez-P{\'e}rez}, {Ramirez-Solano},
  {Rashkovetskyi}, {Ravoux}, {Rocher}, {Rockosi}, {Ross}, {Rossi}, {Ruggeri},
  {Ruhlmann-Kleider}, {Sabiu}, {Said}, {Saintonge}, {Samushia}, {Sanchez},
  {Saulder}, {Schaan}, {Schlafly}, {Schlegel}, {Scholte}, {Schubnell}, {Seo},
  {Shafieloo}, {Sharples}, {Sheu}, {Silber}, {Sinigaglia}, {Siudek}, {Slepian},
  {Smith}, {Soumagnac}, {Sprayberry}, {Stephey}, {Su{\'a}rez-P{\'e}rez}, {Sun},
  {Tan}, {Tarl{\'e}}, {Tojeiro}, {Ure{\~n}a-L{\'o}pez}, {Vaisakh}, {Valcin},
  {Valdes}, {Valluri}, {Vargas-Maga{\~n}a}, {Variu}, {Verde}, {Walther},
  {Wang}, {Wang}, {Weaver}, {Weaverdyck}, {Wechsler}, {White}, {Xie}, {Yang},
  {Y{\`e}che}, {Yu}, {Yuan}, {Zhang}, {Zhang}, {Zhao}, {Zheng}, {Zhou}, {Zhou},
  {Zou}, {Zou}, \& {Zu}}]{DESI2023b.KP1.EDR}
{DESI Collaboration}, {Adame}, A.~G., {Aguilar}, J., {et~al.}
  2024{\natexlab{a}}, \aj, 168, 58, \dodoi{10.3847/1538-3881/ad3217}

\bibitem[{{DESI Collaboration} {et~al.}(2024{\natexlab{b}}){DESI
  Collaboration}, {Adame}, {Aguilar}, {Ahlen}, {Alam}, {Aldering}, {Alexander},
  {Alfarsy}, {Allende Prieto}, {Alvarez}, {Alves}, {Anand}, {Andrade-Oliveira},
  {Armengaud}, {Asorey}, {Avila}, {Aviles}, {Bailey},
  {Balaguera-Antol{\'\i}nez}, {Ballester}, {Baltay}, {Bault}, {Bautista},
  {Behera}, {Beltran}, {BenZvi}, {Beraldo e Silva}, {Bermejo-Climent}, {Berti},
  {Besuner}, {Beutler}, {Bianchi}, {Blake}, {Blum}, {Bolton}, {Brieden},
  {Brodzeller}, {Brooks}, {Brown}, {Buckley-Geer}, {Burtin}, {Cabayol-Garcia},
  {Cai}, {Canning}, {Cardiel-Sas}, {Carnero Rosell}, {Castander},
  {Cervantes-Cota}, {Chabanier}, {Chaussidon}, {Chaves-Montero}, {Chen},
  {Chen}, {Chuang}, {Claybaugh}, {Cole}, {Cooper}, {Cuceu}, {Davis}, {Dawson},
  {de Belsunce}, {de la Cruz}, {de la Macorra}, {de Mattia}, {Demina},
  {Demirbozan}, {DeRose}, {Dey}, {Dey}, {Dhungana}, {Ding}, {Ding}, {Doel},
  {Doshi}, {Douglass}, {Edge}, {Eftekharzadeh}, {Eisenstein}, {Elliott},
  {Escoffier}, {Fagrelius}, {Fan}, {Fanning}, {Fawcett}, {Ferraro}, {Ereza},
  {Flaugher}, {Font-Ribera}, {Forero-S{\'a}nchez}, {Forero-Romero}, {Frenk},
  {G{\"a}nsicke}, {Garc{\'\i}a}, {Garc{\'\i}a-Bellido}, {Garcia-Quintero},
  {Garrison}, {Gil-Mar{\'\i}n}, {Golden-Marx}, {Gontcho A Gontcho},
  {Gonzalez-Morales}, {Gonzalez-Perez}, {Gordon}, {Graur}, {Green}, {Gruen},
  {Guy}, {Hadzhiyska}, {Hahn}, {Han}, {Hanif}, {Herrera-Alcantar}, {Honscheid},
  {Hou}, {Howlett}, {Huterer}, {Ir{\v{s}}i{\v{c}}}, {Ishak}, {Jana}, {Jiang},
  {Jimenez}, {Jing}, {Joudaki}, {Jullo}, {Joyce}, {Juneau}, {Kizhuprakkat},
  {Kara{\c{c}}ayl{\i}}, {Karim}, {Kehoe}, {Kent}, {Khederlarian}, {Kim},
  {Kirkby}, {Kisner}, {Kitaura}, {Kneib}, {Koposov}, {Kov{\'a}cs}, {Kremin},
  {Krolewski}, {L'Huillier}, {Lahav}, {Lambert}, {Lamman}, {Lan}, {Landriau},
  {Lang}, {Lange}, {Lasker}, {Le Guillou}, {Leauthaud}, {Levi}, {Li}, {Linder},
  {Lyons}, {Magneville}, {Manera}, {Manser}, {Margala}, {Martini}, {McDonald},
  {Medina}, {Medina-Varela}, {Meisner}, {Mena-Fern{\'a}ndez}, {Meneses-Rizo},
  {Mezcua}, {Miquel}, {Montero-Camacho}, {Moon}, {Moore}, {Moustakas},
  {Mueller}, {Mundet}, {Mu{\~n}oz-Guti{\'e}rrez}, {Myers}, {Nadathur},
  {Napolitano}, {Neveux}, {Newman}, {Nie}, {Niz}, {Norberg}, {Noriega},
  {Paillas}, {Palanque-Delabrouille}, {Palmese}, {Zhiwei}, {Parkinson},
  {Penmetsa}, {Percival}, {P{\'e}rez-Fern{\'a}ndez}, {P{\'e}rez-R{\`a}fols},
  {Pieri}, {Poppett}, {Porredon}, {Prada}, {Pucha}, {Raichoor},
  {Ram{\'\i}rez-P{\'e}rez}, {Ramirez-Solano}, {Rashkovetskyi}, {Ravoux},
  {Rocher}, {Rockosi}, {Ross}, {Rossi}, {Ruggeri}, {Ruhlmann-Kleider}, {Sabiu},
  {Said}, {Saintonge}, {Samushia}, {Sanchez}, {Saulder}, {Schaan}, {Schlafly},
  {Schlegel}, {Scholte}, {Schubnell}, {Seo}, {Shafieloo}, {Sharples}, {Sheu},
  {Silber}, {Sinigaglia}, {Siudek}, {Slepian}, {Smith}, {Sprayberry},
  {Stephey}, {Su{\'a}rez-P{\'e}rez}, {Sun}, {Tan}, {Tarl{\'e}}, {Tojeiro},
  {Ure{\~n}a-L{\'o}pez}, {Vaisakh}, {Valcin}, {Valdes}, {Valluri},
  {Vargas-Maga{\~n}a}, {Variu}, {Verde}, {Walther}, {Wang}, {Wang}, {Weaver},
  {Weaverdyck}, {Wechsler}, {White}, {Xie}, {Yang}, {Y{\`e}che}, {Yu}, {Yuan},
  {Zhang}, {Zhang}, {Zhao}, {Zheng}, {Zhou}, {Zhou}, {Zou}, {Zou}, {Zu}, \&
  {DESI Collaboration}}]{DESI2023a.KP1.SV}
---. 2024{\natexlab{b}}, \aj, 167, 62, \dodoi{10.3847/1538-3881/ad0b08}

\bibitem[{{DESI Collaboration} {et~al.}(2025{\natexlab{a}}){DESI
  Collaboration}, {Abdul-Karim}, {Adame}, {Aguado}, {Aguilar}, {Ahlen}, {Alam},
  {Aldering}, {Alexander}, {Alfarsy}, {Allen}, {Allende Prieto}, {Alves},
  {Anand}, {Andrade}, {Armengaud}, {Avila}, {Aviles}, {Awan}, {Bailey},
  {Baleato Lizancos}, {Ballester}, {Bault}, {Bautista}, {BenZvi}, {Beraldo e
  Silva}, {Bermejo-Climent}, {Beutler}, {Bianchi}, {Blake}, {Blum}, {Bolton},
  {Bonici}, {Brieden}, {Brodzeller}, {Brooks}, {Buckley-Geer}, {Burtin},
  {Canning}, {Carnero Rosell}, {Carr}, {Carrilho}, {Casas}, {Castander},
  {Cereskaite}, {Cervantes-Cota}, {Chaussidon}, {Chaves-Montero}, {Chen},
  {Chen}, {Claybaugh}, {Cole}, {Cooper}, {Cousinou}, {Cuceu}, {Davis},
  {Dawson}, {de Belsunce}, {de la Cruz}, {de la Macorra}, {de Mattia},
  {Deiosso}, {Della Costa}, {Demina}, {Demirbozan}, {DeRose}, {Dey}, {Dey},
  {Ding}, {Ding}, {Doel}, {Douglass}, {Dowicz}, {Ebina}, {Edelstein},
  {Eisenstein}, {Elbers}, {Emas}, {Escoffier}, {Fagrelius}, {Fan}, {Fanning},
  {Fawcett}, {Fern{\'a}ndez-Garc{\'\i}a}, {Ferraro}, {Findlay}, {Font-Ribera},
  {Forero-Romero}, {Forero-S{\'a}nchez}, {Frenk}, {G{\"a}nsicke}, {Galbany},
  {Garc{\'\i}a-Bellido}, {Garcia-Quintero}, {Garrison}, {Gazta{\~n}aga},
  {Gil-Mar{\'\i}n}, {Gnedin}, {Gontcho}, {Gonzalez-Morales}, {Gonzalez-Perez},
  {Gordon}, {Graur}, {Green}, {Gruen}, {Gsponer}, {Guandalin}, {Gutierrez},
  {Guy}, {Hahn}, {Han}, {Han}, {He}, {Herrera-Alcantar}, {Honscheid}, {Hou},
  {Howlett}, {Huterer}, {Ir{\v{s}}i{\v{c}}}, {Ishak}, {Jacques}, {Jimenez},
  {Jing}, {Joachimi}, {Joudaki}, {Joyce}, {Jullo}, {Juneau},
  {Kara{\c{c}}ayl{\i}}, {Karim}, {Kehoe}, {Kent}, {Khederlarian}, {Kirkby},
  {Kisner}, {Kitaura}, {Kizhuprakkat}, {Kong}, {Koposov}, {Kremin},
  {Krolewski}, {Lahav}, {Lai}, {Lamman}, {Lan}, {Landriau}, {Lang}, {Lange},
  {Lasker}, {Le Goff}, {Le Guillou}, {Leauthaud}, {Levi}, {Li}, {Li}, {Lodha},
  {Lokken}, {Luo}, {Magneville}, {Manera}, {Manser}, {Margala}, {Martini},
  {Maus}, {McCullough}, {McDonald}, {Medina}, {Medina-Varela}, {Meisner},
  {Mena-Fern{\'a}ndez}, {Menegas}, {Mezcua}, {Miquel}, {Montero-Camacho},
  {Moon}, {Moustakas}, {Mu{\~n}oz-Guti{\'e}rrez}, {Mu{\~n}oz-Santos}, {Myers},
  {Myles}, {Nadathur}, {Najita}, {Napolitano}, {Newman}, {Nikakhtar},
  {Nikutta}, {Niz}, {Noriega}, {Padmanabhan}, {Paillas},
  {Palanque-Delabrouille}, {Palmese}, {Pan}, {Pan}, {Parkinson}, {Peacock},
  {Percival}, {P{\'e}rez-Fern{\'a}ndez}, {P{\'e}rez-R{\`a}fols}, \&
  {Peterson}}]{DESI2024.I.DR1}
{DESI Collaboration}, {Abdul-Karim}, M., {Adame}, A.~G., {et~al.}
  2025{\natexlab{a}}, arXiv e-prints, arXiv:2503.14745,
  \dodoi{10.48550/arXiv.2503.14745}

\bibitem[{{DESI Collaboration} {et~al.}(2025{\natexlab{b}}){DESI
  Collaboration}, {Abdul Karim}, {Aguilar}, {Ahlen}, {Alam}, {Allen}, {Prieto},
  {Alves}, {Anand}, {Andrade}, {Armengaud}, {Aviles}, {Bailey}, {Baltay},
  {Bansal}, {Bault}, {Behera}, {BenZvi}, {Bianchi}, {Blake}, {Brieden},
  {Brodzeller}, {Brooks}, {Buckley-Geer}, {Burtin}, {Calderon}, {Canning},
  {Rosell}, {Carrilho}, {Casas}, {Castander}, {Charles}, {Chaussidon},
  {Chaves-Montero}, {Chebat}, {Chen}, {Claybaugh}, {Cole}, {Cooper}, {Cuceu},
  {Dawson}, {de la Macorra}, {de Mattia}, {Deiosso}, {Della Costa}, {Demina},
  {Dey}, {Dey}, {Ding}, {Doel}, {Edelstein}, {Eisenstein}, {Elbers},
  {Fagrelius}, {Fanning}, {Fern{\'a}ndez-Garc{\'\i}a}, {Ferraro},
  {Font-Ribera}, {Forero-Romero}, {Frenk}, {Garcia-Quintero}, {Garrison},
  {Gazta{\~n}aga}, {Gil-Mar{\'\i}n}, {Gontcho A Gontcho}, {Gonzalez},
  {Gonzalez-Morales}, {Gordon}, {Green}, {Gutierrez}, {Guy}, {Hadzhiyska},
  {Hahn}, {He}, {Herbold}, {Herrera-Alcantar}, {Ho}, {Honscheid}, {Howlett},
  {Huterer}, {Ishak}, {Juneau}, {Kamble}, {Kara{\c{c}}ayl{\i}}, {Kehoe},
  {Kent}, {Kim}, {Kirkby}, {Kisner}, {Koposov}, {Kremin}, {Krolewski}, {Lahav},
  {Lamman}, {Landriau}, {Lang}, {Lasker}, {Le Goff}, {Le Guillou}, {Leauthaud},
  {Levi}, {Li}, {Li}, {Lodha}, {Lokken}, {Lozano-Rodr{\'\i}guez}, {Magneville},
  {Manera}, {Martini}, {Matthewson}, {Meisner}, {Mena-Fern{\'a}ndez},
  {Menegas}, {Mergulh{\~a}o}, {Miquel}, {Moustakas}, {Mu{\~n}oz-Guti{\'e}rrez},
  {Mu{\~n}oz-Santos}, {Myers}, {Nadathur}, {Naidoo}, {Napolitano}, {Newman},
  {Niz}, {Noriega}, {Paillas}, {Palanque-Delabrouille}, {Pan}, {Peacock},
  {Pellejero Ibanez}, {Percival}, {P{\'e}rez-Fern{\'a}ndez},
  {P{\'e}rez-R{\`a}fols}, {Pieri}, {Poppett}, {Prada}, {Rabinowitz},
  {Raichoor}, {Ram{\'\i}rez-P{\'e}rez}, {Rashkovetskyi}, {Ravoux}, {Rich},
  {Rocher}, {Rockosi}, {Rohlf}, {Rom{\'a}n-Herrera}, {Ross}, {Rossi},
  {Ruggeri}, {Ruhlmann-Kleider}, {Samushia}, {Sanchez}, {Sanders}, {Schlegel},
  {Schubnell}, {Seo}, {Shafieloo}, {Sharples}, {Silber}, {Sinigaglia},
  {Sprayberry}, {Tan}, {Tarl{\'e}}, {Taylor}, {Turner}, {Ure{\~n}a-L{\'o}pez},
  {Vaisakh}, {Valdes}, {Valogiannis}, {Vargas-Maga{\~n}a}, {Verde}, {Walther},
  {Weaver}, {Weinberg}, {White}, {Wolfson}, {Y{\`e}che}, {Yu}, {Zaborowski},
  {Zarrouk}, {Zhai}, {Zhang}, {Zhao}, {Zhao}, {Zhou}, {Zou}, \& {DESI
  Collaboration}}]{DESI.DR2.BAO.cosmo}
{DESI Collaboration}, {Abdul Karim}, M., {Aguilar}, J., {et~al.}
  2025{\natexlab{b}}, \prd, 112, 083515, \dodoi{10.1103/tr6y-kpc6}

\bibitem[{{DESI Collaboration} {et~al.}(2025{\natexlab{c}}){DESI
  Collaboration}, {Adame}, {Aguilar}, {Ahlen}, {Alam}, {Alexander}, {Allende
  Prieto}, {Alvarez}, {Alves}, {Anand}, {Andrade}, {Armengaud}, {Avila},
  {Aviles}, {Awan}, {Bahr-Kalus}, {Bailey}, {Baltay}, {Bault}, {Behera},
  {BenZvi}, {Beutler}, {Bianchi}, {Blake}, {Blum}, {Bonici}, {Brieden},
  {Brodzeller}, {Brooks}, {Buckley-Geer}, {Burtin}, {Calderon}, {Canning},
  {Carnero Rosell}, {Cereskaite}, {Cervantes-Cota}, {Chabanier}, {Chaussidon},
  {Chaves-Montero}, {Chebat}, {Chen}, {Chen}, {Claybaugh}, {Cole}, {Cuceu},
  {Davis}, {Dawson}, {de la Macorra}, {de Mattia}, {Deiosso}, {Dey}, {Dey},
  {Ding}, {Doel}, {Edelstein}, {Eftekharzadeh}, {Eisenstein}, {Elbers},
  {Elliott}, {Fagrelius}, {Fanning}, {Ferraro}, {Ereza}, {Findlay}, {Flaugher},
  {Font-Ribera}, {Forero-S{\'a}nchez}, {Forero-Romero}, {Frenk},
  {Garcia-Quintero}, {Garrison}, {Gazta{\~n}aga}, {Gil-Mar{\'\i}n}, {Gontcho},
  {Gonzalez-Morales}, {Gonzalez-Perez}, {Gordon}, {Green}, {Gruen}, {Gsponer},
  {Gutierrez}, {Guy}, {Hadzhiyska}, {Hahn}, {Hanif}, {Herrera-Alcantar},
  {Honscheid}, {Howlett}, {Huterer}, {Ir{\v{s}}i{\v{c}}}, {Ishak}, {Joyce},
  {Juneau}, {Kara{\c{c}}ayl{\i}}, {Kehoe}, {Kent}, {Kirkby}, {Kong}, {Koposov},
  {Kremin}, {Krolewski}, {Lahav}, {Lai}, {Lan}, {Landriau}, {Lang}, {Lasker},
  {Le Goff}, {Le Guillou}, {Leauthaud}, {Levi}, {Li}, {Lodha}, {Magneville},
  {Manera}, {Margala}, {Martini}, {Matthewson}, {Maus}, {McDonald},
  {Medina-Varela}, {Meisner}, {Mena-Fern{\'a}ndez}, {Miquel}, {Moon}, {Moore},
  {Moustakas}, {Mudur}, {Mueller}, {Mu{\~n}oz-Guti{\'e}rrez}, {Myers},
  {Nadathur}, {Napolitano}, {Neveux}, {Newman}, {Nguyen}, {Nie}, {Niz},
  {Noriega}, {Padmanabhan}, {Paillas}, {Palanque-Delabrouille}, {Pan},
  {Penmetsa}, {Percival}, {Pieri}, {Pinon}, {Poppett}, {Porredon}, {Prada},
  {P{\'e}rez-Fern{\'a}ndez}, {P{\'e}rez-R{\`a}fols}, {Rabinowitz}, {Raichoor},
  {Ram{\'\i}rez-P{\'e}rez}, {Ramirez-Solano}, {Rashkovetskyi}, {Ravoux},
  {Rezaie}, {Rich}, {Rocher}, {Rockosi}, {Roe}, {Rosado-Marin}, {Ross},
  {Rossi}, {Ruggeri}, {Ruhlmann-Kleider}, {Samushia}, {Sanchez}, {Saulder},
  {Schlafly}, {Schlegel}, {Schubnell}, {Seo}, {Shafieloo}, {Sharples},
  {Silber}, {Slosar}, {Smith}, {Sprayberry}, {Tan}, {Tarl{\'e}}, {Taylor},
  {Trusov}, {Vaisakh}, {Valcin}, {Valdes}, {Valogiannis}, {Vargas-Maga{\~n}a},
  {Verde}, {Walther}, {Wang}, {Wang}, {Weaver}, {Weaverdyck}, {Wechsler},
  {Weinberg}, {White}, {Wilson}, \& {Yi}}]{DESI2024.VII.KP7B}
{DESI Collaboration}, {Adame}, A.~G., {Aguilar}, J., {et~al.}
  2025{\natexlab{c}}, \jcap, 2025, 028, \dodoi{10.1088/1475-7516/2025/07/028}

\bibitem[{{DESI Collaboration} {et~al.}(2025{\natexlab{d}}){DESI
  Collaboration}, {Adame}, {Aguilar}, {Ahlen}, {Alam}, {Alexander}, {Alvarez},
  {Alves}, {Anand}, {Andrade}, {Armengaud}, {Avila}, {Aviles}, {Awan},
  {Bailey}, {Baltay}, {Bault}, {Behera}, {BenZvi}, {Beutler}, {Bianchi},
  {Blake}, {Blum}, {Brieden}, {Brodzeller}, {Brooks}, {Brown}, {Buckley-Geer},
  {Burtin}, {Calderon}, {Canning}, {Carnero Rosell}, {Cereskaite},
  {Cervantes-Cota}, {Chabanier}, {Chaussidon}, {Chaves-Montero}, {Chen},
  {Chen}, {Claybaugh}, {Cole}, {Cuceu}, {Davis}, {Dawson}, {de la Macorra}, {de
  Mattia}, {Deiosso}, {Demina}, {Dey}, {Dey}, {Ding}, {Doel}, {Edelstein},
  {Eftekharzadeh}, {Eisenstein}, {Elliott}, {Fagrelius}, {Fanning}, {Ferraro},
  {Ereza}, {Findlay}, {Flaugher}, {Font-Ribera}, {Forero-S{\'a}nchez},
  {Forero-Romero}, {Frenk}, {Garcia-Quintero}, {Gazta{\~n}aga},
  {Gil-Mar{\'\i}n}, {Gontcho}, {Gonzalez-Morales}, {Gonzalez-Perez}, {Gordon},
  {Green}, {Gruen}, {Gsponer}, {Gutierrez}, {Guy}, {Hadzhiyska}, {Hahn},
  {Hanif}, {Herrera-Alcantar}, {Honscheid}, {Hou}, {Howlett}, {Huterer},
  {Ir{\v{s}}i{\v{c}}}, {Ishak}, {Juneau}, {Kara{\c{c}}ayl{\i}}, {Kehoe},
  {Kent}, {Kirkby}, {Kitaura}, {Kong}, {Kremin}, {Krolewski}, {Lai}, {Lan},
  {Landriau}, {Lang}, {Lasker}, {Le Goff}, {Le Guillou}, {Leauthaud}, {Levi},
  {Li}, {Lodha}, {Magneville}, {Manera}, {Margala}, {Martini}, {Maus},
  {McDonald}, {Medina-Varela}, {Meisner}, {Mena-Fern{\'a}ndez}, {Miquel},
  {Moon}, {Moore}, {Moustakas}, {Mudur}, {Mueller}, {Mu{\~n}oz-Guti{\'e}rrez},
  {Myers}, {Nadathur}, {Napolitano}, {Neveux}, {Newman}, {Nguyen}, {Nie},
  {Niz}, {Noriega}, {Padmanabhan}, {Paillas}, {Palanque-Delabrouille}, {Pan},
  {Penmetsa}, {Percival}, {Pieri}, {Pinon}, {Poppett}, {Porredon}, {Prada},
  {P{\'e}rez-Fern{\'a}ndez}, {P{\'e}rez-R{\`a}fols}, {Rabinowitz}, {Raichoor},
  {Ram{\'\i}rez-P{\'e}rez}, {Ramirez-Solano}, {Rashkovetskyi}, {Ravoux},
  {Rezaie}, {Rich}, {Rocher}, {Rockosi}, {Roe}, {Rosado-Marin}, {Ross},
  {Rossi}, {Ruggeri}, {Ruhlmann-Kleider}, {Samushia}, {Sanchez}, {Saulder},
  {Schlafly}, {Schlegel}, {Scholte}, {Schubnell}, {Seo}, {Sharples}, {Silber},
  {Slosar}, {Smith}, {Sprayberry}, {Tan}, {Tarl{\'e}}, {Trusov}, {Vaisakh},
  {Valcin}, {Valdes}, {Vargas-Maga{\~n}a}, {Verde}, {Walther}, {Wang}, {Wang},
  {Weaver}, {Weaverdyck}, {Wechsler}, {Weinberg}, {White}, {Wilson}, {Yu},
  {Yu}, {Yuan}, {Y{\`e}che}, {Zaborowski}, {Zarrouk}, {Zhang}, {Zhao}, \&
  {Zhao}}]{DESI2024.II.KP3}
---. 2025{\natexlab{d}}, \jcap, 2025, 017,
  \dodoi{10.1088/1475-7516/2025/07/017}

\bibitem[{{DESI Collaboration} {et~al.}(2025{\natexlab{e}}){DESI
  Collaboration}, {Adame}, {Aguilar}, {Ahlen}, {Alam}, {Alexander}, {Alvarez},
  {Alves}, {Anand}, {Andrade}, {Armengaud}, {Avila}, {Aviles}, {Awan},
  {Bailey}, {Baltay}, {Bault}, {Bautista}, {Behera}, {BenZvi}, {Beutler},
  {Bianchi}, {Blake}, {Blum}, {Brieden}, {Brodzeller}, {Brooks},
  {Buckley-Geer}, {Burtin}, {Calderon}, {Canning}, {Carnero Rosell},
  {Cereskaite}, {Cervantes-Cota}, {Chabanier}, {Chaussidon}, {Chaves-Montero},
  {Chen}, {Chen}, {Claybaugh}, {Cole}, {Cuceu}, {Davis}, {Dawson}, {de la
  Cruz}, {de la Macorra}, {de Mattia}, {Deiosso}, {Dey}, {Dey}, {Ding}, {Ding},
  {Doel}, {Edelstein}, {Eftekharzadeh}, {Eisenstein}, {Elliott}, {Fagrelius},
  {Fanning}, {Ferraro}, {Ereza}, {Findlay}, {Flaugher}, {Font-Ribera},
  {Forero-S{\'a}nchez}, {Forero-Romero}, {Garcia-Quintero}, {Gazta{\~n}aga},
  {Gil-Mar{\'\i}n}, {Gontcho}, {Gonzalez-Morales}, {Gonzalez-Perez}, {Gordon},
  {Green}, {Gruen}, {Gsponer}, {Gutierrez}, {Guy}, {Hadzhiyska}, {Hahn},
  {Hanif}, {Herrera-Alcantar}, {Honscheid}, {Howlett}, {Huterer},
  {Ir{\v{s}}i{\v{c}}}, {Ishak}, {Juneau}, {Kara{\c{c}}ayl{\i}}, {Kehoe},
  {Kent}, {Kirkby}, {Kremin}, {Krolewski}, {Lai}, {Lan}, {Landriau}, {Lang},
  {Lasker}, {Le Goff}, {Le Guillou}, {Leauthaud}, {Levi}, {Li}, {Linder},
  {Lodha}, {Magneville}, {Manera}, {Margala}, {Martini}, {Maus}, {McDonald},
  {Medina-Varela}, {Meisner}, {Mena-Fern{\'a}ndez}, {Miquel}, {Moon}, {Moore},
  {Moustakas}, {Mueller}, {Mu{\~n}oz-Guti{\'e}rrez}, {Myers}, {Nadathur},
  {Napolitano}, {Neveux}, {Newman}, {Nguyen}, {Nie}, {Niz}, {Noriega},
  {Padmanabhan}, {Paillas}, {Palanque-Delabrouille}, {Pan}, {Penmetsa},
  {Percival}, {Pieri}, {Pinon}, {Poppett}, {Porredon}, {Prada},
  {P{\'e}rez-Fern{\'a}ndez}, {P{\'e}rez-R{\`a}fols}, {Rabinowitz}, {Raichoor},
  {Ram{\'\i}rez-P{\'e}rez}, {Ramirez-Solano}, {Rashkovetskyi}, {Ravoux},
  {Rezaie}, {Rich}, {Rocher}, {Rockosi}, {Roe}, {Rosado-Marin}, {Ross},
  {Rossi}, {Ruggeri}, {Ruhlmann-Kleider}, {Samushia}, {Sanchez}, {Saulder},
  {Schlafly}, {Schlegel}, {Schubnell}, {Seo}, {Sharples}, {Silber},
  {Sinigaglia}, {Slosar}, {Smith}, {Sprayberry}, {Tan}, {Tarl{\'e}}, {Trusov},
  {Vaisakh}, {Valcin}, {Valdes}, {Vargas-Maga{\~n}a}, {Verde}, {Walther},
  {Wang}, {Wang}, {Weaver}, {Weaverdyck}, {Wechsler}, {Weinberg}, {White},
  {Yu}, {Yu}, {Yuan}, {Y{\`e}che}, {Zaborowski}, {Zarrouk}, {Zhang}, {Zhao},
  {Zhao}, {Zhou}, {Zou}, \& {DESI Collaboration}}]{DESI2024.IV.KP6}
---. 2025{\natexlab{e}}, \jcap, 2025, 124,
  \dodoi{10.1088/1475-7516/2025/01/124}

\bibitem[{{DESI Collaboration} {et~al.}(2025{\natexlab{f}}){DESI
  Collaboration}, {Adame}, {Aguilar}, {Ahlen}, {Alam}, {Alexander}, {Alvarez},
  {Alves}, {Anand}, {Andrade}, {Armengaud}, {Avila}, {Aviles}, {Awan},
  {Bailey}, {Baltay}, {Bault}, {Behera}, {BenZvi}, {Beutler}, {Bianchi},
  {Blake}, {Blum}, {Brieden}, {Brodzeller}, {Brooks}, {Buckley-Geer}, {Burtin},
  {Calderon}, {Canning}, {Carnero Rosell}, {Cereskaite}, {Cervantes-Cota},
  {Chabanier}, {Chaussidon}, {Chaves-Montero}, {Chen}, {Chen}, {Claybaugh},
  {Cole}, {Cuceu}, {Davis}, {Dawson}, {de la Macorra}, {de Mattia}, {Deiosso},
  {Dey}, {Dey}, {Ding}, {Doel}, {Edelstein}, {Eftekharzadeh}, {Eisenstein},
  {Elliott}, {Fagrelius}, {Fanning}, {Ferraro}, {Ereza}, {Findlay}, {Flaugher},
  {Font-Ribera}, {Forero-S{\'a}nchez}, {Forero-Romero}, {Garcia-Quintero},
  {Garrison}, {Gazta{\~n}aga}, {Gil-Mar{\'\i}n}, {Gontcho}, {Gonzalez-Morales},
  {Gonzalez-Perez}, {Gordon}, {Green}, {Gruen}, {Gsponer}, {Gutierrez}, {Guy},
  {Hadzhiyska}, {Hahn}, {Hanif}, {Herrera-Alcantar}, {Honscheid}, {Howlett},
  {Huterer}, {Ir{\v{s}}i{\v{c}}}, {Ishak}, {Juneau}, {Kara{\c{c}}ayl{\i}},
  {Kehoe}, {Kent}, {Kirkby}, {Kong}, {Koposov}, {Kremin}, {Krolewski}, {Lai},
  {Lan}, {Landriau}, {Lang}, {Lasker}, {Le Goff}, {Le Guillou}, {Leauthaud},
  {Levi}, {Li}, {Lodha}, {Magneville}, {Manera}, {Margala}, {Martini}, {Maus},
  {McDonald}, {Medina-Varela}, {Meisner}, {Mena-Fern{\'a}ndez}, {Miquel},
  {Moon}, {Moore}, {Moustakas}, {Mueller}, {Mu{\~n}oz-Guti{\'e}rrez}, {Myers},
  {Nadathur}, {Napolitano}, {Neveux}, {Newman}, {Nguyen}, {Nie}, {Niz},
  {Noriega}, {Padmanabhan}, {Paillas}, {Palanque-Delabrouille}, {Pan},
  {Penmetsa}, {Percival}, {Pieri}, {Pinon}, {Poppett}, {Porredon}, {Prada},
  {P{\'e}rez-Fern{\'a}ndez}, {P{\'e}rez-R{\`a}fols}, {Rabinowitz}, {Raichoor},
  {Ram{\'\i}rez-P{\'e}rez}, {Ramirez-Solano}, {Rashkovetskyi}, {Ravoux},
  {Rezaie}, {Rich}, {Rocher}, {Rockosi}, {Rodr{\'\i}guez-Mart{\'\i}nez}, {Roe},
  {Rosado-Marin}, {Ross}, {Rossi}, {Ruggeri}, {Ruhlmann-Kleider}, {Samushia},
  {Sanchez}, {Saulder}, {Schlafly}, {Schlegel}, {Schubnell}, {Seo}, {Sharples},
  {Silber}, {Slosar}, {Smith}, {Sprayberry}, {Tan}, {Tarl{\'e}}, {Trusov},
  {Vaisakh}, {Valcin}, {Valdes}, {Vargas-Maga{\~n}a}, {Verde}, {Walther},
  {Wang}, {Wang}, {Weaver}, {Weaverdyck}, {Wechsler}, {Weinberg}, {White},
  {Wilson}, {Yu}, {Yu}, {Yuan}, {Y{\`e}che}, {Zaborowski}, {Zarrouk}, {Zhang},
  {Zhao}, {Zhao}, {Zhou}, {Zou}, \& {The DESI collaboration}}]{DESI2024.V.KP5}
---. 2025{\natexlab{f}}, \jcap, 2025, 008,
  \dodoi{10.1088/1475-7516/2025/09/008}

\bibitem[{{Djorgovski} \& {Davis}(1987)}]{Djorgovski1987}
{Djorgovski}, S., \& {Davis}, M. 1987, \apj, 313, 59, \dodoi{10.1086/164948}

\bibitem[{{Douglass} {et~al.}(2025){Douglass}, {BenZvi}, {Kim}, {Moore},
  {Carr}, {Largett}, {Ravi}, {Aguilar}, {Ahlen}, {Amsellem}, {Bautista},
  {Bianchi}, {Blake}, {Brooks}, {Claybaugh}, {Cuceu}, {de la Macorra},
  {Demina}, {Doel}, {Ferraro}, {Font-Ribera}, {Forero-Romero}, {Gaztanaga},
  {Gontcho}, {Gutierrez}, {Guy}, {Herrera-Alcantar}, {Honscheid}, {Howlett},
  {Huterer}, {Ishak}, {Joyce}, {Kremin}, {Lahav}, {Lamman}, {Landriau}, {Le
  Guillou}, {Leauthaud}, {Levi}, {Manera}, {Martini}, {Meisner}, {Miquel},
  {Moustakas}, {Munoz-Gutierrez}, {Nadathur}, {Palanque-Delabrouille},
  {Palmese}, {Percival}, {Poppett}, {Prada}, {Perez-Rafols}, {Qin}, {Ross},
  {Rossi}, {Said}, {Sanchez}, {Schlegel}, {Schubnell}, {Seo}, {Silber},
  {Sprayberry}, {Tarle}, {Turner}, {Weaver}, {Zhou}, \&
  {Zou}}]{DESIPV_Douglass}
{Douglass}, K., {BenZvi}, S., {Kim}, A.~G., {et~al.} 2025, arXiv e-prints,
  arXiv:2512.03227, \dodoi{10.48550/arXiv.2512.03227}

\bibitem[{{Dressler} {et~al.}(1987){Dressler}, {Lynden-Bell}, {Burstein},
  {Davies}, {Faber}, {Terlevich}, \& {Wegner}}]{Dressler1987}
{Dressler}, A., {Lynden-Bell}, D., {Burstein}, D., {et~al.} 1987, \apj, 313,
  42, \dodoi{10.1086/164947}

\bibitem[{{Dupuy} \& {Courtois}(2023)}]{Dupuy2023}
{Dupuy}, A., \& {Courtois}, H.~M. 2023, \aap, 678, A176,
  \dodoi{10.1051/0004-6361/202346802}

\bibitem[{{Dupuy} {et~al.}(2019){Dupuy}, {Courtois}, \& {Kubik}}]{Dupuy2019}
{Dupuy}, A., {Courtois}, H.~M., \& {Kubik}, B. 2019, \mnras, 486, 440,
  \dodoi{10.1093/mnras/stz901}

\bibitem[{{Erdo{\v{g}}du} {et~al.}(2006){Erdo{\v{g}}du}, {Lahav}, {Huchra},
  {Colless}, {Cutri}, {Falco}, {George}, {Jarrett}, {Jones}, {Macri}, {Mader},
  {Martimbeau}, {Pahre}, {Parker}, {Rassat}, \& {Saunders}}]{Erdogdu2006}
{Erdo{\v{g}}du}, P., {Lahav}, O., {Huchra}, J.~P., {et~al.} 2006, \mnras, 373,
  45, \dodoi{10.1111/j.1365-2966.2006.11049.x}

\bibitem[{{Feldman} {et~al.}(1994){Feldman}, {Kaiser}, \&
  {Peacock}}]{Feldman1994}
{Feldman}, H.~A., {Kaiser}, N., \& {Peacock}, J.~A. 1994, \apj, 426, 23,
  \dodoi{10.1086/174036}

\bibitem[{{Feldman} {et~al.}(2010){Feldman}, {Watkins}, \&
  {Hudson}}]{Feldman2010}
{Feldman}, H.~A., {Watkins}, R., \& {Hudson}, M.~J. 2010, \mnras, 407, 2328,
  \dodoi{10.1111/j.1365-2966.2010.17052.x}

\bibitem[{{Ganeshaiah Veena} {et~al.}(2023){Ganeshaiah Veena}, {Lilow}, \&
  {Nusser}}]{Ganeshaiah2023}
{Ganeshaiah Veena}, P., {Lilow}, R., \& {Nusser}, A. 2023, \mnras, 522, 5291,
  \dodoi{10.1093/mnras/stad1222}

\bibitem[{{Gorski} {et~al.}(1989){Gorski}, {Davis}, {Strauss}, {White}, \&
  {Yahil}}]{Gorski1989}
{Gorski}, K.~M., {Davis}, M., {Strauss}, M.~A., {White}, S. D.~M., \& {Yahil},
  A. 1989, \apj, 344, 1, \dodoi{10.1086/167771}

\bibitem[{{Guy} {et~al.}(2023){Guy}, {Bailey}, {Kremin}, {Alam}, {Alexander},
  {Allende Prieto}, {BenZvi}, {Bolton}, {Brooks}, {Chaussidon}, {Cooper},
  {Dawson}, {de la Macorra}, {Dey}, {Dey}, {Dhungana}, {Eisenstein},
  {Font-Ribera}, {Forero-Romero}, {Gazta{\~n}aga}, {Gontcho A Gontcho},
  {Green}, {Honscheid}, {Ishak}, {Kehoe}, {Kirkby}, {Kisner}, {Koposov}, {Lan},
  {Landriau}, {Le Guillou}, {Levi}, {Magneville}, {Manser}, {Martini},
  {Meisner}, {Miquel}, {Moustakas}, {Myers}, {Newman}, {Nie},
  {Palanque-Delabrouille}, {Percival}, {Poppett}, {Prada}, {Raichoor},
  {Ravoux}, {Ross}, {Schlafly}, {Schlegel}, {Schubnell}, {Sharples},
  {Tarl{\'e}}, {Weaver}, {Y{\'e}che}, {Zhou}, {Zhou}, \&
  {Zou}}]{Spectro.Pipeline.Guy.2023}
{Guy}, J., {Bailey}, S., {Kremin}, A., {et~al.} 2023, \aj, 165, 144,
  \dodoi{10.3847/1538-3881/acb212}

\bibitem[{{Hahn} {et~al.}(2023){Hahn}, {Wilson}, {Ruiz-Macias}, {Cole},
  {Weinberg}, {Moustakas}, {Kremin}, {Tinker}, {Smith}, {Wechsler}, {Ahlen},
  {Alam}, {Bailey}, {Brooks}, {Cooper}, {Davis}, {Dawson}, {Dey}, {Dey},
  {Eftekharzadeh}, {Eisenstein}, {Fanning}, {Forero-Romero}, {Frenk},
  {Gazta{\~n}aga}, {A Gontcho}, {Guy}, {Honscheid}, {Ishak}, {Juneau}, {Kehoe},
  {Kisner}, {Lan}, {Landriau}, {Le Guillou}, {Levi}, {Magneville}, {Martini},
  {Meisner}, {Myers}, {Nie}, {Norberg}, {Palanque-Delabrouille}, {Percival},
  {Poppett}, {Prada}, {Raichoor}, {Ross}, {Gaines}, {Saulder}, {Schlafly},
  {Schlegel}, {Sierra-Porta}, {Tarle}, {Weaver}, {Y{\`e}che}, {Zarrouk},
  {Zhou}, {Zhou}, \& {Zou}}]{Hahn2023}
{Hahn}, C., {Wilson}, M.~J., {Ruiz-Macias}, O., {et~al.} 2023, \aj, 165, 253,
  \dodoi{10.3847/1538-3881/accff8}

\bibitem[{{Hartlap} {et~al.}(2007){Hartlap}, {Simon}, \&
  {Schneider}}]{Hartlap2007}
{Hartlap}, J., {Simon}, P., \& {Schneider}, P. 2007, \aap, 464, 399,
  \dodoi{10.1051/0004-6361:20066170}

\bibitem[{{Heath}(1977)}]{Heath1977}
{Heath}, D.~J. 1977, \mnras, 179, 351, \dodoi{10.1093/mnras/179.3.351}

\bibitem[{{Heinesen}(2023)}]{Heinesen2023}
{Heinesen}, A. 2023, \prd, 108, 103530, \dodoi{10.1103/PhysRevD.108.103530}

\bibitem[{Hojjati {et~al.}(2011)Hojjati, Pogosian, \& Zhao}]{Hojjati:2011ix}
Hojjati, A., Pogosian, L., \& Zhao, G.-B. 2011, JCAP, 08, 005,
  \dodoi{10.1088/1475-7516/2011/08/005}

\bibitem[{{Hollinger} \& {Hudson}(2024)}]{Hollinger2024}
{Hollinger}, A.~M., \& {Hudson}, M.~J. 2024, \mnras, 531, 788,
  \dodoi{10.1093/mnras/stae1042}

\bibitem[{{Hong} {et~al.}(2021){Hong}, {Jeong}, {Hwang}, \&
  {Kim}}]{Sungwook2021}
{Hong}, S.~E., {Jeong}, D., {Hwang}, H.~S., \& {Kim}, J. 2021, \apj, 913, 76,
  \dodoi{10.3847/1538-4357/abf040}

\bibitem[{{Hong} {et~al.}(2014){Hong}, {Springob}, {Staveley-Smith},
  {Scrimgeour}, {Masters}, {Macri}, {Koribalski}, {Jones}, \&
  {Jarrett}}]{Hong2014}
{Hong}, T., {Springob}, C.~M., {Staveley-Smith}, L., {et~al.} 2014, \mnras,
  445, 402, \dodoi{10.1093/mnras/stu1774}

\bibitem[{{Hong} {et~al.}(2019){Hong}, {Staveley-Smith}, {Masters}, {Springob},
  {Macri}, {Koribalski}, {Jones}, {Jarrett}, {Crook}, {Howlett}, \&
  {Qin}}]{Hong2019}
{Hong}, T., {Staveley-Smith}, L., {Masters}, K.~L., {et~al.} 2019, \mnras, 487,
  2061, \dodoi{10.1093/mnras/stz1413}

\bibitem[{{Howlett}(2019)}]{Howlett2019}
{Howlett}, C. 2019, \mnras, 487, 5209, \dodoi{10.1093/mnras/stz1403}

\bibitem[{{Howlett} {et~al.}(2015){Howlett}, {Ross}, {Samushia}, {Percival}, \&
  {Manera}}]{Howlett2015correlationfunction}
{Howlett}, C., {Ross}, A.~J., {Samushia}, L., {Percival}, W.~J., \& {Manera},
  M. 2015, \mnras, 449, 848, \dodoi{10.1093/mnras/stu2693}

\bibitem[{{Howlett} {et~al.}(2022){Howlett}, {Said}, {Lucey}, {Colless}, {Qin},
  {Lai}, {Tully}, \& {Davis}}]{Howlett2022}
{Howlett}, C., {Said}, K., {Lucey}, J.~R., {et~al.} 2022, \mnras, 515, 953,
  \dodoi{10.1093/mnras/stac1681}

\bibitem[{{Howlett} {et~al.}(2017){Howlett}, {Staveley-Smith}, {Elahi}, {Hong},
  {Jarrett}, {Jones}, {Koribalski}, {Macri}, {Masters}, \&
  {Springob}}]{Howlett2017velocitypower}
{Howlett}, C., {Staveley-Smith}, L., {Elahi}, P.~J., {et~al.} 2017, \mnras,
  471, 3135, \dodoi{10.1093/mnras/stx1521}

\bibitem[{{Ishak}(2019)}]{Ishak2019}
{Ishak}, M. 2019, Living Reviews in Relativity, 22, 1,
  \dodoi{10.1007/s41114-018-0017-4}

\bibitem[{{Ishak} {et~al.}(2025){Ishak}, {Pan}, {Calderon}, {Lodha},
  {Valogiannis}, {Aviles}, {Niz}, {Yi}, {Zheng}, {Garcia-Quintero}, {de
  Mattia}, {Medina-Varela}, {Cervantes-Cota}, {Andrade}, {Huterer}, {Noriega},
  {Zhao}, {Shafieloo}, {Fang}, {Ahlen}, {Bianchi}, {Brooks}, {Burtin},
  {Chaussidon}, {Claybaugh}, {Cole}, {de la Macorra}, {Dey}, {Fanning},
  {Ferraro}, {Font-Ribera}, {Forero-Romero}, {Gazta{\~n}aga}, {Gil-Mar{\'\i}n},
  {Gontcho A. Gontcho}, {Gutierrez}, {Hahn}, {Honscheid}, {Howlett}, {Juneau},
  {Kirkby}, {Kisner}, {Kremin}, {Landriau}, {Le Guillou}, {Leauthaud}, {Levi},
  {Meisner}, {Miquel}, {Moustakas}, {Newman}, {Palanque-Delabrouille},
  {Percival}, {Poppett}, {Prada}, {P{\'e}rez-R{\`a}fols}, {Ross}, {Rossi},
  {Sanchez}, {Schlegel}, {Schubnell}, {Seo}, {Sprayberry}, {Tarl{\'e}},
  {Vargas-Maga{\~n}a}, {Weaver}, {Wechsler}, {Y{\`e}che}, {Zarrouk}, {Zhou}, \&
  {Zou}}]{Ishak2025}
{Ishak}, M., {Pan}, J., {Calderon}, R., {et~al.} 2025, \jcap, 2025, 053,
  \dodoi{10.1088/1475-7516/2025/09/053}

\bibitem[{{Ivanov} {et~al.}(2020){Ivanov}, {Simonovi{\'c}}, \&
  {Zaldarriaga}}]{Ivanov2020}
{Ivanov}, M.~M., {Simonovi{\'c}}, M., \& {Zaldarriaga}, M. 2020, \jcap, 2020,
  042, \dodoi{10.1088/1475-7516/2020/05/042}

\bibitem[{{Jaffe} \& {Kaiser}(1995)}]{Jaffe1995}
{Jaffe}, A.~H., \& {Kaiser}, N. 1995, \apj, 455, 26, \dodoi{10.1086/176551}

\bibitem[{{Johnson} {et~al.}(2014){Johnson}, {Blake}, {Koda}, {Ma}, {Colless},
  {Crocce}, {Davis}, {Jones}, {Magoulas}, {Lucey}, {Mould}, {Scrimgeour}, \&
  {Springob}}]{Johnson2014}
{Johnson}, A., {Blake}, C., {Koda}, J., {et~al.} 2014, \mnras, 444, 3926,
  \dodoi{10.1093/mnras/stu1615}

\bibitem[{{Kaiser}(1984)}]{Kaiser1984correlation}
{Kaiser}, N. 1984, \apjl, 284, L9, \dodoi{10.1086/184341}

\bibitem[{{Kaiser}(1987)}]{Kaiser1987}
---. 1987, \mnras, 227, 1, \dodoi{10.1093/mnras/227.1.1}

\bibitem[{{Kaiser}(1988)}]{Kaiser1988CosFlow}
---. 1988, \mnras, 231, 149, \dodoi{10.1093/mnras/231.2.149}

\bibitem[{{Kaiser} \& {Peacock}(1991)}]{Kaiser1991PS}
{Kaiser}, N., \& {Peacock}, J.~A. 1991, \apj, 379, 482, \dodoi{10.1086/170523}

\bibitem[{{Karamanis} \& {Beutler}(2021)}]{Karamanis2021}
{Karamanis}, M., \& {Beutler}, F. 2021, arXiv e-prints, arXiv:2106.06331,
  \dodoi{10.48550/arXiv.2106.06331}

\bibitem[{{Kitaura} {et~al.}(2012){Kitaura}, {Angulo}, {Hoffman}, \&
  {Gottl{\"o}ber}}]{Kitaura2012}
{Kitaura}, F.-S., {Angulo}, R.~E., {Hoffman}, Y., \& {Gottl{\"o}ber}, S. 2012,
  \mnras, 425, 2422, \dodoi{10.1111/j.1365-2966.2012.21589.x}

\bibitem[{{Koda} {et~al.}(2014){Koda}, {Blake}, {Davis}, {Magoulas},
  {Springob}, {Scrimgeour}, {Johnson}, {Poole}, \& {Staveley-Smith}}]{Koda2014}
{Koda}, J., {Blake}, C., {Davis}, T., {et~al.} 2014, \mnras, 445, 4267,
  \dodoi{10.1093/mnras/stu1610}

\bibitem[{{Kudlicki} {et~al.}(2000){Kudlicki}, {Chodorowski}, {Plewa}, \&
  {R{\'o}{\.z}yczka}}]{Kudlicki2000}
{Kudlicki}, A., {Chodorowski}, M., {Plewa}, T., \& {R{\'o}{\.z}yczka}, M. 2000,
  \mnras, 316, 464, \dodoi{10.1046/j.1365-8711.2000.03463.x}

\bibitem[{{Lai} {et~al.}(2023){Lai}, {Howlett}, \& {Davis}}]{Lai2023}
{Lai}, Y., {Howlett}, C., \& {Davis}, T.~M. 2023, \mnras, 518, 1840,
  \dodoi{10.1093/mnras/stac3252}

\bibitem[{{Lai} {et~al.}(2025){Lai}, {Howlett}, {Aguilar}, {Ahlen}, {Amsellem},
  {Bautista}, {BenZvi}, {Bianchi}, {Blake}, {Brooks}, {Carr}, {Claybaugh},
  {Davis}, {de la Macorra}, {Doel}, {Douglass}, {Ferraro}, {Font-Ribera},
  {Forero-Romero}, {Gazta{\~n}aga}, {Gutierrez}, {Guy}, {Herrera-Alcantar},
  {Huterer}, {Ishak}, {Joyce}, {Kim}, {Kirkby}, {Kisner}, {Kremin}, {Lahav},
  {Lamman}, {Landriau}, {Le Guillou}, {Leauthaud}, {Levi}, {Manera}, {Martini},
  {Meisner}, {Miquel}, {Moustakas}, {Mu{\~n}oz-Guti{\'e}rrez}, {Nadathur},
  {Percival}, {Poppett}, {Prada}, {P{\'e}rez-R{\`a}fols}, {Qin}, {Ross},
  {Rossi}, {Said}, {Sanchez}, {Schlegel}, {Schubnell}, {Seo}, {Silber},
  {Sprayberry}, {Tarl{\'e}}, {Turner}, {Weaver}, {Zarrouk}, {Zhou}, \&
  {Zou}}]{DESIPV_Lai}
{Lai}, Y., {Howlett}, C., {Aguilar}, J., {et~al.} 2025, arXiv e-prints,
  arXiv:2512.03229, \dodoi{10.48550/arXiv.2512.03229}

\bibitem[{{Levi} {et~al.}(2013){Levi}, {Bebek}, {Beers}, {Blum}, {Cahn},
  {Eisenstein}, {Flaugher}, {Honscheid}, {Kron}, {Lahav}, {McDonald}, {Roe},
  {Schlegel}, \& {representing the DESI collaboration}}]{Snowmass2013.Levi}
{Levi}, M., {Bebek}, C., {Beers}, T., {et~al.} 2013, arXiv e-prints,
  arXiv:1308.0847.
\newblock \doarXiv{1308.0847}

\bibitem[{Lewis {et~al.}(2000)Lewis, Challinor, \& Lasenby}]{Lewis:1999bs}
Lewis, A., Challinor, A., \& Lasenby, A. 2000, \apj, 538, 473,
  \dodoi{10.1086/309179}

\bibitem[{{Lilow} {et~al.}(2024){Lilow}, {Ganeshaiah Veena}, \&
  {Nusser}}]{Lilow2024}
{Lilow}, R., {Ganeshaiah Veena}, P., \& {Nusser}, A. 2024, \aap, 689, A226,
  \dodoi{10.1051/0004-6361/202450219}

\bibitem[{Linder(2005)}]{Linder:2005in}
Linder, E.~V. 2005, Phys. Rev. D, 72, 043529,
  \dodoi{10.1103/PhysRevD.72.043529}

\bibitem[{{Lopes} {et~al.}(2024){Lopes}, {Bernui}, {Franco}, \&
  {Avila}}]{Lopes2024}
{Lopes}, M., {Bernui}, A., {Franco}, C., \& {Avila}, F. 2024, \apj, 967, 47,
  \dodoi{10.3847/1538-4357/ad3735}

\bibitem[{{Lyall} {et~al.}(2023){Lyall}, {Blake}, {Turner}, {Ruggeri}, \&
  {Winther}}]{Lyall2023}
{Lyall}, S., {Blake}, C., {Turner}, R., {Ruggeri}, R., \& {Winther}, H. 2023,
  \mnras, 518, 5929, \dodoi{10.1093/mnras/stac3323}

\bibitem[{{Lyall} {et~al.}(2024){Lyall}, {Blake}, \& {Turner}}]{Lyall2024}
{Lyall}, S., {Blake}, C., \& {Turner}, R.~J. 2024, \mnras, 532, 3972,
  \dodoi{10.1093/mnras/stae1718}

\bibitem[{{Ma} {et~al.}(2012){Ma}, {Branchini}, \& {Scott}}]{Ma2012}
{Ma}, Y.-Z., {Branchini}, E., \& {Scott}, D. 2012, \mnras, 425, 2880,
  \dodoi{10.1111/j.1365-2966.2012.21671.x}

\bibitem[{Maksimova {et~al.}(2021)Maksimova, Garrison, Eisenstein, Hadzhiyska,
  Bose, \& Satterthwaite}]{ABACUS2021}
Maksimova, N.~A., Garrison, L.~H., Eisenstein, D.~J., {et~al.} 2021, Monthly
  Notices of the Royal Astronomical Society, 508, 4017,
  \dodoi{10.1093/mnras/stab2484}

\bibitem[{{Mao} {et~al.}(2021){Mao}, {Wang}, {Li}, {Cai}, {Falck}, {Neyrinck},
  \& {Szalay}}]{Mao2021}
{Mao}, T.-X., {Wang}, J., {Li}, B., {et~al.} 2021, \mnras, 501, 1499,
  \dodoi{10.1093/mnras/staa3741}

\bibitem[{{McDonald} \& {Roy}(2009)}]{McDonald2009}
{McDonald}, P., \& {Roy}, A. 2009, \jcap, 2009, 020,
  \dodoi{10.1088/1475-7516/2009/08/020}

\bibitem[{{Miller} {et~al.}(2024){Miller}, {Doel}, {Gutierrez}, {Besuner},
  {Brooks}, {Gallo}, {Heetderks}, {Jelinsky}, {Kent}, {Lampton}, {Levi},
  {Liang}, {Meisner}, {Sholl}, {Silber}, {Sprayberry}, {Aguilar}, {de la
  Macorra}, {Eisenstein}, {Fanning}, {Font-Ribera}, {Gazta{\~n}aga}, {Gontcho A
  Gontcho}, {Honscheid}, {Jimenez}, {Joyce}, {Kehoe}, {Kisner}, {Kremin},
  {Landriau}, {Le Guillou}, {Magneville}, {Martini}, {Miquel}, {Moustakas},
  {Nie}, {Percival}, {Poppett}, {Prada}, {Rossi}, {Schlegel}, {Schubnell},
  {Seo}, {Sharples}, {Tarl{\'e}}, {Vargas-Maga{\~n}a}, {Zhou}, \& {the DESI
  Collaboration}}]{Corrector.Miller.2023}
{Miller}, T.~N., {Doel}, P., {Gutierrez}, G., {et~al.} 2024, \aj, 168, 95,
  \dodoi{10.3847/1538-3881/ad45fe}

\bibitem[{{Nguyen} {et~al.}(2025){Nguyen}, {Blake}, {Turner}, {Aronica},
  {Bautista}, {Aguilar}, {Ahlen}, {BenZvi}, {Bianchi}, {Brooks}, {Carr},
  {Claybaugh}, {Cuceu}, {de la Macorra}, {Dey}, {Doel}, {Douglass}, {Ferraro},
  {Forero-Romero}, {Gazta{\~n}aga}, {Gontcho}, {Gutierrez}, {Guy}, {Honscheid},
  {Howlett}, {Huterer}, {Ishak}, {Joyce}, {Kehoe}, {Kim}, {Kremin}, {Lahav},
  {Landriau}, {Le Guillou}, {Leauthaud}, {Levi}, {Manera}, {Martini},
  {Meisner}, {Miquel}, {Mueller}, {Nadathur}, {Palanque-Delabrouille},
  {Percival}, {Poppett}, {Prada}, {Qin}, {Ross}, {Ross}, {Rossi}, {Sanchez},
  {Schlegel}, {Schubnell}, {Sprayberry}, {Tarl{\'e}}, {Weaver}, {Zarrouk},
  {Zhou}, \& {Zou}}]{DESIPV_Nguyen}
{Nguyen}, A., {Blake}, C., {Turner}, R.~J., {et~al.} 2025, arXiv e-prints,
  arXiv:2510.07673, \dodoi{10.48550/arXiv.2510.07673}

\bibitem[{{Nguyen} {et~al.}(2023){Nguyen}, {Huterer}, \& {Wen}}]{Nguyen2023}
{Nguyen}, N.-M., {Huterer}, D., \& {Wen}, Y. 2023, \prl, 131, 111001,
  \dodoi{10.1103/PhysRevLett.131.111001}

\bibitem[{{Nusser}(2014)}]{Nusser2014}
{Nusser}, A. 2014, \apj, 795, 3, \dodoi{10.1088/0004-637X/795/1/3}

\bibitem[{{Nusser} {et~al.}(1991){Nusser}, {Dekel}, {Bertschinger}, \&
  {Blumenthal}}]{Nusser1991}
{Nusser}, A., {Dekel}, A., {Bertschinger}, E., \& {Blumenthal}, G.~R. 1991,
  \apj, 379, 6, \dodoi{10.1086/170480}

\bibitem[{{Okumura} {et~al.}(2014){Okumura}, {Seljak}, {Vlah}, \&
  {Desjacques}}]{Okumura2014}
{Okumura}, T., {Seljak}, U., {Vlah}, Z., \& {Desjacques}, V. 2014, \jcap, 2014,
  003, \dodoi{10.1088/1475-7516/2014/05/003}

\bibitem[{{Park}(2000)}]{Park2000}
{Park}, C. 2000, \mnras, 319, 573, \dodoi{10.1046/j.1365-8711.2000.03886.x}

\bibitem[{{Park} \& {Park}(2006)}]{Park2006}
{Park}, C.-G., \& {Park}, C. 2006, \apj, 637, 1, \dodoi{10.1086/498258}

\bibitem[{{Parnovsky} {et~al.}(2001){Parnovsky}, {Kudrya}, {Karachentseva}, \&
  {Karachentsev}}]{Parnovsky2001}
{Parnovsky}, S.~L., {Kudrya}, Y.~N., {Karachentseva}, V.~E., \& {Karachentsev},
  I.~D. 2001, Astronomy Letters, 27, 765, \dodoi{10.1134/1.1424358}

\bibitem[{{Peery} {et~al.}(2018){Peery}, {Watkins}, \& {Feldman}}]{Peery2018}
{Peery}, S., {Watkins}, R., \& {Feldman}, H.~A. 2018, \mnras, 481, 1368,
  \dodoi{10.1093/mnras/sty2332}

\bibitem[{{Pike} \& {Hudson}(2005)}]{Pike2005}
{Pike}, R.~W., \& {Hudson}, M.~J. 2005, \apj, 635, 11, \dodoi{10.1086/497359}

\bibitem[{{Planck Collaboration} {et~al.}(2020){Planck Collaboration},
  {Aghanim}, {Akrami}, {Ashdown}, {Aumont}, {Baccigalupi}, {Ballardini},
  {Banday}, {Barreiro}, {Bartolo}, {Basak}, {Battye}, {Benabed}, {Bernard},
  {Bersanelli}, {Bielewicz}, {Bock}, {Bond}, {Borrill}, {Bouchet}, {Boulanger},
  {Bucher}, {Burigana}, {Butler}, {Calabrese}, {Cardoso}, {Carron},
  {Challinor}, {Chiang}, {Chluba}, {Colombo}, {Combet}, {Contreras}, {Crill},
  {Cuttaia}, {de Bernardis}, {de Zotti}, {Delabrouille}, {Delouis}, {Di
  Valentino}, {Diego}, {Dor{\'e}}, {Douspis}, {Ducout}, {Dupac}, {Dusini},
  {Efstathiou}, {Elsner}, {En{\ss}lin}, {Eriksen}, {Fantaye}, {Farhang},
  {Fergusson}, {Fernandez-Cobos}, {Finelli}, {Forastieri}, {Frailis},
  {Fraisse}, {Franceschi}, {Frolov}, {Galeotta}, {Galli}, {Ganga},
  {G{\'e}nova-Santos}, {Gerbino}, {Ghosh}, {Gonz{\'a}lez-Nuevo}, {G{\'o}rski},
  {Gratton}, {Gruppuso}, {Gudmundsson}, {Hamann}, {Handley}, {Hansen},
  {Herranz}, {Hildebrandt}, {Hivon}, {Huang}, {Jaffe}, {Jones}, {Karakci},
  {Keih{\"a}nen}, {Keskitalo}, {Kiiveri}, {Kim}, {Kisner}, {Knox},
  {Krachmalnicoff}, {Kunz}, {Kurki-Suonio}, {Lagache}, {Lamarre}, {Lasenby},
  {Lattanzi}, {Lawrence}, {Le Jeune}, {Lemos}, {Lesgourgues}, {Levrier},
  {Lewis}, {Liguori}, {Lilje}, {Lilley}, {Lindholm}, {L{\'o}pez-Caniego},
  {Lubin}, {Ma}, {Mac{\'\i}as-P{\'e}rez}, {Maggio}, {Maino}, {Mandolesi},
  {Mangilli}, {Marcos-Caballero}, {Maris}, {Martin}, {Martinelli},
  {Mart{\'\i}nez-Gonz{\'a}lez}, {Matarrese}, {Mauri}, {McEwen}, {Meinhold},
  {Melchiorri}, {Mennella}, {Migliaccio}, {Millea}, {Mitra},
  {Miville-Desch{\^e}nes}, {Molinari}, {Montier}, {Morgante}, {Moss}, {Natoli},
  {N{\o}rgaard-Nielsen}, {Pagano}, {Paoletti}, {Partridge}, {Patanchon},
  {Peiris}, {Perrotta}, {Pettorino}, {Piacentini}, {Polastri}, {Polenta},
  {Puget}, {Rachen}, {Reinecke}, {Remazeilles}, {Renzi}, {Rocha}, {Rosset},
  {Roudier}, {Rubi{\~n}o-Mart{\'\i}n}, {Ruiz-Granados}, {Salvati}, {Sandri},
  {Savelainen}, {Scott}, {Shellard}, {Sirignano}, {Sirri}, {Spencer},
  {Sunyaev}, {Suur-Uski}, {Tauber}, {Tavagnacco}, {Tenti}, {Toffolatti},
  {Tomasi}, {Trombetti}, {Valenziano}, {Valiviita}, {Van Tent}, {Vibert},
  {Vielva}, {Villa}, {Vittorio}, {Wandelt}, {Wehus}, {White}, {White},
  {Zacchei}, \& {Zonca}}]{Planck2020}
{Planck Collaboration}, {Aghanim}, N., {Akrami}, Y., {et~al.} 2020, \aap, 641,
  A6, \dodoi{10.1051/0004-6361/201833910}

\bibitem[{{Poppett} {et~al.}(2024){Poppett}, {Tyas}, {Aguilar}, {Bebek},
  {Bramall}, {Claybaugh}, {Edelstein}, {Fagrelius}, {Heetderks}, {Jelinsky},
  {Jelinsky}, {Lafever}, {Lambert}, {Lampton}, {Levi}, {Martini}, {Rockosi},
  {Schmoll}, {Sharples}, {Sirk}, {Wishnow}, {Yu}, {Ahlen}, {Bault}, {BenZvi},
  {Brooks}, {Cole}, {de la Macorra}, {Dey}, {Doel}, {Fanning}, {Font-Ribera},
  {Forero-Romero}, {Gazta{\~n}aga}, {Gontcho A Gontcho}, {Gonzalez-Morales},
  {Hahn}, {Honscheid}, {Jimenez}, {Juneau}, {Kirkby}, {Kremin}, {Landriau}, {Le
  Guillou}, {Manera}, {Meisner}, {Miquel}, {Moustakas}, {Mueller},
  {Mu{\~n}oz-Guti{\'e}rrez}, {Myers}, {Nie}, {Niz}, {Palanque-Delabrouille},
  {Percival}, {Prada}, {Rabinowitz}, {Rezaie}, {Rossi}, {Sanchez}, {Schlafly},
  {Schlegel}, {Schubnell}, {Seo}, {Sprayberry}, {Tarl{\'e}},
  {Vargas-Maga{\~n}a}, {Weaver}, \& {Zhou}}]{FiberSystem.Poppett.2024}
{Poppett}, C., {Tyas}, L., {Aguilar}, J., {et~al.} 2024, \aj, 168, 245,
  \dodoi{10.3847/1538-3881/ad76a4}

\bibitem[{{Qin}(2021)}]{Qin2021CosFlowBoxCox}
{Qin}, F. 2021, Research in Astronomy and Astrophysics, 21, 242,
  \dodoi{10.1088/1674-4527/21/10/242}

\bibitem[{{Qin} {et~al.}(2025){Qin}, {Howlett}, \& {Parkinson}}]{Qin2025}
{Qin}, F., {Howlett}, C., \& {Parkinson}, D. 2025, \apj, 978, 7,
  \dodoi{10.3847/1538-4357/ad9391}

\bibitem[{{Qin} {et~al.}(2019{\natexlab{a}}){Qin}, {Howlett}, \&
  {Staveley-Smith}}]{Qin2019PS}
{Qin}, F., {Howlett}, C., \& {Staveley-Smith}, L. 2019{\natexlab{a}}, \mnras,
  487, 5235, \dodoi{10.1093/mnras/stz1576}

\bibitem[{{Qin} {et~al.}(2018){Qin}, {Howlett}, {Staveley-Smith}, \&
  {Hong}}]{Qin2018}
{Qin}, F., {Howlett}, C., {Staveley-Smith}, L., \& {Hong}, T. 2018, \mnras,
  477, 5150, \dodoi{10.1093/mnras/sty928}

\bibitem[{{Qin} {et~al.}(2019{\natexlab{b}}){Qin}, {Howlett}, {Staveley-Smith},
  \& {Hong}}]{Qin2019CosFlow}
---. 2019{\natexlab{b}}, \mnras, 482, 1920, \dodoi{10.1093/mnras/sty2826}

\bibitem[{{Qin} {et~al.}(2022){Qin}, {Howlett}, {Stevens}, \&
  {Parkinson}}]{Qin2022}
{Qin}, F., {Howlett}, C., {Stevens}, A. R.~H., \& {Parkinson}, D. 2022, \apj,
  937, 113, \dodoi{10.3847/1538-4357/ac8b6f}

\bibitem[{{Qin} {et~al.}(2023{\natexlab{a}}){Qin}, {Parkinson}, {Hong}, \&
  {Sabiu}}]{Qin2023Recon}
{Qin}, F., {Parkinson}, D., {Hong}, S.~E., \& {Sabiu}, C.~G.
  2023{\natexlab{a}}, \jcap, 2023, 062, \dodoi{10.1088/1475-7516/2023/06/062}

\bibitem[{{Qin} {et~al.}(2021){Qin}, {Parkinson}, {Howlett}, \&
  {Said}}]{Qin2021CosFlowCF4}
{Qin}, F., {Parkinson}, D., {Howlett}, C., \& {Said}, K. 2021, \apj, 922, 59,
  \dodoi{10.3847/1538-4357/ac249d}

\bibitem[{{Qin} {et~al.}(2023{\natexlab{b}}){Qin}, {Parkinson}, {Stevens}, \&
  {Howlett}}]{Qin2023HaloProf}
{Qin}, F., {Parkinson}, D., {Stevens}, A. R.~H., \& {Howlett}, C.
  2023{\natexlab{b}}, \apj, 957, 40, \dodoi{10.3847/1538-4357/acfda5}

\bibitem[{{Ravoux} {et~al.}(2025){Ravoux}, {Carreres}, {Rosselli}, {Bautista},
  {Carr}, {Dumerchat}, {Kim}, {Parkinson}, {Racine}, {Fouchez}, \&
  {Feinstein}}]{Ravoux2025}
{Ravoux}, C., {Carreres}, B., {Rosselli}, D., {et~al.} 2025, \aap, 698, A273,
  \dodoi{10.1051/0004-6361/202554319}

\bibitem[{{Rocher} {et~al.}(2023){Rocher}, {Ruhlmann-Kleider}, {Burtin},
  {Yuan}, {de Mattia}, {Ross}, {Aguilar}, {Ahlen}, {Alam}, {Bianchi}, {Brooks},
  {Cole}, {Dawson}, {de la Macorra}, {Doel}, {Eisenstein}, {Fanning},
  {Forero-Romero}, {Garrison}, {Gontcho A Gontcho}, {Gonzalez-Perez}, {Guy},
  {Hadzhiyska}, {Hahn}, {Honscheid}, {Kisner}, {Landriau}, {Lasker}, {E. Levi},
  {Manera}, {Meisner}, {Miquel}, {Moustakas}, {Mueller}, {Newman}, {Nie},
  {Percival}, {Poppett}, {Qin}, {Rossi}, {Samushia}, {Sanchez}, {Schlegel},
  {Schubnell}, {Seo}, {Tarl{\'e}}, {Vargas-Maga{\~n}a}, {Weaver}, {Yu},
  {Zhang}, {Zheng}, {Zhou}, \& {Zou}}]{Rocher2023}
{Rocher}, A., {Ruhlmann-Kleider}, V., {Burtin}, E., {et~al.} 2023, \jcap, 2023,
  016, \dodoi{10.1088/1475-7516/2023/10/016}

\bibitem[{{Ross} {et~al.}(2025){Ross}, {Howlett}, {Lucey}, {Said}, {Davis},
  {Aguilar}, {Ahlen}, {Amsellem}, {Bautista}, {BenZvi}, {Bianchi}, {Blake},
  {Brooks}, {Carr}, {Claybaugh}, {Cuceu}, {de la Macorra}, {Dey}, {Doel},
  {Douglass}, {Ferraro}, {Font-Ribera}, {Forero-Romero}, {Gazta{\~n}aga},
  {Gontcho}, {Gutierrez}, {Guy}, {Honscheid}, {Huterer}, {Ishak}, {Joyce},
  {Kim}, {Kremin}, {Lahav}, {Lamman}, {Landriau}, {Le Guillou}, {Leauthaud},
  {Levi}, {Martini}, {Meisner}, {Miquel}, {Moustakas}, {Mu\textbackslash
  noz-Guti{\'e}rrez}, {Nadathur}, {Palanque-Delabrouille}, {Percival},
  {Poppett}, {Prada}, {P{\'e}rez-R{\`a}fols}, {Qin}, {Rossi}, {Sanchez},
  {Schlegel}, {Schubnell}, {Sprayberry}, {Tarl{\'e}}, {Turner}, {Weaver},
  {Zhou}, \& {Zou}}]{DESIPV_Ross}
{Ross}, C.~E., {Howlett}, C., {Lucey}, J.~R., {et~al.} 2025, arXiv e-prints,
  arXiv:2512.03226, \dodoi{10.48550/arXiv.2512.03226}

\bibitem[{{Said} {et~al.}(2020){Said}, {Colless}, {Magoulas}, {Lucey}, \&
  {Hudson}}]{Said2020}
{Said}, K., {Colless}, M., {Magoulas}, C., {Lucey}, J.~R., \& {Hudson}, M.~J.
  2020, \mnras, 497, 1275, \dodoi{10.1093/mnras/staa2032}

\bibitem[{{Saito} {et~al.}(2014){Saito}, {Baldauf}, {Vlah}, {Seljak},
  {Okumura}, \& {McDonald}}]{Saito2014}
{Saito}, S., {Baldauf}, T., {Vlah}, Z., {et~al.} 2014, \prd, 90, 123522,
  \dodoi{10.1103/PhysRevD.90.123522}

\bibitem[{{S{\'a}nchez} \& {Cole}(2008)}]{Sanchez2008}
{S{\'a}nchez}, A.~G., \& {Cole}, S. 2008, \mnras, 385, 830,
  \dodoi{10.1111/j.1365-2966.2007.12787.x}

\bibitem[{{Schlafly} {et~al.}(2023){Schlafly}, {Kirkby}, {Schlegel}, {Myers},
  {Raichoor}, {Dawson}, {Aguilar}, {Allende Prieto}, {Bailey}, {BenZvi},
  {Bermejo-Climent}, {Brooks}, {de la Macorra}, {Dey}, {Doel}, {Fanning},
  {Font-Ribera}, {Forero-Romero}, {Garc{\'\i}a-Bellido}, {Gontcho A Gontcho},
  {Guy}, {Hahn}, {Honscheid}, {Ishak}, {Juneau}, {Kehoe}, {Kisner}, {Kremin},
  {Landriau}, {Lang}, {Lasker}, {Levi}, {Magneville}, {Manser}, {Martini},
  {Meisner}, {Miquel}, {Moustakas}, {Newman}, {Nie}, {Palanque-Delabrouille},
  {Percival}, {Poppett}, {Rockosi}, {Ross}, {Rossi}, {Tarl{\'e}}, {Weaver},
  {Y{\`e}che}, {Zhou}, \& {DESI Collaboration}}]{SurveyOps.Schlafly.2023}
{Schlafly}, E.~F., {Kirkby}, D., {Schlegel}, D.~J., {et~al.} 2023, \aj, 166,
  259, \dodoi{10.3847/1538-3881/ad0832}

\bibitem[{{Scoccimarro}(2015)}]{Scoccimarro2015}
{Scoccimarro}, R. 2015, \prd, 92, 083532, \dodoi{10.1103/PhysRevD.92.083532}

\bibitem[{{Scrimgeour} {et~al.}(2016){Scrimgeour}, {Davis}, {Blake},
  {Staveley-Smith}, {Magoulas}, {Springob}, {Beutler}, {Colless}, {Johnson},
  {Jones}, {Koda}, {Lucey}, {Ma}, {Mould}, \& {Poole}}]{Scrimgeour2016}
{Scrimgeour}, M.~I., {Davis}, T.~M., {Blake}, C., {et~al.} 2016, \mnras, 455,
  386, \dodoi{10.1093/mnras/stv2146}

\bibitem[{{Sellentin} \& {Heavens}(2016)}]{Sellentin2016}
{Sellentin}, E., \& {Heavens}, A.~F. 2016, \mnras, 456, L132,
  \dodoi{10.1093/mnrasl/slv190}

\bibitem[{{Shi} {et~al.}(2018){Shi}, {Yang}, {Wang}, {Zhang}, {Mo}, {van den
  Bosch}, {Luo}, {Tweed}, {Li}, {Liu}, {Lu}, \& {Yang}}]{ShiFeng2018}
{Shi}, F., {Yang}, X., {Wang}, H., {et~al.} 2018, \apj, 861, 137,
  \dodoi{10.3847/1538-4357/aacb20}

\bibitem[{{Shi} {et~al.}(2025){Shi}, {Wang}, {Yang}, {Gu}, {Wei}, {Li}, {Han},
  {Wang}, {Zhang}, {Hong}, {Wang}, \& {Li}}]{Shi2025}
{Shi}, F., {Wang}, Z., {Yang}, X., {et~al.} 2025, arXiv e-prints,
  arXiv:2501.12621, \dodoi{10.48550/arXiv.2501.12621}

\bibitem[{{Shi} {et~al.}(2024){Shi}, {Zhang}, {Mao}, \& {Gu}}]{Shi2024}
{Shi}, Y., {Zhang}, P., {Mao}, S., \& {Gu}, Q. 2024, \mnras, 528, 4922,
  \dodoi{10.1093/mnras/stae274}

\bibitem[{{Silber} {et~al.}(2023){Silber}, {Fagrelius}, {Fanning}, {Schubnell},
  {Aguilar}, {Ahlen}, {Ameel}, {Ballester}, {Baltay}, {Bebek}, {Benton Beard},
  {Besuner}, {Cardiel-Sas}, {Casas}, {Castander}, {Claybaugh}, {Dobson},
  {Duan}, {Dunlop}, {Edelstein}, {Emmet}, {Elliott}, {Evatt}, {Gershkovich},
  {Guy}, {Harris}, {Heetderks}, {Heetderks}, {Honscheid}, {Illa}, {Jelinsky},
  {Jelinsky}, {Jimenez}, {Karcher}, {Kent}, {Kirkby}, {Kneib}, {Lambert},
  {Lampton}, {Leitner}, {Levi}, {McCauley}, {Meisner}, {Miller}, {Miquel},
  {Mundet}, {Poppett}, {Rabinowitz}, {Reil}, {Roman}, {Schlegel}, {Serrano},
  {Van Shourt}, {Sprayberry}, {Tarl{\'e}}, {Tie}, {Weaverdyck}, {Zhang},
  {Azzaro}, {Bailey}, {Becerril}, {Blackwell}, {Bouri}, {Brooks},
  {Buckley-Geer}, {Castro}, {Derwent}, {Dey}, {Dhungana}, {Doel}, {Eisenstein},
  {Fahim}, {Garcia-Bellido}, {Gazta{\~n}aga}, {A Gontcho}, {Gutierrez},
  {H{\"o}rler}, {Kehoe}, {Kisner}, {Kremin}, {Kronig}, {Landriau}, {Le
  Guillou}, {Martini}, {Moustakas}, {Palanque-Delabrouille}, {Peng},
  {Percival}, {Prada}, {Allende Prieto}, {de Rivera}, {Sanchez}, {Sanchez},
  {Sharples}, {Soares-Santos}, {Schlafly}, {Weaver}, {Zhou}, {Zhu}, {Zou}, \&
  {DESI Collaboration}}]{FocalPlane.Silber.2023}
{Silber}, J.~H., {Fagrelius}, P., {Fanning}, K., {et~al.} 2023, \aj, 165, 9,
  \dodoi{10.3847/1538-3881/ac9ab1}

\bibitem[{{Smith}(2009)}]{Smith2009}
{Smith}, R.~E. 2009, \mnras, 400, 851, \dodoi{10.1111/j.1365-2966.2009.15490.x}

\bibitem[{{Springob} {et~al.}(2014){Springob}, {Magoulas}, {Colless}, {Mould},
  {Erdo{\u{g}}du}, {Jones}, {Lucey}, {Campbell}, \& {Fluke}}]{Springob2014}
{Springob}, C.~M., {Magoulas}, C., {Colless}, M., {et~al.} 2014, \mnras, 445,
  2677, \dodoi{10.1093/mnras/stu1743}

\bibitem[{{Strauss} \& {Willick}(1995)}]{Strauss1995}
{Strauss}, M.~A., \& {Willick}, J.~A. 1995, \physrep, 261, 271,
  \dodoi{10.1016/0370-1573(95)00013-7}

\bibitem[{Torrado \& Lewis(2021)}]{Torrado:2020dgo}
Torrado, J., \& Lewis, A. 2021, JCAP, 05, 057,
  \dodoi{10.1088/1475-7516/2021/05/057}

\bibitem[{{Tully} \& {Fisher}(1977)}]{Tully1977}
{Tully}, R.~B., \& {Fisher}, J.~R. 1977, \aap, 54, 661

\bibitem[{{Turner} \& {Blake}(2023)}]{Turner2023Recons}
{Turner}, R.~J., \& {Blake}, C. 2023, \mnras, 526, 337,
  \dodoi{10.1093/mnras/stad2713}

\bibitem[{{Turner} {et~al.}(2021){Turner}, {Blake}, \& {Ruggeri}}]{Turner2021}
{Turner}, R.~J., {Blake}, C., \& {Ruggeri}, R. 2021, \mnras, 502, 2087,
  \dodoi{10.1093/mnras/stab212}

\bibitem[{{Turner} {et~al.}(2023){Turner}, {Blake}, \&
  {Ruggeri}}]{Turner2023correlation}
---. 2023, \mnras, 518, 2436, \dodoi{10.1093/mnras/stac3256}

\bibitem[{{Turner} {et~al.}(2025){Turner}, {Blake}, {Qin}, {Aguilar}, {Ahlen},
  {Amsellem}, {Bautista}, {BenZvi}, {Bianchi}, {Brooks}, {Carr}, {Chaussidon},
  {Claybaugh}, {Cuceu}, {de la Macorra}, {Doel}, {Douglass}, {Ferraro},
  {Font-Ribera}, {Forero-Romero}, {Gazta{\~n}aga}, {Gontcho}, {Gutierrez},
  {Guy}, {Herrera-Alcantar}, {Honscheid}, {Howlett}, {Huterer}, {Ishak},
  {Joyce}, {Kehoe}, {Kim}, {Kirkby}, {Kremin}, {Lahav}, {Lai}, {Lamman},
  {Landriau}, {Le Guillou}, {Leauthaud}, {Levi}, {Manera}, {Meisner}, {Miquel},
  {Moustakas}, {Mu{\~n}oz-Guti{\'e}rrez}, {Nadathur}, {Palanque-Delabrouille},
  {Percival}, {Poppett}, {Prada}, {P{\'e}rez-R{\`a}fols}, {Ross}, {Rossi},
  {Said}, {Sanchez}, {Schlegel}, {Schubnell}, {Silber}, {Sprayberry},
  {Tarl{\'e}}, {Weaver}, {Zarrouk}, \& {Zou}}]{DESIPV_Turner}
{Turner}, R.~J., {Blake}, C., {Qin}, F., {et~al.} 2025, arXiv e-prints,
  arXiv:2512.03230, \dodoi{10.48550/arXiv.2512.03230}

\bibitem[{{Vlah} {et~al.}(2012){Vlah}, {Seljak}, {McDonald}, {Okumura}, \&
  {Baldauf}}]{Vlah2012denPS}
{Vlah}, Z., {Seljak}, U., {McDonald}, P., {Okumura}, T., \& {Baldauf}, T. 2012,
  \jcap, 2012, 009, \dodoi{10.1088/1475-7516/2012/11/009}

\bibitem[{{Vlah} {et~al.}(2013){Vlah}, {Seljak}, {Okumura}, \&
  {Desjacques}}]{Vlah2013denPSHaloBias}
{Vlah}, Z., {Seljak}, U., {Okumura}, T., \& {Desjacques}, V. 2013, \jcap, 2013,
  053, \dodoi{10.1088/1475-7516/2013/10/053}

\bibitem[{{Wang} {et~al.}(2012){Wang}, {Mo}, {Yang}, \& {van den
  Bosch}}]{Wang2012Recons}
{Wang}, H., {Mo}, H.~J., {Yang}, X., \& {van den Bosch}, F.~C. 2012, \mnras,
  420, 1809, \dodoi{10.1111/j.1365-2966.2011.20174.x}

\bibitem[{{Wang} {et~al.}(2019){Wang}, {Percival}, {Avila}, {Crittenden}, \&
  {Bianchi}}]{Wang2018PSBOXCOX}
{Wang}, M.~S., {Percival}, W.~J., {Avila}, S., {Crittenden}, R., \& {Bianchi},
  D. 2019, \mnras, 786, \dodoi{10.1093/mnras/stz829}

\bibitem[{{Wang} {et~al.}(2021){Wang}, {Peery}, {Feldman}, \&
  {Watkins}}]{YuyuWang2021}
{Wang}, Y., {Peery}, S., {Feldman}, H.~A., \& {Watkins}, R. 2021, \apj, 918,
  49, \dodoi{10.3847/1538-4357/ac0e37}

\bibitem[{{Wang} {et~al.}(2018){Wang}, {Rooney}, {Feldman}, \&
  {Watkins}}]{YuyuWang2018}
{Wang}, Y., {Rooney}, C., {Feldman}, H.~A., \& {Watkins}, R. 2018, \mnras, 480,
  5332, \dodoi{10.1093/mnras/sty2224}

\bibitem[{{Wang} \& {Yang}(2024)}]{yuyuWang2024}
{Wang}, Y., \& {Yang}, X. 2024, \apj, 969, 76, \dodoi{10.3847/1538-4357/ad4d84}

\bibitem[{Wang {et~al.}(2023)Wang, Mirpoorian, Pogosian, Silvestri, \&
  Zhao}]{Wang:2023tjj}
Wang, Z., Mirpoorian, S.~H., Pogosian, L., Silvestri, A., \& Zhao, G.-B. 2023,
  JCAP, 08, 038, \dodoi{10.1088/1475-7516/2023/08/038}

\bibitem[{{Watkins} \& {Feldman}(2015)}]{Watkins2015}
{Watkins}, R., \& {Feldman}, H.~A. 2015, \mnras, 450, 1868,
  \dodoi{10.1093/mnras/stv651}

\bibitem[{{Watkins} {et~al.}(2009){Watkins}, {Feldman}, \&
  {Hudson}}]{Watkins2009}
{Watkins}, R., {Feldman}, H.~A., \& {Hudson}, M.~J. 2009, \mnras, 392, 743,
  \dodoi{10.1111/j.1365-2966.2008.14089.x}

\bibitem[{{Watkins} {et~al.}(2023){Watkins}, {Allen}, {Bradford}, {Ramon},
  {Walker}, {Feldman}, {Cionitti}, {Al-Shorman}, {Kourkchi}, \&
  {Tully}}]{Watkins2023}
{Watkins}, R., {Allen}, T., {Bradford}, C.~J., {et~al.} 2023, \mnras, 524,
  1885, \dodoi{10.1093/mnras/stad1984}

\bibitem[{{Whitford} {et~al.}(2023){Whitford}, {Howlett}, \&
  {Davis}}]{Whitford2023}
{Whitford}, A.~M., {Howlett}, C., \& {Davis}, T.~M. 2023, \mnras, 526, 3051,
  \dodoi{10.1093/mnras/stad2764}

\bibitem[{{Wu} {et~al.}(2021){Wu}, {Zhang}, {Pan}, {Miao}, {Luo}, {Wang},
  {Sabiu}, {Forero-Romero}, {Wang}, \& {Li}}]{WU2021AI}
{Wu}, Z., {Zhang}, Z., {Pan}, S., {et~al.} 2021, \apj, 913, 2,
  \dodoi{10.3847/1538-4357/abf3bb}

\bibitem[{{Wu} {et~al.}(2023){Wu}, {Xiao}, {Xiao}, {Wang}, {Kang}, {Wang},
  {Wang}, {Le Zhang}, \& {Li}}]{Wu2023AI}
{Wu}, Z., {Xiao}, L., {Xiao}, X., {et~al.} 2023, \mnras, 522, 4748,
  \dodoi{10.1093/mnras/stad1290}

\bibitem[{{Yamamoto}(2003)}]{Yamamoto2003}
{Yamamoto}, K. 2003, \apj, 595, 577, \dodoi{10.1086/377488}

\bibitem[{{Yamamoto} {et~al.}(2006){Yamamoto}, {Nakamichi}, {Kamino},
  {Bassett}, \& {Nishioka}}]{Yamamoto2006}
{Yamamoto}, K., {Nakamichi}, M., {Kamino}, A., {Bassett}, B.~A., \& {Nishioka},
  H. 2006, \pasj, 58, 93, \dodoi{10.1093/pasj/58.1.93}

\bibitem[{{Zaroubi} {et~al.}(1995){Zaroubi}, {Hoffman}, {Fisher}, \&
  {Lahav}}]{Zaroubi1995}
{Zaroubi}, S., {Hoffman}, Y., {Fisher}, K.~B., \& {Lahav}, O. 1995, \apj, 449,
  446, \dodoi{10.1086/176070}

\bibitem[{{Zhang} {et~al.}(2017){Zhang}, {Qin}, \& {Wang}}]{Zhang2017}
{Zhang}, Y., {Qin}, F., \& {Wang}, B. 2017, \prd, 96, 103523,
  \dodoi{10.1103/PhysRevD.96.103523}

\end{thebibliography}
 
\appendix  

\section{Peculiar velocity estimator}\label{sec:PVest}
By employing the low-redshift approximation of the log-distance ratio, and under the assumption that the true peculiar velocities of galaxies are significantly smaller than their observed redshifts, the line-of-sight peculiar velocities can then be inferred from $\eta$ through \citep{Johnson2014,Watkins2015, Adams2017, Howlett2017velocitypower, Qin2018,Carreres2023}
\be\label{watvp}
v=\frac{cz_{\mathrm{mod}} \ln10}{1+z_{\mathrm{mod}}}\eta,
\ee
where $z_{\mathrm{mod}}$ is given by \citep{Davis2014,Watkins2015}
\be  \label{zmod}
z_{\mathrm{mod}}=z\left[1+\frac{1}{2}(1-q_0)z-\frac{1}{6}(1-q_0-3q_0^2+1)z^2\right]\,,
\ee
and where $z$ is the observed redshift of a galaxy. The acceleration parameter is $q_0=0.5(\Omega_m-2\Omega_{\Lambda})$.  

Following the methodology outlined in \cite{Watkins2015} and \papII, we adopt the following for Eq.\ref{woptp} 
\be\label{stdvs}
\langle v^2({\bf r}) \rangle=\left( \frac{\ln(10)cz}{1+z}\sigma_{\eta}\right)^2+300^2\,\mathrm{km^{2}\,s^{-2}},
\ee
where $\sigma_{\eta}$ denotes the measurement error in the log-distance ratio $\eta$. $300^2$ km$^{2}$ s$^{-2}$ represents the intrinsic scatter in peculiar velocities caused by non-linear galactic motions. As discussed in Appendix A of \papII, variations in the assumed value of this intrinsic scatter have minimal impact on our results. The $\langle v^2({\bf r}) \rangle$ term for the DESI-PV random catalog is computed by \cite{DESIPV_Bautista}.

\section{Verifying $P^{\delta p}_{\ell}=-P^{\delta p *}_{\ell}$ for cross-power spectrum }\label{sec:symetric}

In Section \ref{sec:PSest}, we indicated that the estimator of the cross-power spectrum given in Eq.\ref{crsps} should satisfy the condition $P^{\delta p}_{\ell}=-P^{\delta p *}_{\ell}$. This property will be verified in the present section. To begin with, let $i$ denote the imaginary unit, such that $i^2=-1$. Upon disregarding the shot-noise term and referring to Eq.\ref{crsps}, we obtain
\be  
\begin{split}
P^{\delta p}_{\ell} \rightarrow ~ &  F^{p}F_{\ell}^{\delta *}-F^{\delta}F_{\ell}^{p *}\\
=&(\Res\{F^{p}\}+\Ims\{F^{p}\}i)\times(\Res\{F^{\delta}_{\ell}\}-\Ims\{F^{\delta}_{\ell}\}i)\\
&-(\Res\{F^{\delta}\}+\Ims\{F^{\delta}\}i)\times(\Res\{F^{p}_{\ell}\}-\Ims\{F^{p}_{\ell}\}i)\\
=& \Res\{F^{p}\}\Res\{F^{\delta}_{\ell}\}-\Res\{F^{p}\}\Ims\{F^{\delta}_{\ell}\}i\\
&+\Ims\{F^{p}\}\Res\{F^{\delta}_{\ell}\}i+\Ims\{F^{p}\}\Ims\{F^{\delta}_{\ell}\}\\
&-\Res\{F^{\delta}\}\Res\{F^{p}_{\ell}\}+\Res\{F^{\delta}\}\Ims\{F^{p}_{\ell}\}i\\
&-\Ims\{F^{\delta}\}\Res\{F^{p}_{\ell}\}i-\Ims\{F^{\delta}\}\Ims\{F^{p}_{\ell}\}\\
=&\Big(\Res\{F^{p}\}\Res\{F^{\delta}_{\ell}\}+\Ims\{F^{p}\}\Ims\{F^{\delta}_{\ell}\}\\
&-\Res\{F^{\delta}\}\Res\{F^{p}_{\ell}\}-\Ims\{F^{\delta}\}\Ims\{F^{p}_{\ell}\}\Big)\\
&+ \Big(\Ims\{F^{p}\}\Res\{F^{\delta}_{\ell}\}+\Res\{F^{\delta}\}\Ims\{F^{p}_{\ell}\}\\
&-\Res\{F^{p}\}\Ims\{F^{\delta}_{\ell}\}-\Ims\{F^{\delta}\}\Res\{F^{p}_{\ell}\}\Big)~i
\end{split}
\ee 
which subsequently implies
\be \label{symtrix1}
\begin{split}
P^{\delta p}_{\ell} &\propto \Ims\{F^{p}F_{\ell}^{\delta *}-F^{\delta}F_{\ell}^{p *}\}\\
&=\Ims\{F^{p}\}\Res\{F^{\delta}_{\ell}\}+\Res\{F^{\delta}\}\Ims\{F^{p}_{\ell}\}\\
&~~~~~-\Res\{F^{p}\}\Ims\{F^{\delta}_{\ell}\}-\Ims\{F^{\delta}\}\Res\{F^{p}_{\ell}\}
\end{split}
\ee 
On the other hand, we have
\be  
\begin{split}
P^{\delta p*}_{\ell} \rightarrow &~  F^{p*}F_{\ell}^{\delta}-F^{\delta*}F_{\ell}^{p}\\
=&(\Res\{F^{p}\}-\Ims\{F^{p}\}i)\times(\Res\{F^{\delta}_{\ell}\}+\Ims\{F^{\delta}_{\ell}\}i)\\
&-(\Res\{F^{\delta}\}-\Ims\{F^{\delta}\}i)\times(\Res\{F^{p}_{\ell}\}+\Ims\{F^{p}_{\ell}\}i)\\
=& \Res\{F^{p}\}\Res\{F^{\delta}_{\ell}\}+\Res\{F^{p}\}\Ims\{F^{\delta}_{\ell}\}i\\
&-\Ims\{F^{p}\}\Res\{F^{\delta}_{\ell}\}i+\Ims\{F^{p}\}\Ims\{F^{\delta}_{\ell}\}\\
&-\Res\{F^{\delta}\}\Res\{F^{p}_{\ell}\}-\Res\{F^{\delta}\}\Ims\{F^{p}_{\ell}\}i\\
&+\Ims\{F^{\delta}\}\Res\{F^{p}_{\ell}\}i-\Ims\{F^{\delta}\}\Ims\{F^{p}_{\ell}\}
\end{split}
\ee 
which subsequently implies
\be 
\begin{split}
P^{\delta p*}_{\ell} &\propto ~\Ims\{F^{p*}F_{\ell}^{\delta}-F^{\delta*}F_{\ell}^{p}\}\\
&=\Res\{F^{p}\}\Ims\{F^{\delta}_{\ell}\}-\Ims\{F^{p}\}\Res\{F^{\delta}_{\ell}\}\\
&-\Res\{F^{\delta}\}\Ims\{F^{p}_{\ell}\}+\Ims\{F^{\delta}\}\Res\{F^{p}_{\ell}\}\\
&=-\Big(\Ims\{F^{p}\}\Res\{F^{\delta}_{\ell}\}+\Res\{F^{\delta}\}\Ims\{F^{p}_{\ell}\}\\
&~~~~~~~~~~-\Res\{F^{p}\}\Ims\{F^{\delta}_{\ell}\}-\Ims\{F^{\delta}\}\Res\{F^{p}_{\ell}\}\Big)~.
\end{split}
\ee 
Comparing the above equation to Eq.\ref{symtrix1}, one can logically infer that $P^{\delta p}_{\ell}=-P^{\delta p *}_{\ell}$. 

\section{The linear models of power spectrum and the models for the correlation functions. }\label{sec:app1}

\cite{Kaiser1987} developed a linear model for the density power spectrum, commonly referred to as the Kaiser Formula, given by
\be \label{kasierden}
P^{\delta}(k,\mu) = (b_1+f\mu^2)^2P_L(k) 
\ee
which has been extensively utilized over the past decades. Building upon this foundation, \cite{Koda2014} extended the analysis to the power spectrum of peculiar velocities as well as cross-power, they modelling the velocity power spectrum and density-velocity cross power spectrum using
{\setstretch{0.3}
\be \label{kasiermom}
P^{p}(k,\mu) = \frac{(aHf\mu)^2}{k^2} P_L(k)
\ee
\be \label{kasiercrs}
P^{\delta p}(k,\mu) = (b_1+f\mu^2)\frac{iaHf\mu}{k} P_L(k)
\ee}
Furthermore, they incorporated damping terms into the above linear models in order to more accurately account for non-linear motions of galaxies, given by
{\setstretch{0.4}
\be 
P^{\delta}(k,\mu) = (b_1+f\mu^2)^2P_L(k) D^2_g 
\ee
\be 
P^{p}(k,\mu) = \frac{(aHf\mu)^2}{k^2} P_L(k) D^2_p 
\ee
\be 
P^{\delta p}(k,\mu) = (b_1+f\mu^2)\frac{iaHf\mu}{k} P_L(k) D_g  D_p 
\ee}
where the damping terms are defined by 
\be 
D_g=\frac{1}{\sqrt{1+\frac{1}{2}(k\mu\sigma_g)^2}}~,~~D_p=\frac{\sin(k\sigma_v)}{k\sigma_v}
\ee 
to account for non-linear motions of galaxies. The $\sigma_g$ and $\sigma_v$ are the non-linear velocity dispersion parameters for the density and velocity fields, respectively. 

\begin{figure} 
\centering
  \includegraphics[width=\columnwidth]{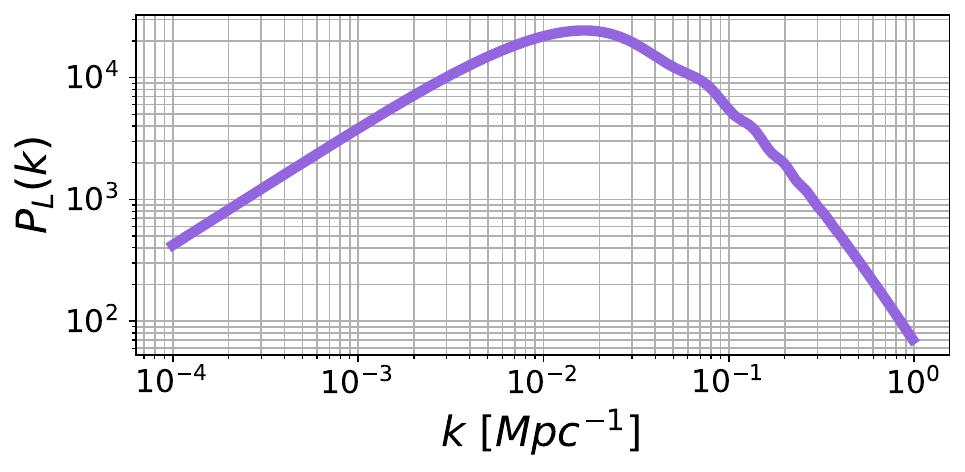}
 \caption{ The linear matter power spectrum $P_L(k)$.}
 \label{pltPL}
\end{figure}

\begin{figure} 
\centering
 \includegraphics[width=\columnwidth]{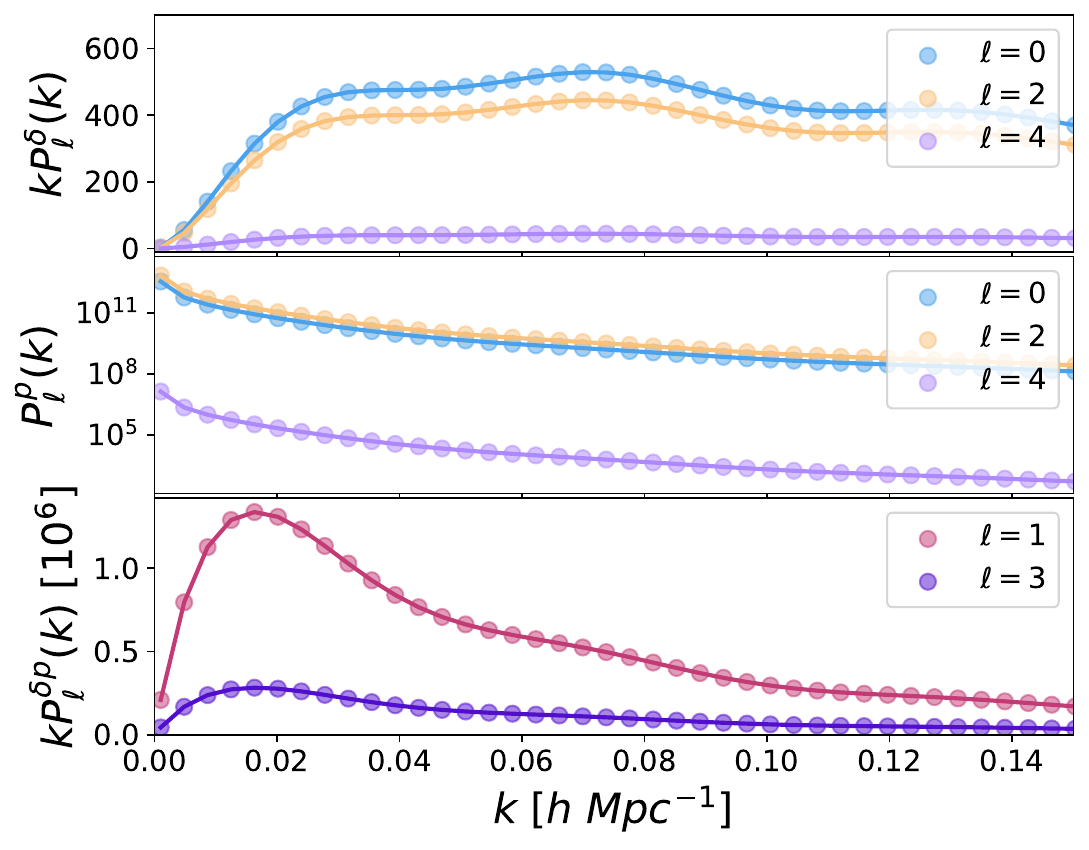}
 \caption{ Comparing the Kaiser formulas (filled circles), Eq.\ref{kasierden},\ref{kasiermom} and \ref{kasiercrs} to the linearized perturbation models (curves) Eq. \ref{psmodd}, \ref{psmodp} and \ref{psmoddp}.  }
 \label{pltKaiser}
\end{figure}

We initialize all loop terms—$I_{mn}$, $K_{mn}$, $K^s_{mn}$, $\sigma_3$ and $\sigma_4$—to zero, and similarly set the higher-order biasing and velocity dispersion parameters to zeros. In essence, we retain only $f\sigma_8$ and $b_1\sigma_8$. Consequently, Eq. \ref{psmodd}, \ref{psmodp} and \ref{psmoddp} reduce to the classic Kaiser formulae expressed in Eq.\ref{kasierden},\ref{kasiermom} and \ref{kasiercrs}, respectively, as illustrated in Fig.\ref{pltKaiser}. For simplicity, we manually set $f\sigma_8=0.437$, $b_1\sigma_8=0.564$ and effective redshift $z=0$ to generate this plot. 
In the middel panel of Fig. \ref{pltKaiser}, it becomes evident that the momentum power spectrum quadrupole $P^p_2$ (represented by yellow dots and the yellow curve) surpasses the monopole $P^p_0$ (depicted by blue dots and its curve). This arises due to the absence of window function convolution effects in our current consideration. 
Illustrated in Fig.\ref{pltPL} is the linear matter power spectrum utilized in our analysis.

\begin{figure} 
 \centering
 \includegraphics[width=\columnwidth]{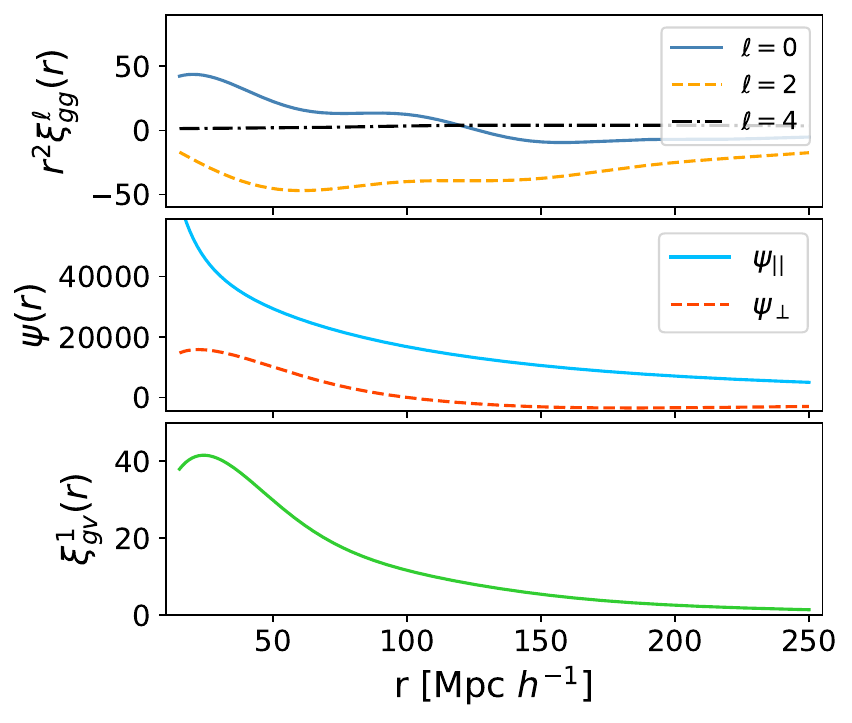}
 \caption{ The two-point correlation function models are derived from the power spectrum models introduced in Section \ref{sec:psmodel}, using the parameter values listed in Table \ref{tabs2last2}. The top panel illustrates the galaxy auto-two-point correlation function $\xi^{\ell}_{gg}$, the middle panel presents the monopole of the galaxy auto-velocity correlation function $\Psi$, and the bottom panel displays the dipole component of the galaxy-velocity cross-correlation function $\xi^{1}_{gv}$. }
 \label{pltCFgal}
\end{figure}

The mathematical framework for converting power spectrum to correlation functions primarily draws upon \cite{DESIPV_Turner}.   
In Fig.\ref{pltCFgal}, we present  the model correlation functions converted from the power spectrum models presented in Section \ref{sec:psmodel} with parameters from Table \ref{tabs2last2}.

\section{The multipoles of the field function}\label{sec:Tl}

The Eq.\ref{sdgdef123} can be further expressed in a more refined form as
 \be   \label{jkh864357}
T_{\ell}({\bf k})=\int \bigg(\frac{k_xr_x+k_yr_y+k_zr_z}{kr}\bigg)^{\ell} F({\bf r}) e^{i {\bf k} \cdot {\bf r} } d^3r
\ee
Thus, by directly expanding the term $\bigg(\frac{k_xr_x+k_yr_y+k_zr_z}{kr}\bigg)^{\ell}$ for $\ell=0,~2$ and 4, 
the even-multipole components of the Fourier-transformed field function are explicitly presented in  
\be
\begin{split}
T_0({\bf k})&=\int   F({\bf r}) e^{i {\bf k} \cdot {\bf r} }    d^3r= V\times F({\bf k})
\end{split}
\ee 
\be 
\begin{split}
T_2({\bf k})=&\frac{1}{k^2}\Big(k^2_xU_{xx}+k^2_yU_{yy}+k^2_zU_{zz} \\
&+2(k_xk_yU_{xy}+k_xk_zU_{xz}+k_yk_zU_{yz})\Big)
\end{split}
\ee 
and 
\be
\begin{split}
T_4({\bf k})=&\frac{1}{k^4}\Big( k^4_xB_{xxx}+k^4_yB_{yyy}+k^4_zB_{zzz}+4(k^3_xk_yB_{xxy}+k^3_xk_zB_{xxz} \\
&+k^3_yk_xB_{yyx} +k^3_yk_zB_{yyz}+k^3_zk_xB_{zzx}+k^3_zk_yB_{zzy}) \\
&+ 6(k^2_xk^2_yB_{xyy}+k^2_xk^2_zB_{xzz}+k^2_yk^2_zB_{yzz})\\
&+12k_xk_yk_z(k_xB_{xyz}+k_yB_{yxz}+k_zB_{zxy})\Big)
\end{split}
\ee
respectively, and where
\be 
U_{ij}=\int \frac{r_ir_j}{r^2} F({\bf r}) e^{i {\bf k} \cdot {\bf r} }  d^3r~,~~ 
B_{ijn}=\int \frac{r^2_ir_jr_n}{r^4} F({\bf r}) e^{i {\bf k} \cdot {\bf r} }  d^3r  
\ee
where $i,j,n=x,y,z$.

Furthermore, guided by the theoretical framework proposed in \papIII, based on Eq.\ref{jkh864357}, the odd-multipole components of the Fourier-transformed field function are formulated as described in
\be  
T_{1}({\bf k})=\frac{1}{k}\left( k_xE_x+k_yE_y+k_zE_z \right)
\ee 
\be  
\begin{split}
T_{3}({\bf k})=\frac{1}{k^3} (& k^3_xM_{xxx}+k^3_yM_{yyy}+k^3_zM_{zzz}\\
+& 3k^2_xk_yM_{xxy}+3k^2_xk_zM_{xxz}
+ 3k^2_yk_xM_{yyx}+3k^2_yk_zM_{yyz}\\
+& 3k^2_zk_xM_{zzx}+3k^2_zk_yM_{zzy}+6k_xk_yk_zM_{xyz})
\end{split}
\ee 
and where
\be 
E_i=\int \frac{r_i}{r} F({\bf r}) e^{i {\bf k} \cdot {\bf r} }  d^3r~,~~ 
M_{ijn}=\int \frac{r_ir_jr_n}{r^3} F({\bf r}) e^{i {\bf k} \cdot {\bf r} }  d^3r 
\ee
 where $i,j,n=x,y,z$.

  \section{More tests using mocks}\label{sec:moretest}

\begin{figure} 
\includegraphics[width=\columnwidth]{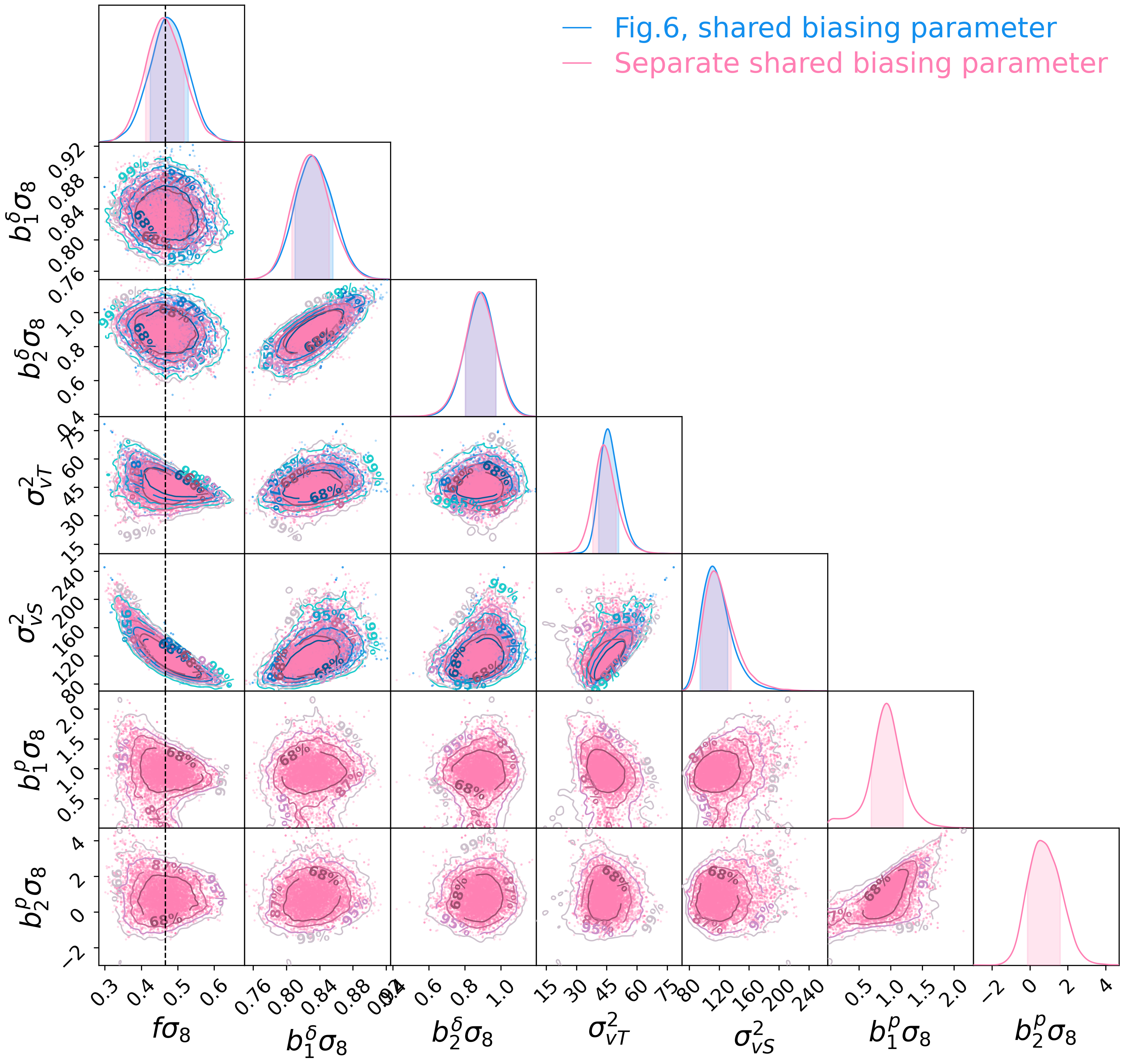}
 \caption{ {  
 The pink histograms and 2D contour plots show the fitting results obtained with  $[~f\sigma_8,~b^{\delta}_{1}\sigma_8,~b^{\delta}_{2}\sigma_8,~b^p_{1}\sigma_8,~b^p_{2}\sigma_8,~\sigma_{vT},~\sigma_{vS}~]$, while the blue counterparts are identical to those in Fig.\ref{pltfsig8mock}.}}
 \label{pltfsig7ck4334}
 \end{figure}

{ 
In Section \ref{sec:5.2}, we demonstrate the adaptability of the power spectrum models by performing fits on $[P^{\delta}_0, P^{\delta}_2, P^{p}_0, P^{\delta p}_1]$ using five free parameters $[~f\sigma_8,~b_{1}\sigma_8,~b_{2}\sigma_8,~\sigma_{vT},~\sigma_{vS}~]$, which suffices to maintain robustness and precision in the fitting process. Alternatively, we can separate the biasing parameters between the density and momentum fields—i.e., adopt seven free parameters $[~f\sigma_8,~b^{\delta}_{1}\sigma_8,~b^{\delta}_{2}\sigma_8,~b^p_{1}\sigma_8,~b^p_{2}\sigma_8,~\sigma_{vT},~\sigma_{vS}~]$—without significantly altering the resulting $f\sigma_8$, see Fig.\ref{pltfsig7ck4334}. Consequently, to maintain model simplicity and reduce parameters, we adopt shared biasing parameters for both fields. 
}

\section{Fitting $f\sigma_8$ of mocks  from $P^{\delta}_{0,2}+P^{p}_{0,2}+P^{\delta p}_1$ }\label{gtrnew}

In Section \ref{sec:5.2}, we demonstrate the adaptability of the power spectrum models by performing fits on $[P^{\delta}_0, P^{\delta}_2, P^{p}_0, P^{\delta p}_1]$. Given that the momentum power spectrum quadrupole $P^p_{2}$ derived from the FP mock average displays a sufficiently smooth behavior to support a robust fitting procedure, we extend our analysis to include $P^p_{2}$ in the parameter estimation, thereby enabling a more comprehensive determination of $f\sigma_8$ based on $[P^{\delta}_0, P^{\delta}_2, P^{p}_0, P^{p}_2, P^{\delta p}_1]$.   

As we incorporate additional data into the fitting process, it becomes essential to enhance the flexibility of the power spectrum models accordingly. To achieve this, we introduce two additional free parameters. Specifically, we distinguish the linear biasing parameters between the density and momentum fields, and similarly differentiate the velocity dispersion parameters across these two fields. Consequently, the set of free parameters is defined as $\theta = [f\sigma_{8}, b^{\delta}_{1}\sigma_{8}, b_{2}\sigma_{8}, b^p_{1}\sigma_{8}, \sigma^2_{\delta,vT}, \sigma^2_{vS}, \sigma^2_{p,vT}]$, resulting in a total of seven independent parameters. All other parameters remain shared between the density and momentum fields. Furthermore, to ensure model stability and prevent breakdown at nonlinear  scales, we adopt a reduced $k_{max}$ value of 0.25 $h$ Mpc$^{-1}$, in contrast to that used in Section \ref{sec:5.2}.

\begin{figure}  
\centering
\includegraphics[width=\columnwidth]{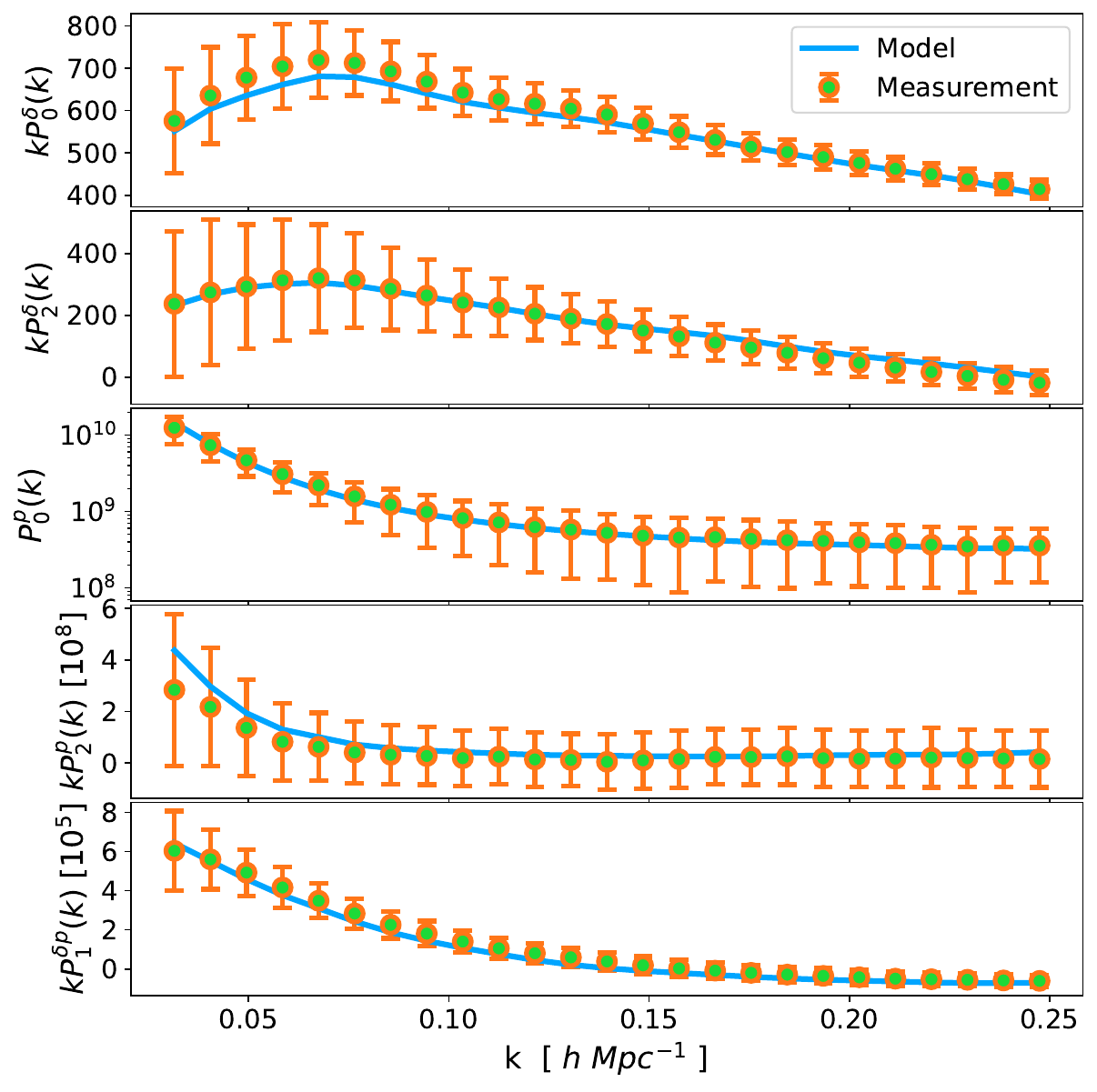}
 \caption{ Same as Fig.\ref{pltfsig8mock}, but for  including the  momentum power spectrum quadrupole $P^p_{2}$ in the fitting.   }
 \label{pltfsig8mock02}
\end{figure}

\begin{figure} 
\centering
\includegraphics[width=\columnwidth]{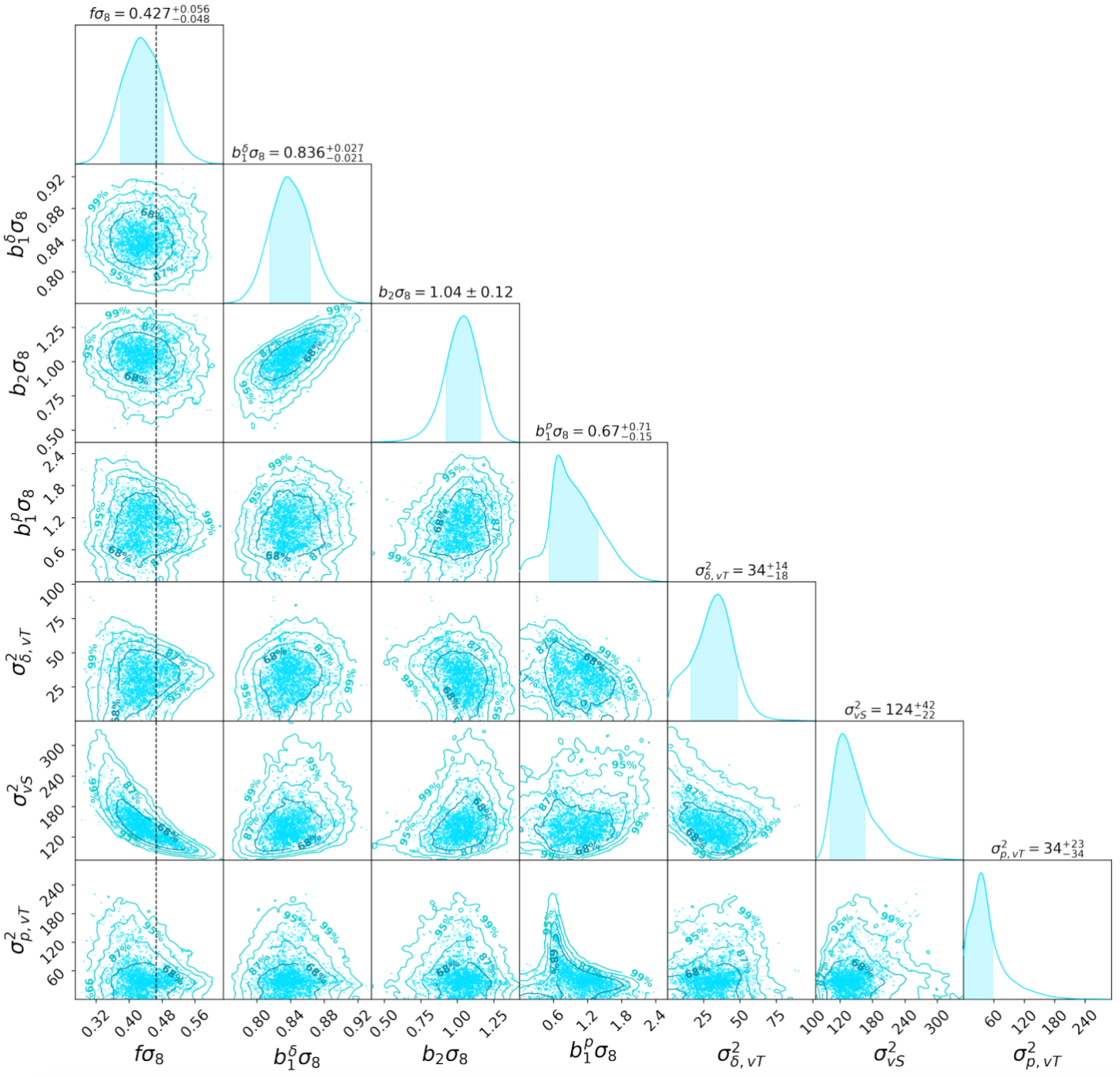}
 \caption{ Same as Fig.\ref{pltfsig8mock}, but for  including the  momentum power spectrum quadrupole $P^p_{2}$ in the fitting.  The corresponding marginalized parameter values are summarized in Table \ref{tabs2last2apped}. }
 \label{pltfsig8mock02ss}
\end{figure}

As illustrated in Fig. \ref{pltfsig8mock02}, the filled circles represent the average of the power spectrum measured from 675 mocks, while the solid curves depict the theoretical models fit to the measurements. The MCMC resulting parameter estimates are show in Fig.\ref{pltfsig8mock02ss} and summarized in Table \ref{tabs2last2apped}. The  fit value of $f\sigma_{8}=0.427^{+0.056}_{-0.048}$, which well aligns with the fiducial simulation value of $f\sigma_{8,fid}=0.466$ at effective redshift $z_{eff}=0.2$.  This agreement demonstrates that our fitting methodology effectively recovers the true growth rate embedded in the mocks, and that the power spectrum models perform robustly up to the non-linear scale of $k_{max}$=0.25 $h$ Mpc$^{-1}$ under current conditions.

 \begin{table}   \centering
\caption{ The MCMC-fit cosmological parameter estimates.
  }
\begin{tabular}{cc|cc    }
\hline \noalign{\vskip 2pt}
  $f\sigma_8$  &$0.427^{+0.056}_{-0.048}$ & $\sigma^2_{\delta,vT}$ & $34^{+14}_{-18}$\\ \noalign{\vskip 2pt}
\hline \noalign{\vskip 2pt}
 $b^{\delta}_1\sigma_8$ & $0.836^{+0.027}_{-0.021}$ & $\sigma^2_{vS}$ & $124^{+42}_{-22}$ \\ \noalign{\vskip 2pt}
 \hline \noalign{\vskip 2pt}
 $b_2\sigma_8$ & $1.040^{+0.120}_{-0.120}$& $\sigma^2_{p,vT}$ & $34^{+23}_{-34}$ \\  \noalign{\vskip 2pt}
 \hline \noalign{\vskip 2pt}
 $b^p_1\sigma_8$ & $0.670^{+0.710}_{-0.150}$ &
 $\chi^2/$d.o.f & $10.445/(125-7)$ \\ \noalign{\vskip 2pt}
\hline
\end{tabular}
\tablefoot{Corresponding to Fig.\ref{pltfsig8mock02ss}, the MCMC-fit cosmological parameter estimates are presented here.}
 \label{tabs2last2apped}
\end{table}

\section{The Galilean transformation of power spectrum}\label{sec:gali}

As elaborated in \papIII, the momentum power spectrum $P^{p}({\bf k})$ and cross-power spectrum $P^{\delta p}({\bf k})$ are not invariant under the Galilean transformation; that is, their values depend on the bulk motion of the sample relative to the observational frame. Specifically, for the momentum power spectrum, if a constant bulk velocity shift $\epsilon$ is introduced to the line-of-sight peculiar velocities, i.e., by substituting $v({\bf r}) \rightarrow v({\bf r})+\epsilon$,  the momentum correlation yields
\be  
\begin{split}
 & \langle    p_G({\bf r})   p_G({\bf r}')     \rangle \\
 \equiv&\langle (1+\delta({\bf r}))(v({\bf r})+\epsilon)(1+\delta({\bf r}'))(v({\bf r}')+\epsilon)\rangle \\
=&\langle   
(1+\delta({\bf r}))v({\bf r}) (1+\delta({\bf r}'))v({\bf r}') \rangle\\
&
+\langle (1+\delta({\bf r}))v({\bf r})\rangle\epsilon+\langle(1+\delta({\bf r}))v({\bf r})\delta({\bf r}') \rangle\epsilon\\
&+\langle (1+\delta({\bf r}'))v({\bf r}')\rangle\epsilon+\langle(1+\delta({\bf r}'))v({\bf r}')\delta({\bf r}) \rangle \epsilon\\
&
 + \epsilon^2 +\langle \delta({\bf r}) \rangle \epsilon^2+\langle\delta({\bf r}') \rangle \epsilon^2+\langle\delta({\bf r})\delta({\bf r}')\rangle \epsilon^2  \\
 =&\langle   
p({\bf r}) p({\bf r}') \rangle +\langle p({\bf r})\rangle\epsilon+\langle \delta({\bf r}') p({\bf r}) \rangle\epsilon +\langle p({\bf r}')\rangle\epsilon+\langle \delta({\bf r})p({\bf r}') \rangle \epsilon\\
&
  +\langle \delta({\bf r}) \rangle \epsilon^2+\langle\delta({\bf r}') \rangle \epsilon^2+\langle\delta({\bf r})\delta({\bf r}')\rangle \epsilon^2  +  \epsilon^2 
\end{split}
\ee 
In this expression, the average of the density contrast field, i.e.  $\langle \delta({\bf r}) \rangle$ and $\langle\delta({\bf r}') \rangle$ are zeros. The average of the momentum field, i.e. $\langle p({\bf r})\rangle$ and $\langle p({\bf r}')\rangle$ are zeros too. Consequently, the above equation simplifies to
\be   
\begin{split}
\langle    p_G({\bf r})   p_G({\bf r}')     \rangle &= \langle p({\bf r}) p({\bf r}') \rangle  +\langle  \delta({\bf r}') p({\bf r}) \rangle\epsilon +\langle \delta({\bf r})p({\bf r}') \rangle \epsilon \\
& 
  +\langle\delta({\bf r})\delta({\bf r}')\rangle \epsilon^2  +  \epsilon^2
\end{split}
\ee 
Applying the Fourier transform to the aforementioned equation yields
\be  
\begin{split}
&\frac{1}{(2\pi)^6}\int  \langle    p_G({\bf k})   p^*_G({\bf k}')     \rangle e^{-i {\bf k} \cdot {\bf r} } e^{i {\bf k}' \cdot {\bf r}' }    d^3kd^3k'\\
=&\frac{1}{(2\pi)^6}\int\Big[ \langle   p({\bf k}) p^*({\bf k}') \rangle +\langle \delta^*({\bf k}')p({\bf k})   \rangle\epsilon+\langle \delta({\bf k})p^*({\bf k}') \rangle \epsilon \\
&+\langle\delta({\bf k})\delta^*({\bf k}')\rangle \epsilon^2\Big]e^{-i {\bf k} \cdot {\bf r} } e^{i {\bf k}' \cdot {\bf r}' }    d^3kd^3k'  +    \epsilon^2  \delta^D({\bf k}) \delta^D({\bf k}')
\end{split}
\ee 
in the final term, the $\delta^D({\bf k})$ and $\delta^D({\bf k}')$ cannot equal unity simultaneously, as we are exclusively considering the case $k\ne k'$ in the context of two-point statistic. Consequently,  $\delta^D({\bf k}) \delta^D({\bf k}')=0$, and the above equation indicates 
 \be  
 \begin{split}
\langle    p_G({\bf k})   p_G({\bf k}')     \rangle &= \langle   p({\bf k}) p^*({\bf k}') \rangle +\langle \delta^*({\bf k}') p({\bf k}) \rangle\epsilon  \\
&+\langle \delta({\bf k})p^*({\bf k}') \rangle \epsilon+\langle\delta({\bf k})\delta^*({\bf k}')\rangle \epsilon^2\\
\end{split}
 \ee
Substituting the definitions of the power spectrum provided in Eq.\ref{defmomps456}, \ref{defmomps} and \ref{defmomps789} into the above equation leads to
\be  
\begin{split}
 P_G^{p}({\bf k})= P^{p}({\bf k})
 + P^{\delta p *}({\bf k}) \epsilon  + P^{\delta p}({\bf k}) \epsilon+\epsilon^2  P^{\delta}({\bf k})  
\end{split}
\ee 
Given that $P^{\delta p}=-P^{\delta p *}$ (see Appendix \ref{sec:symetric}), the above  equation further reduces to 
\be  \label{Galipp}
\begin{split}
 P_G^{p}({\bf k})= P^{p}({\bf k})+\epsilon^2   P^{\delta}({\bf k}) 
\end{split}
\ee 
which represents the Galilean-transformed form of the momentum power spectrum. Analogously, the corresponding Galilean-transformed cross-power spectrum is expressed as
\be\label{Galipdp}
\begin{split}
P_G^{\delta p}({\bf k})= P^{\delta p}({\bf k})+\epsilon  P^{\delta }({\bf k})
\end{split}
\ee
The bulk velocity $\epsilon$ comprises two possible  components: one stems from coherent bulk motion of galaxies driven by local gravitational fluctuations, and the other originates from potential systematic errors induced by observational or measurement inaccuracies. 

\begin{figure} 
 \centering
 \includegraphics[width=85mm]{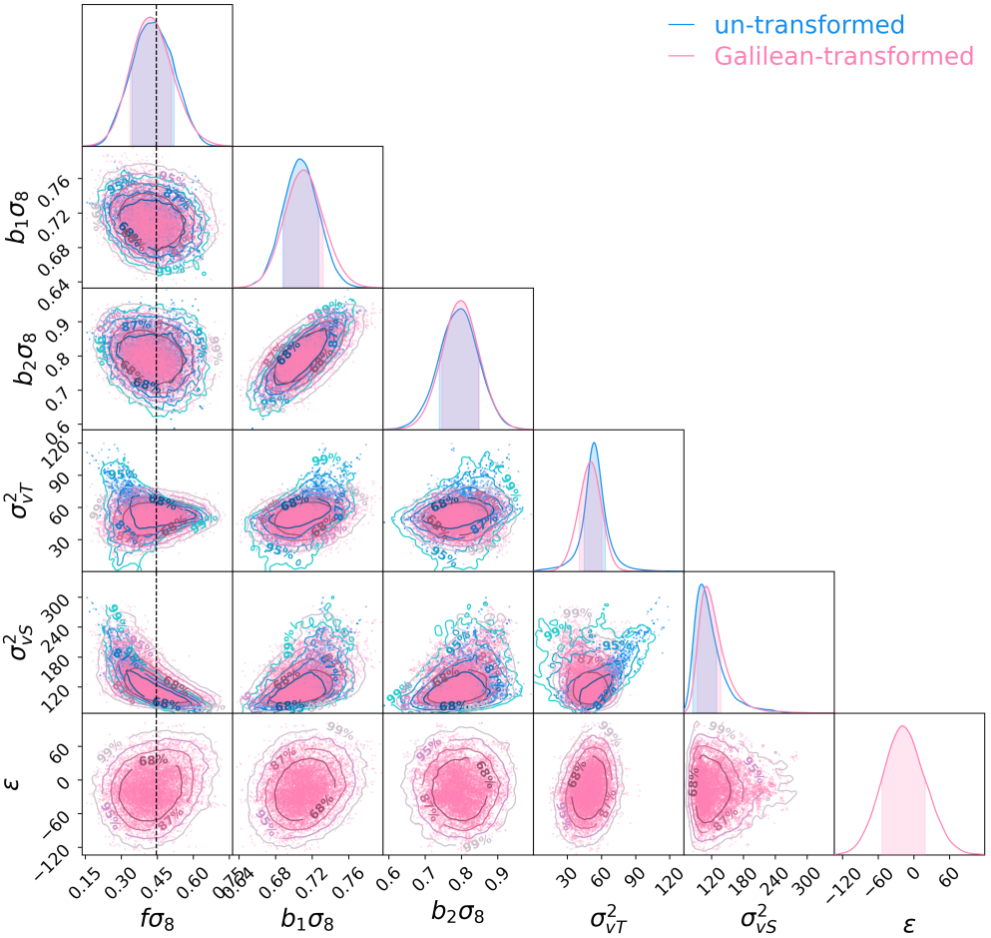}
 \caption{ The pink histograms and 2D contour plots depict the fitting results obtained using the Galilean-transformed power spectrum models, while the blue counterparts illustrate the results derived without the transformation (the same as Fig.\ref{pltfsig8survey}). }
 \label{pltGalii}
\end{figure}

We investigate whether the fitting outcomes exhibit sensitivity to the parameter $\epsilon$. By substituting the power spectrum models described in Section \ref{sec:psmodel} into Eqs.~\ref{Galipp} and \ref{Galipdp}, we obtain the Galilean-transformed versions of these models. These transformed models are then fitted to the measured power spectrum of BGS and DESI-PV data. In this analysis, $\epsilon$ is treated as a free parameter with a flat prior over the broad range of $\epsilon \in [-1000, 1000]$ km s$^{-1}$. The resulting fits are visualized through pink-colored histograms and 2D contour plots in Fig.\ref{pltGalii}. For comparative purposes, the blue-colored histograms and 2D contours represent the fitting results obtained without applying the Galilean transformation (identical to those shown in Fig.\ref{pltfsig8survey}). The results from both approaches are in close agreement, and the fit value for $\epsilon$ is determined to be $\epsilon = -15.94^{+35.95}_{-35.95}$ km s$^{-1}$, which is statistically consistent with zero. Consequently, the influence of the Galilean transformation is found to be negligible within the context of our analysis.

\onecolumn

\section{Deriving the estimator of power spectrum and the window funtion convolution matrix}\label{psest879}

Corresponding to Section \ref{sec:PSest} and Section \ref{sec:windoconv}, in this section, we present a detailed derivation of the estimator for the power spectrum and its associated window function convolution matrix by taking the cross-power spectrum as a representative example. The mathematical framework primarily draws upon Section 3.1 of \papI, Section 3.2 of \papIII, and Section 2.2 of \cite{Blake2018}.

\subsection{Power spectrum estimator}\label{sec:CRSest}

We begin by introducing the Legendre transformation, denoted as \citep{Blake2018}
\be  
F_{\ell}({\bf r})= (2\ell+1)\int F({\bf r})L_{\ell}(\hat{{\bf k}}\cdot\hat{{\bf r}})\frac{d \Omega_k}{4\pi}
\ee 
Following the methodology outlined in \cite{Yamamoto2006} and Section 3.2 of \papIII ~and employing the Fourier transform definition given in Eq.\ref{A0}, we formulate our estimator for the cross-power spectrum under the "local plane-parallel approximation", i.e. $L_{\ell}(\hat{{\bf k}}\cdot\hat{{\bf r}})=L_{\ell}(\hat{{\bf k}}\cdot\hat{{\bf r}'})$, as
\be  \label{jhgd78}
\begin{split}
    |F^{\delta}(k)F^{p *}_{\ell}(k)|=&\frac{1}{V}\int F^{\delta}({\bf r})e^{i {\bf k} \cdot{\bf r}} d^3r\frac{1}{V'} \int  \Big[(2\ell+1)\int F^{p*}({\bf r}')L^*_{\ell}(\hat{{\bf k}}\cdot\hat{{\bf r}'})\frac{d \Omega_k}{4\pi}\Big]e^{-i {\bf k} \cdot{\bf r}'}d^3r'\\
    =&\frac{2\ell+1}{VV'}\int\frac{d \Omega_k}{4\pi}\int d^3r \int  d^3r' F^{\delta}({\bf r})F^{p*}({\bf r}') L^*_{\ell}(\hat{{\bf k}}\cdot\hat{{\bf r}'})  e^{i {\bf k} \cdot ({\bf r}-{\bf r}') }.
\end{split}
\ee
Subsequently, the cross-correlation between the density and momentum fields is expressed as
\be  \label{872ghjf}
\begin{split}
    \langle|F^{\delta}(k)F^{p *}_{\ell}(k)|\rangle 
    =&\frac{2\ell+1}{VV'}\int\frac{d \Omega_k}{4\pi}\int d^3r \int  d^3r'\langle F^{\delta}({\bf r})F^{p*}({\bf r}') \rangle L^*_{\ell}(\hat{{\bf k}}\cdot\hat{{\bf r}'})  e^{i {\bf k} \cdot ({\bf r}-{\bf r}') } 
\end{split}
\ee 
Building upon the methodology outlined in Section 3.2 of \papIII, and employing Eq.\ref{fieldd} and \ref{fieldp}, we derive the cross-correlation between the density and momentum fields as 
\be   \label{872ghjfww}
\begin{split}
\langle F^{\delta}({\bf r}) F^{p*}({\bf r}' ) \rangle=&\bigg\langle \frac{w_{\delta}({\bf r})\left[ n_{\delta}({\bf r})-\alpha n_s({\bf r}) \right]}{A_{\delta}}~\frac{w_p({\bf r}')n_p({\bf r}')v({\bf r}')}{A_p} \bigg\rangle = \frac{1}{ {A_{\delta}A_{p}}}  w_{\delta}({\bf r})w_{p}({\bf r}')\Big(\langle   n_{\delta}({\bf r})
n_p({\bf r}' ) v({\bf r}' ) \rangle 
-\alpha \langle n_s({\bf r})  n_p({\bf r}' ) v({\bf r}' )  \rangle \Big)
\end{split}
\ee  
Drawing from the approach described in \citealt{Park2006} and   \papI, the first term within the brackets of Eq.\ref{872ghjfww} can be rewritten as $\langle   n_{\delta}({\bf r})
n_p({\bf r}' ) v({\bf r}' ) \rangle= \bar{n}_{\delta}({\bf r})\bar{n}_p({\bf r}')\xi_{\delta p}(|{\bf r}-{\bf r}'|)  +  \min\{\bar{n}_{\delta}({\bf r}),\bar{n}_p({\bf r})\} \langle v({\bf r}) \rangle
\delta^D(|{\bf r}-{\bf r}'|) $. Given that the cross-correlation between galaxies and random points is zero, the second term within the brackets of Eq.\ref{872ghjfww} becomes zero, i.e. $\alpha \langle n_s({\bf r})  n_p({\bf r}' ) v({\bf r}' )  \rangle=0$.  Consequently, we obtain 
\be  
\langle F^{\delta}({\bf r}) F^{p*}({\bf r}' ) \rangle= \frac{1}{ {A_{\delta}A_{p}}}  w_{\delta}({\bf r})w_{p}({\bf r}')\big[  \bar{n}_{\delta}({\bf r})\bar{n}_p({\bf r}')\xi_{\delta p}(|{\bf r}-{\bf r}'|)  +  \min\{\bar{n}_{\delta}({\bf r}),\bar{n}_p({\bf r})\} \langle v({\bf r}') \rangle
\delta^D(|{\bf r}-{\bf r}'|)\big]
\ee 
By substituting the above equation into Eq.\ref{872ghjf}  in place of $\langle F^{\delta}({\bf r})F^{p*}({\bf r}')\rangle $, we arrive at
\be  
\begin{split}
    \langle|F^{\delta}(k)F^{p *}_{\ell}(k)|\rangle 
    =&\frac{1}{ {A_{\delta}A_{p}}} \frac{2\ell+1}{V}\int\frac{d \Omega_k}{4\pi}\int d^3r \Big[\frac{1}{V'}\int  d^3r'  w_{\delta}({\bf r})w_{p}({\bf r}')\bar{n}_{\delta}({\bf r})\bar{n}_p({\bf r}')\xi_{\delta p}(|{\bf r}-{\bf r}'|) L^*_{\ell}(\hat{{\bf k}}\cdot\hat{{\bf r}'})  e^{i {\bf k} \cdot ({\bf r}-{\bf r}') }\\
& + \frac{1}{V'} \int  d^3r'w_{\delta}({\bf r})w_{p}({\bf r})\min\{\bar{n}_{\delta}({\bf r}),\bar{n}_p({\bf r})\} \langle v({\bf r}) \rangle
\delta^D({\bf r}-{\bf r}')  L^*_{\ell}(\hat{{\bf k}}\cdot\hat{{\bf r}'})  e^{i {\bf k} \cdot ({\bf r}-{\bf r}') }\Big]
\end{split}
\ee  
Upon integrating the second term over the Dirac delta function about ${\bf r}'$ using the relation  
$
f({\bf r})=\frac{1}{V}\int f({\bf r}) \delta^D({\bf r}-{\bf r}') d^3r'
$, 
we obtain
\be  \label{FFest}
\begin{split}
    \langle|F^{\delta}(k)F^{p *}_{\ell}(k)|\rangle =&\frac{1}{ {A_{\delta}A_{p}}} \frac{2\ell+1}{V}\int\frac{d \Omega_k}{4\pi}\int d^3r \Big[\frac{1}{V'}\int  d^3r'  w_{\delta}({\bf r})w_{p}({\bf r}')\bar{n}_{\delta}({\bf r})\bar{n}_p({\bf r}')\xi_{\delta p}(|{\bf r}-{\bf r}'|) L^*_{\ell}(\hat{{\bf k}}\cdot\hat{{\bf r}'})  e^{i {\bf k} \cdot ({\bf r}-{\bf r}') }\\
& +   w_{\delta}({\bf r})w_{p}({\bf r})\min\{\bar{n}_{\delta}({\bf r}),\bar{n}_p({\bf r})\} \langle v({\bf r}) \rangle
 L^*_{\ell}(\hat{{\bf k}}\cdot\hat{{\bf r}})   \Big]\\
=&\frac{1}{ {A_{\delta}A_{p}}} \frac{2\ell+1}{V}\int\frac{d \Omega_k}{4\pi}\bigg[\int d^3r \frac{1}{V'}\int  d^3r'  w_{\delta}({\bf r})w_{p}({\bf r}')\bar{n}_{\delta}({\bf r})\bar{n}_p({\bf r}')\xi_{\delta p}(|{\bf r}-{\bf r}'|) L^*_{\ell}(\hat{{\bf k}}\cdot\hat{{\bf r}'})  e^{i {\bf k} \cdot ({\bf r}-{\bf r}') }\\
& +   \int w_{\delta}({\bf r})w_{p}({\bf r})\min\{\bar{n}_{\delta}({\bf r}),\bar{n}_p({\bf r})\} \langle v({\bf r}) \rangle
 L^*_{\ell}(\hat{{\bf k}}\cdot\hat{{\bf r}})  d^3r \bigg]
\end{split}
\ee 
We define the function
\be 
N^{\delta p}  \equiv \int   w_{\delta}({\bf r})w_{p}({\bf r})\min\{\bar{n}_{\delta}({\bf r}),\bar{n}_p({\bf r})\} \langle v({\bf r}) \rangle
 L^*_{\ell}(\hat{{\bf k}}\cdot\hat{{\bf r}})  d^3r =  \frac{A_{\delta}A_pV  }{2\ell+1} \times \mathscr{N}^{\delta p}_{\ell}
\ee 
where the shot-noise term $\mathscr{N}^{\delta p}_{\ell}$ is formally introduced as Eq.\ref{Pnoisedp} of this paper, 
allowing us to simplify Eq.\ref{FFest} into
\be \label{FFxi}  
\begin{split}
    \langle|F^{\delta}(k)F^{p *}_{\ell}(k)|\rangle =&\frac{1}{ {A_{\delta}A_{p}}} \frac{2\ell+1}{V}\int\frac{d \Omega_k}{4\pi}\Big[\int d^3r \frac{1}{V'}\int  d^3r'  w_{\delta}({\bf r})w_{p}({\bf r}')\bar{n}_{\delta}({\bf r})\bar{n}_p({\bf r}')\xi_{\delta p}(|{\bf r}-{\bf r}'|) L^*_{\ell}(\hat{{\bf k}}\cdot\hat{{\bf r}'})  e^{i {\bf k} \cdot ({\bf r}-{\bf r}') } +   N^{\delta p}     \Big]
\end{split}
\ee 
The two-point correlation function maintains an intrinsic relationship with the power spectrum through
\be 
\xi_{\delta p}(|{\bf r}-{\bf r}'|)\equiv\frac{1}{(2\pi)^3}\int P^{\delta p}({\bf k})e^{-i {\bf k} \cdot ({\bf r}-{\bf r}') } d^3k
\ee
Incorporating this relationship into Eq.\ref{FFxi} to substitute $\xi_{\delta p}$ leads to
\be  \label{iugh87}
\begin{split}
    &\langle|F^{\delta}(k)F^{p *}_{\ell}(k)|\rangle \\
    =&\frac{1}{ {A_{\delta}A_{p}}} \frac{2\ell+1}{V}\int\frac{d \Omega_k}{4\pi}\Big[\int d^3r \frac{1}{V'}\int  d^3r'  w_{\delta}({\bf r})w_{p}({\bf r}')\bar{n}_{\delta}({\bf r})\bar{n}_p({\bf r}') \Big( \int P^{\delta p}({\bf k}')e^{-i {\bf k}' \cdot ({\bf r}-{\bf r}') } \frac{d^3k'}{(2\pi)^3} \Big) L^*_{\ell}(\hat{{\bf k}}\cdot\hat{{\bf r}'})  e^{i {\bf k} \cdot ({\bf r}-{\bf r}') }  + N^{\delta p}  \Big]\\
    =&\frac{1}{ {A_{\delta}A_{p}}} \frac{2\ell+1}{V}\int\frac{d \Omega_k}{4\pi}\Big[\Big(\int   w_{\delta}({\bf r})\bar{n}_{\delta}({\bf r}) e^{i    ({\bf k}-{\bf k}')\cdot {\bf r} } d^3r \Big)\frac{1}{V'}\int   d^3r'   \int \frac{d^3k'}{(2\pi)^3} P^{\delta p}({\bf k}')     L^*_{\ell}(\hat{{\bf k}}\cdot\hat{{\bf r}'})  e^{-i   ({\bf k}-{\bf k}')\cdot{\bf r}' }w_{p}({\bf r}')\bar{n}_p({\bf r}') + N^{\delta p} \Big]\\
\end{split}
\ee  
We define the function 
\be 
G^{\delta}({\bf k}-{\bf k}')\equiv \int   w_{\delta}({\bf r})\bar{n}_{\delta}({\bf r}) e^{i    ({\bf k}-{\bf k}')\cdot {\bf r} } d^3r
\ee
as proposed in Eq. \ref{nbfft},  
Eq.\ref{iugh87} can be reduced to
\be  
\begin{split}
    \langle|F^{\delta}(k)F^{p *}_{\ell}(k)|\rangle=&\frac{1}{ {A_{\delta}A_{p}}} \frac{2\ell+1}{V}\int\frac{d \Omega_k}{4\pi}  \Big[ \frac{1}{V'}\int  d^3r'   \int \frac{d^3k'}{(2\pi)^3} P^{\delta p}({\bf k}')     L^*_{\ell}(\hat{{\bf k}}\cdot\hat{{\bf r}'})  e^{-i   ({\bf k}-{\bf k}')\cdot{\bf r}' } w_{p}({\bf r}')\bar{n}_p({\bf r}') G^{\delta}({\bf k}-{\bf k}')+ N^{\delta p}  \Big]\\
\end{split}
\ee  
Expanding the power spectrum $P^{\delta p}({\bf k}')$ in terms of Legendre polynomials using Eq.\ref{plkest} of this paper, we derive
\be  \label{r232gg}
\begin{split}
    &\langle|F^{\delta}(k)F^{p *}_{\ell}(k)|\rangle\\
    =&\frac{1}{ {A_{\delta}A_{p}}} \frac{2\ell+1}{V}\int\frac{d \Omega_k}{4\pi}  \Big[\frac{1}{V'} \int  d^3r'   \int \frac{d^3k'}{(2\pi)^3} \Big(\sum_{\ell'}P^{\delta p}_{\ell'}(k' )L_{\ell'}(\hat{{\bf k}}'\cdot\hat{{\bf r}'})\Big)    L^*_{\ell}(\hat{{\bf k}}\cdot\hat{{\bf r}'})  e^{-i   ({\bf k}-{\bf k}')\cdot{\bf r}' } w_{p}({\bf r}')\bar{n}_p({\bf r}') G^{\delta}({\bf k}-{\bf k}')+ N^{\delta p}  \Big]\\
    =&\frac{2\ell+1}{ {A_{\delta}A_{p}}V} \int\frac{d \Omega_k}{4\pi}  \bigg[    \sum_{\ell'} \int \frac{d^3k'}{(2\pi)^3}   P^{\delta p}_{\ell'}(k' ) G^{\delta}({\bf k}-{\bf k}') \Big(\frac{1}{V'}\int  d^3r' w_{p}({\bf r}')\bar{n}_p({\bf r}') L_{\ell'}(\hat{{\bf k}}'\cdot\hat{{\bf r}'}) L^*_{\ell}(\hat{{\bf k}}\cdot\hat{{\bf r}'})  e^{-i   ({\bf k}-{\bf k}')\cdot{\bf r}' } \Big)+ N^{\delta p}  \bigg] \\
\end{split}
\ee  
Further defining the function
\be \label{yui789}
S^{p}_{\ell\ell'}({\bf k},{\bf k}')\equiv\frac{1}{V}\int   w_{p}({\bf r})\bar{n}_p({\bf r})  L_{\ell}(\hat{{\bf k}}\cdot\hat{{\bf r} })L^*_{\ell'}(\hat{{\bf k}}'\cdot\hat{{\bf r} })  e^{i   ({\bf k}-{\bf k}')\cdot{\bf r}  } d^3r
\ee 
 the conjugate form of the above equation enables us to reformulate Eq.\ref{r232gg} as
\be  \label{estpsss123}
\begin{split}
    \langle|F^{\delta}(k)F^{p *}_{\ell}(k)|\rangle=\frac{2\ell+1}{ {A_{\delta}A_{p}}V} \int\frac{d \Omega_k}{4\pi} \Big[      \sum_{\ell'} \int \frac{d^3k'}{(2\pi)^3}   P^{\delta p}_{\ell'}(k' ) G^{\delta}({\bf k}-{\bf k}') S^{p*}_{\ell\ell'}({\bf k},{\bf k}')  + N^{\delta p}  \Big]
\end{split}
\ee
which is similar to Equation 15 of \papIII\footnote{Equation 15 of \papIII ~is not aligns precisely with Eq.\ref{estpsss123} due to their definition of Fourier transformation diverges from our
definition Eq.\ref{A0}}. We define the window function as
\be  \label{windfwer}
 {\bf W} \equiv  \frac{2\ell+1}{ {A_{\delta}A_{p}}V} \int\frac{d \Omega_k}{4\pi} \Big[      \sum_{\ell'} \int \frac{d^3k'}{(2\pi)^3}    G^{\delta}({\bf k}-{\bf k}') S^{p*}_{\ell\ell'}({\bf k},{\bf k}')   \Big] 
\ee
in the absence of this component, Eq.\ref{estpsss123} collapses 
to  
\be  
\langle |F^{\delta}(k)F_{\ell}^{p *}(k)| \rangle=
P^{\delta p}_{\ell}+\frac{(2\ell+1)N^{\delta p} }{ {A_{\delta}A_{p}}V} = P^{\delta p}_{\ell}+\mathscr{N}^{\delta p}_{\ell}
\ee
i.e. we can formally express our estimator of the
density-momentum cross power spectrum as $   
P^{\delta p}_{\ell} =   |F^{\delta}(k)F_{\ell}^{p *}(k)|  - \mathscr{N}^{\delta p}_{\ell} 
$ with $|F^{\delta}(k)F_{\ell}^{p *}(k)|$ given by Eq.\ref{jhgd78}. Consequently, we re-formulate our estimator  as 
\be  \label{xf788q} 
P^{\delta p}_{\ell}=\frac{1}{2}\mathrm{Im}\{|F^{p}(k)F_{\ell}^{\delta *}(k)|-|F^{\delta}(k)F_{\ell}^{p *}(k)| \}-\mathscr{N}^{\delta p}_{\ell}
\ee  
ensuring its symmetry under $P^{\delta p}_{\ell}=-P^{\delta p *}_{\ell}$ as established in Appendix \ref{sec:symetric}. 

\subsection{Window function}\label{sec:CRSest2}

 We can expand the Legendre functions in terms of spherical harmonics following the approach outlined in \cite{Blake2018}
\be  
L_{\ell}(\hat{{\bf k}}\cdot\hat{{\bf r} })=\frac{4\pi}{2\ell+1}\sum^{\ell}_{m=-\ell}Y^{\ell *}_{m}(\hat{{\bf k}})Y^{\ell}_{m}(\hat{{\bf r}})
\ee 
Substituting this expansion and its conjugate into Eq.\ref{yui789} yields
\be 
S^{p}_{\ell\ell'}({\bf k},{\bf k}')=\frac{4\pi}{2\ell+1}\frac{4\pi}{2\ell'+1}\sum^{\ell}_{m=-\ell}Y^{\ell *}_{m}(\hat{{\bf k}})\sum^{\ell'}_{m'=-\ell'} Y^{\ell'}_{m'}(\hat{{\bf k}}') \tilde{S}^{p,~\ell\ell'  }_{mm'}( {\bf k} ) 
\ee 
where $\tilde{S}^{\ell\ell'  }_{mm'}( {\bf k} )$ is defined by Eq.\ref{Smmll} of this paper. Substituting this result back into Eq.\ref{windfwer} leads to
\be \label{aoirzxauz678} 
  {\bf W} =\frac{1}{ {A_{\delta}A_{p}}} \int\frac{d \Omega_k}{4\pi} \sum^{\ell}_{m=-\ell}Y^{\ell }_{m}(\hat{{\bf k}})   \sum_{\ell'}\frac{(4\pi)^2}{2\ell'+1} \sum^{\ell'}_{m'=-\ell'}\Big[   \frac{1}{V}    \int \frac{d^3k'}{(2\pi)^3}  Y^{\ell'*}_{m'}(\hat{{\bf k}}')  G^{\delta}({\bf k}-{\bf k}')    \tilde{S}^{p,~\ell\ell' \ast}_{mm'}( {\bf k} )  \Big] 
\ee
which corresponds to Eq.\ref{winmat} of this paper, and serves as the foundation for computing the window function convolution matrix. 
The aforementioned expression of ${\bf W}$ corresponds to the term $|F^{\delta}(k)F^{p *}_{\ell}(k)|$. However, as suggested by \cite{Blake2018}, in order to derive the corresponding ${\bf W}$ that conforms to the format specified in Eq.\ref{xf788q}, we just need to substitute the term $G^{\delta}({\bf k}-{\bf k}')\tilde{S}^{p,~\ell\ell' \ast}_{mm'}( {\bf k} ) $ in the above equation with Eq.\ref{symictrinS}.
Notably, the factor $V$ in Equation 19 of \cite{Blake2018} (and Equation 38 of \papIII) is differ from the above Eq.\ref{aoirzxauz678}  due to their definition of Fourier transformation and $G^{\delta}$ diverges from our definition presented in Eq.\ref{A0} and \ref{nbfft}. 

 \subsection{The expression of multipoles}\label{sec:CRSest3}
To illustrate the methodology for computing Eq.\ref{P0k} to \ref{P3k}, we consider the $\ell=4$ auto-power spectrum as a representative example for computing the estimators defined in Eq.\ref{P4k}. In this specific case, Eq.\ref{jhgd78} simplifies to 
\be   
\begin{split}
    &|F(k)F^{*}_4(k)| \\
    =&\frac{9}{V}\int\frac{d \Omega_k}{4\pi}\int d^3r \frac{1}{V'}\int  d^3r' F({\bf r})F^{*}({\bf r}') \times \frac{1}{8} \Big(35 (\hat{{\bf k}}\cdot\hat{{\bf r}'})^4-30 (\hat{{\bf k}}\cdot\hat{{\bf r}'})^2+3  (\hat{{\bf k}}\cdot\hat{{\bf r}'})^0\Big) \times e^{i {\bf k} \cdot ({\bf r}-{\bf r}') }\\
    =&\frac{9}{8}\int\frac{d \Omega_k}{4\pi}\bigg(\frac{1}{V}\int F({\bf r})e^{i {\bf k} \cdot {\bf r}} d^3r \bigg) \frac{1}{V'}\int  d^3r'  F^{*}({\bf r}')  \Big(35 (\hat{{\bf k}}\cdot\hat{{\bf r}'})^4-30 (\hat{{\bf k}}\cdot\hat{{\bf r}'})^2+3  (\hat{{\bf k}}\cdot\hat{{\bf r}'})^0\Big)  e^{-i {\bf k} \cdot {\bf r}' }
\end{split}
\ee
By applying Eq.\ref{A0}, the resulting expression becomes
\be   
\begin{split}
    &|F(k)F^{*}_4(k)| \\
    =&\frac{9}{8}\int\frac{d \Omega_k}{4\pi}F({\bf k})\frac{1}{V'} \int  d^3r'  F^{*}({\bf r}')  \Big(35 (\hat{{\bf k}}\cdot\hat{{\bf r}'})^4-30 (\hat{{\bf k}}\cdot\hat{{\bf r}'})^2+3  (\hat{{\bf k}}\cdot\hat{{\bf r}'})^0\Big)  e^{-i {\bf k} \cdot {\bf r}' }\\
    =&\frac{9}{8V'}\int\frac{d \Omega_k}{4\pi}F({\bf k})      \bigg[35\int   (\hat{{\bf k}}\cdot\hat{{\bf r}'})^4F^{*}({\bf r}')e^{-i {\bf k} \cdot {\bf r}'  }d^3r'-30\int  (\hat{{\bf k}}\cdot\hat{{\bf r}'})^2F^{*}({\bf r}')e^{-i {\bf k} \cdot {\bf r}'  }d^3r'+3  \int (\hat{{\bf k}}\cdot\hat{{\bf r}'})^0F^{*}({\bf r}') e^{-i {\bf k} \cdot {\bf r}'  }d^3r'\bigg ] 
\end{split}
\ee
Further incorporating the conjugate form of Eq.\ref{sdgdef123} for the cases where $\ell=0,2,4$, this expression reduces to
\be   
\begin{split}
    |F(k)F^{*}_4(k)| =\frac{9}{8V'}\int\frac{d \Omega_k}{4\pi}F({\bf k}) [35T^*_4({\bf k})-30 T^*_2({\bf k})+3T^*_0({\bf k})]
\end{split}
\ee
Substituting this expansion  
into Eq.\ref{autops} or Eq.\ref{autopss} yields
\be  
P_{4}(k)=\frac{9}{8V'}\int\frac{d \Omega_k}{4\pi}F({\bf k}) [35T^*_4({\bf k})-30 T^*_2({\bf k})+3T^*_0({\bf k})]-\mathscr{N}_{4} 
\ee 
with $\mathscr{N}_4\approx0$, the above equation collapses precisely to Eq.\ref{P4k}.

\section{Data and Code Availability}\label{sec:appsdf1233}

We developed a set of code for the power spectrum measurements, modeling and window function convolution. This package is called `{\color{cyan}CosmPS}', the GitHub link is  \url{https://github.com/FeiQin-cosmologist/Galaxy_Power_Spectrum} . We have PYTHON version `CosmPSPy' and FORTRAN version `CosmPSFt', respectively. We also have a NERSC directory for the `CosmPSPy'  package, which is {\color{cyan} /global/cfs/cdirs/desi/science/td/pv/PS/CosmPSPy } .  

\ 

1: The `{\color{cyan}CosmPSPy}' package can be found in \url{https://github.com/FeiQin-cosmologist/Galaxy_Power_Spectrum/tree/main/CosmPSPy}  The Jupyter Notebook for the examples of the utilization of the `CosmPSPy' package (including power spectrum measurements, modeling and window function convolution)  can be found in 
\url{https://github.com/FeiQin-cosmologist/Galaxy_Power_Spectrum/blob/main/CosmPSPy/Code/Examp_PS.ipynb}

\

2: The `{\color{cyan}CosmPSFt}' package, i.e. the FORTRAN  code used to compute the measured power spectrum can be found in 
\url{https://github.com/FeiQin-cosmologist/Galaxy_Power_Spectrum/tree/main/CosmPSFt} 

\ 

3: The hand-written draft for the derivation of the power spectrum estimator and window function convolution in Appendix \ref{psest879}
can be found in 
\url{https://github.com/FeiQin-cosmologist/Galaxy_Power_Spectrum/blob/main/Math/1%20PS%20estimator.pdf} 

\

4: The Mathematica code for the derivation of the perturbation models of power spectrum in Section \ref{sec:psmodel}  can be found in 
\url{https://github.com/FeiQin-cosmologist/Galaxy_Power_Spectrum/blob/main/Math/2%20modPS.nb} or see the `.pdf' version \url{https://github.com/FeiQin-cosmologist/Galaxy_Power_Spectrum/blob/main/Math/3%20modPS.pdf} 

\ 

5: The PYTHON module used to compute the perturbation models of power spectrum and window function convolution can be found in 
\url{https://github.com/FeiQin-cosmologist/Galaxy_Power_Spectrum/blob/main/CosmPSPy/Code/PSmodFun.py}  

\

6: The PYTHON module used to compute the measured power spectrum can be found in 
\url{https://github.com/FeiQin-cosmologist/Galaxy_Power_Spectrum/blob/main/CosmPSPy/Code/PSestFun.py}  

\

7: The example PYTHON  code for fiting the growth-rate $f\sigma_8$ from galaxy mocks can be found in 
\url{https://github.com/FeiQin-cosmologist/Galaxy_Power_Spectrum/tree/main/FitExamp}

\end{document}